\documentclass[aps,prd,onecolumn,groupedaddress]{revtex4-2}

\usepackage[T1]{fontenc}        
\usepackage{textcomp}           

\usepackage{graphicx}           
\usepackage{subfigure}          
\usepackage{dcolumn}            
\usepackage{bm}                 
\usepackage{color}              
\usepackage{multirow} 
\usepackage{longtable}          
\usepackage{supertabular}       
\usepackage{epstopdf}           
\usepackage{amsmath}            
\usepackage{float}              
\usepackage{hyperref}           
\usepackage{booktabs}
\usepackage{url}
\usepackage{makecell}
\usepackage{upgreek}
\usepackage{appendix}
\usepackage{xcolor}
\usepackage{colortbl}
\usepackage{array}
\makeatletter

\newcommand{\Rmnum}[1]{\expandafter\@slowromancap\romannumeral #1@}
\makeatother
\begin{document}
\title{Testing the equivalence principle with lunar laser ranging residuals from different planetary and lunar ephemerides}
\author{Jun Ke$^{1}$}
\author{Jie Luo$^{2}$}\email[E-mail: ]{luojiethanks@126.com}
\author{Qin Li$^{1}$}
\author{Yu-Jie Tan$^{1}$}
\author{Cheng-Gang Shao$^{3}$}\email[E-mail: ]{cgshao@hust.edu.cn}

\affiliation{
$^{1}$ National Gravitation Laboratory, MOE Key Laboratory of Fundamental Physical Quantities Measurement, School of Physics, Huazhong University of Science and Technology, Wuhan 430074,  People's Republic of China\\
$^{2}$ School of Mechanical Engineering and Electronic Information, China University of Geosciences, Wuhan 430074, People's Republic of China\\
$^{3}$ School of Physics and Optoelectronic, Yangtze University, Jingzhou 434023, People's Republic of China\\
}

\date{\today}
\begin{abstract}
Lunar laser ranging (LLR) provides a sensitive test of the equivalence principle (EP) through a possible synodic perturbation in the Earth--Moon distance. In this work, we develop an independent LLR data-reduction framework and investigate this signature in the post-fit O--C residuals obtained with the EPM21, INPOP21, and DE430 planetary and lunar ephemerides. The same observation model and parameter-adjustment procedure are applied to the three ephemerides, allowing their residual-based results to be compared within a consistent framework. To account for the nonuniform distribution of LLR observations over the synodic angle $D$, the residuals are fitted with cosine and sine harmonics up to the third order. Signal-injection tests are also performed to quantify the attenuation of a possible $\cos D$ signal during the parameter adjustment. After correcting for signal attenuation, solar radiation pressure, and the thermal response of the lunar retroreflectors, the preferred EPM21, INPOP21, and DE430 solutions yield $\Delta(m_g/m_i)_{\rm EM}=(0.900\pm3.183)\times10^{-14}$, $(0.496\pm3.624)\times10^{-14}$, and $(2.669\pm7.048)\times10^{-14}$, respectively. The EPM21 and INPOP21 results are based on observations from 1970 to 2024, whereas the DE430 solution was obtained using observations up to 2016. All three estimates are consistent with zero within their uncertainties. This work demonstrates that residual-based analysis can provide an independent and transparent EP test, while also offering guidance for assessing sampling and ephemeris-dependent effects in future high-precision LLR studies.
\end{abstract}
\maketitle


\section{INTRODUCTION}\label{section1}
The equivalence of gravitational mass and inertial mass, generally called the equivalence principle (EP), is the foundation of a wide class of gravitational theories including Einstein’s theory of general relativity \cite{Will2014}. However, the EP is predicted to be violated in the presence of possible new particles or interactions, such as composition-dependent forces \cite{Fischbach1986}, dilaton in string theory \cite{Damour2012}, bosons in supersymmetric theories \cite{Fayet2018}, and dark matter \cite{Graham2016,Carroll2009}. Therefore, any experimental test of EP violation has the potential to provide insight into new physics. A possible violation of the EP is usually characterized by the difference between the free-fall accelerations of two test bodies in the same external gravitational field. Up to now, the EP has been tested through a wide variety of laboratory, space-based, and astronomical experiments with continuously improving precision \cite{Schlamminger2008,Wagner2012,Zhu2018,Ross2026,Touboul2017,Touboul2022,Shapiro1976,Williams1976,Dickey1994,Williams1996,Anderson2001, Muller1998,Williams2004a,Williams2004b,Williams2009,Muller2012a,Williams2012,Merkowitz2010,Hofmann2010,Muller2014,Biskupek2015,Pitjeva2014,Biskupek2015,Fienga2015,Hofmann2017,Viswanathan2018,Hofmann2018,Muller2019,
ViswanathanPhd,Pitjeva2019,Zhang2020,Pavlov2020,Pavlov2024,Biskupek2021,Zhang2022,Singh2023,Pavlov2023,Zhang2024,Fienga2024,Mariani2024}. Rotating torsion-balance experiments have constrained the EP violation parameter at the level of ${10^{-13}}$ \cite{Schlamminger2008,Wagner2012,Zhu2018,Ross2026}, while the MICROSCOPE space mission has improved the precision to the ${10^{-15}}$ level \cite{Touboul2017,Touboul2022}. For laboratory-sized bodies, the contribution of gravitational self-energy is negligible compared with their total mass-energy. Consequently, the above experiments are primarily sensitive to possible violations arising from differences in composition and internal structure, i.e., the weak equivalence principle (WEP) \cite{Nordtvedt1968a}.

In contrast, Lunar Laser Ranging (LLR) provides a unique opportunity to test both the WEP and the strong equivalence principle (SEP) in the Earth–Moon system, owing to the different compositions and gravitational self-energies of the Earth and Moon \cite{Nordtvedt1968a,Nordtvedt1995,Nordtvedt1968b}. Since the first LLR-based EP tests were reported in 1976 \cite{Shapiro1976,Williams1976}, continued improvements in ranging precision and dynamical modeling have increased the sensitivity of these tests by more than two orders of magnitude \cite{Williams2012,Viswanathan2018,Biskupek2021,Pavlov2023}. Major contributions have been made by several independent groups, including the Jet Propulsion Laboratory (JPL) \cite{Williams1976,Dickey1994,Williams1996,Anderson2001,Williams2004a,Williams2004b,Williams2009,Williams2012}, Institut für Erdmessung (IfE) \cite{Muller1998,Hofmann2010,Muller2012a,Muller2014,Biskupek2015,Hofmann2017,Hofmann2018,Zhang2020,Biskupek2021,Zhang2022,Muller2019,Singh2023,Zhang2024},  Institute of Celestial Mechanics and Ephemeris Computations (IMCCE) and Observatoire de la Côte d’Azur (OCA) \cite{Fienga2015,ViswanathanPhd,Viswanathan2018,Fienga2024,Mariani2024}, and Institute of Applied Astronomy of the Russian Academy of Sciences (IAA RAS) \cite{Pitjeva2014,Pitjeva2019,Pavlov2020,Pavlov2023,Pavlov2024}. The latest LLR analysis has constrained the EP violation parameter to ${(-2.1 \pm 2.4) \times 10^{-14}}$, corresponding to a millimeter-level signature in the Earth–Moon range \cite{Biskupek2021}. 

At present, LLR-based tests of the EP are generally performed using two complementary approaches, which differ in whether the EP parameter is included among the solve-for quantities in the least-squares adjustment. The first approach, which is the most widely adopted, estimates the EP-violation parameter simultaneously with other dynamical model parameters in a global least-squares adjustment \cite{Williams2012, Hofmann2018,Viswanathan2018,Pavlov2023}. In this approach, the correlations between the EP parameter and the other solve-for parameters can be directly evaluated from the resulting covariance or correlation matrix. However, the implementation of this approach requires careful treatment of the correlations between the EP-violation parameter and the other solve-for parameters to avoid unstable or physically implausible estimates. Most published LLR constraints on EP violations, including those reported by the JPL, IMCCE, OCA, IfE, and IAA RAS groups, have been derived using this strategy. The second approach directly extracts the characteristic synodic signature of a possible EP violation from post-fit LLR residuals obtained under the assumption that the EP is valid. The underlying basis is that an EP-violating differential acceleration of the Earth and Moon toward the Sun would polarize the lunar orbit and induce a synodic periodic radial variation in the Earth–Moon distance \cite{Nordtvedt1968b}. This approach therefore provides a complementary means of investigating possible EP violations through the analysis of post-fit residuals. Previous studies have shown that, although part of the signal may be absorbed by correlated model parameters, a detectable synodic signature still remains in the post-fit residuals \cite{Hofmann2018,Pitjeva2019}. The IfE group obtained a synodic-period signature associated with galactic dark matter from post-fit LLR residuals, further demonstrating the feasibility of the residual-based strategy \cite{Zhang2020}. 

In addition, current LLR constraints on EP violations are generally obtained by individual research groups using their own analysis frameworks. Although these results are all consistent with no detectable violation, their mutual consistency within a common framework has not been fully assessed. This issue becomes increasingly important as the sensitivity of modern LLR tests approaches the millimeter level \cite{Murphy2012,Murphy2013}. Because the expected EP signal itself corresponds to only a millimeter-scale perturbation in the Earth--Moon range, even small differences among modern lunar ephemerides may affect the inferred result \cite{Vokrouhlicky1997,Nordtvedt1998,Anderson2001}. A residual-based analysis therefore provides a useful way to identify ephemeris-dependent signatures and other unmodeled systematic effects. Three widely used planetary and lunar ephemeris series are the Development Ephemeris (DE) maintained by JPL, the Ephemerides of Planets and the Moon (EPM) developed by IAA RAS, and the Intégrateur Numérique Planétaire de l’Observatoire de Paris (INPOP) developed by IMCCE and OCA. Since these ephemerides are constructed by different groups using different observational data, dynamical models, and parameter-estimation strategies, a comparison among them within a common LLR framework provides a direct test of the robustness of the derived EP constraints.

Motivated by these considerations, we apply a common LLR data-reduction and residual-analysis procedure to three independently developed ephemerides: DE430 \cite{Folkner2014}, EPM21 \cite{Kan2021}, and INPOP21 \cite{Fienga2021}. For each solution, the post-fit O--C residuals are generated using the same observation model, weighting strategy, and parameter-adjustment procedure, and the synodic $\cos D$ signature is extracted using harmonic analysis. This allows the consistency of the EP-related results to be assessed directly and reveals how strongly they depend on the adopted ephemeris. As shown in the following analysis, the residuals obtained from different ephemerides exhibit millimeter-level differences in the synodic $\cos D$ term. These differences are largely related to the different treatments of solar radiation pressure and other non-EP synodic effects in the ephemerides. After applying the corresponding corrections, the three solutions give mutually consistent results and show no evidence for an EP-violating signal. This consistency supports the usefulness of the common residual-based framework for identifying ephemeris-dependent effects in LLR analyses. More broadly, the resulting framework is relevant to the analysis of future high-precision LLR observations  \cite{Murphy2012,Murphy2013,Huang2024}, including those from the Chang'e program \cite{Zhou2022}, the NGLR-1 retroreflector deployed by the Blue Ghost mission \cite{Williams2023,Yu2026}, and next-generation hollow corner-cube reflectors \cite{He2018,Cao2024}.

The remainder of this paper is organized as follows. Sec. \ref{section2} introduces the theoretical background of the EP test and the LLR observations used in this study. Sec. \ref{section3} describes the data reduction procedure and the adopted dynamical and measurement models. Sec. \ref{section4} presents the residual analysis and the extraction of the EP signature using different lunar ephemerides, together with a comparison of the resulting constraints. Finally, Sec. \ref{section5} summarizes the main conclusions of this work.

\section{Theoretical Background and LLR Observations}\label{section2}
\subsection{Equivalence Principle in the Earth–Moon System}\label{section2.1}
The EP states that gravitational and inertial properties of a body are equivalent, implying that the free-fall acceleration in an external gravitational field is independent of its physical properties. Gravitational mass ${m_g}$ characterizes the coupling of a body to gravity, while inertial mass ${m_i}$ is introduced through Newton’s second law as a measure of a body’s inertia. Therefore, the validity of the EP requires that the ratio $m_g/m_i $ is equal to 1 for all bodies. In general, a possible violation of the EP can be tested by comparing the free-fall accelerations ${a_1}$ and ${a_2}$ of two bodies (1 and 2) \cite{Nordtvedt1968b}:
\begin{equation}\label{N1}
\frac{{\Delta a}}{a} = \frac{{{a_1} - {a_2}}}{{({a_1} + {a_2})/2}} = \frac{{{{({m_g}/{m_i})}_1} - {{({m_g}/{m_i})}_2}}}{{[{{({m_g}/{m_i})}_1} + {{({m_g}/{m_i})}_2}]/2}} \approx {\left( {\frac{{{m_g}}}{{{m_i}}}} \right)_1} - {\left( {\frac{{{m_g}}}{{{m_i}}}} \right)_2} = \Delta \left(\frac{{{m_g}}}{{{m_i}}}\right),
\end{equation}

If the EP is violated, different test bodies will experience different free-fall accelerations in the same gravitational field, resulting in a non-zero value of ${\Delta a /a}$. In its weak form, the WEP states that the free-fall acceleration of a body is independent of its composition and internal properties. The relevant differences between test bodies include their fractional nuclear-binding energies, neutron-to-proton ratios, atomic charges, etc. The SEP extends the WEP by including the contribution of gravitational self-energy. While this effect is negligible for laboratory-sized bodies, it becomes significant for astronomical bodies. The ratio ${m_g/m_i}$ of a body with mass $M$ can be expressed as:
\begin{equation}\label{N2}
\frac{{{m_g}}}{{{m_i}}} = 1 + \eta \frac{U}{{M{c^2}}},
\end{equation}
where ${\eta}$ is the Nordtvedt parameter, ${U}$ is the body’s gravitational self-energy, and $c$ is the speed of light in vacuum. General relativity predicts ${\eta = 0}$. 

In LLR experiments, the Earth and the Moon are treated as test bodies in the gravitational field of the Sun. Since their compositions and gravitational self-energies are different, LLR provides a combined test of the WEP and SEP. A violation of the EP would imply a difference in the  ${m_g/m_i}$ ratios of the Earth and Moon:
\begin{equation}\label{N3}
\Delta {\left( {\frac{{{m_g}}}{{{m_i}}}} \right)_{EM}} = {\left( {\frac{{{m_g}}}{{{m_i}}}} \right)_E} - {\left( {\frac{{{m_g}}}{{{m_i}}}} \right)_M},
\end{equation}
which results in a differential acceleration toward the Sun. By treating this EP-violating acceleration as a perturbation to the lunar orbit, Nordtvedt demonstrated that it produces a polarization of the lunar orbit along the Sun–Earth direction, leading to a periodic variation in the Earth–Moon range \cite{Nordtvedt1968b}:
\begin{equation}\label{N4}
\Delta {r_{EM}} = S\Delta {\left( {\frac{{{m_g}}}{{{m_i}}}} \right)_{EM}}\cos D,
\end{equation}
where $S$ is the dynamical response coefficient of about ${-2.943 \times 10^{10} }$ m \cite{Williams2009}, $D$ is the synodic angle with a signal period of 29.53 days, which differs from the Moon's sidereal orbital period of 27.3 days. If only the SEP violation is considered, the above expression can be further simplified as:
\begin{equation}\label{N5}
\Delta {r_{EM}} = S\eta \left[ {{{\left( {\frac{U}{{M{c^2}}}} \right)}_E} - {{\left( {\frac{U}{{M{c^2}}}} \right)}_M}} \right]\cos D.
\end{equation}
The difference in the gravitational self-energy between the Earth and Moon is:
\begin{equation}\label{N6}
{\left( {\frac{U}{{M{c^2}}}} \right)_E} - {\left( {\frac{U}{{M{c^2}}}} \right)_M} =  - 4.45 \times {10^{ - 10}},
\end{equation}
Combining Eqs. (\ref{N5}) and (\ref{N6}) yields:
\begin{equation}\label{N7}
\Delta {r_{EM}} = 13.1 \, \mathrm{m} \, \eta \cos D.
\end{equation}

Therefore, the presence of an EP-violation signal is expected to appear as a $\cos D$ signature in the Earth–Moon range, which provides the basis for the subsequent analysis in this work.
\subsection{LLR Observations and Data Description}
\begin{figure*}[htbp]
\includegraphics[width=0.50\textwidth]{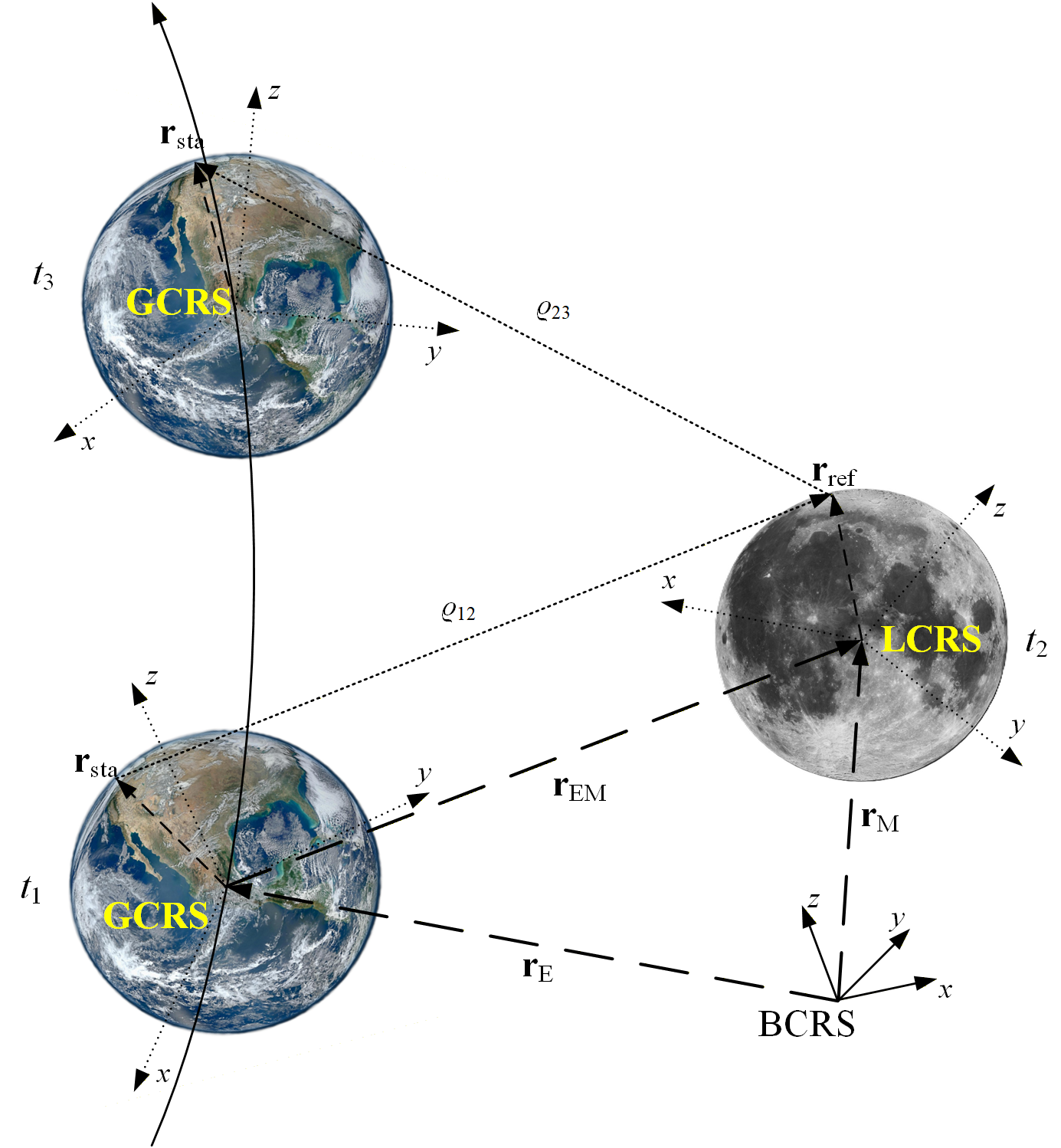}
\caption{Scheme of a Lunar Laser Ranging (LLR) measurement.}
\label{Fig1}
\end{figure*}

LLR is a high-precision technique that measures the Earth–Moon distance by recording the round-trip travel time of laser pulses exchanged between ground stations and lunar retroreflectors. The first successful LLR measurements were obtained shortly after the deployment of retroreflector arrays on the lunar surface during the Apollo missions in the late 1960s and early 1970s \cite{Williams1976}. As illustrated in Fig. \ref{Fig1}, a laser pulse is emitted from a ground station at time $t_1$, reflected by a lunar retroreflector at time $t_2$, and received back at the station at time $t_3$. The measured round-trip light time $t_3- t_1$ constitutes the basic LLR observable. 

\begin{figure*}[htbp]
\centering
\subfigure{%
    \includegraphics[width=3.05in]{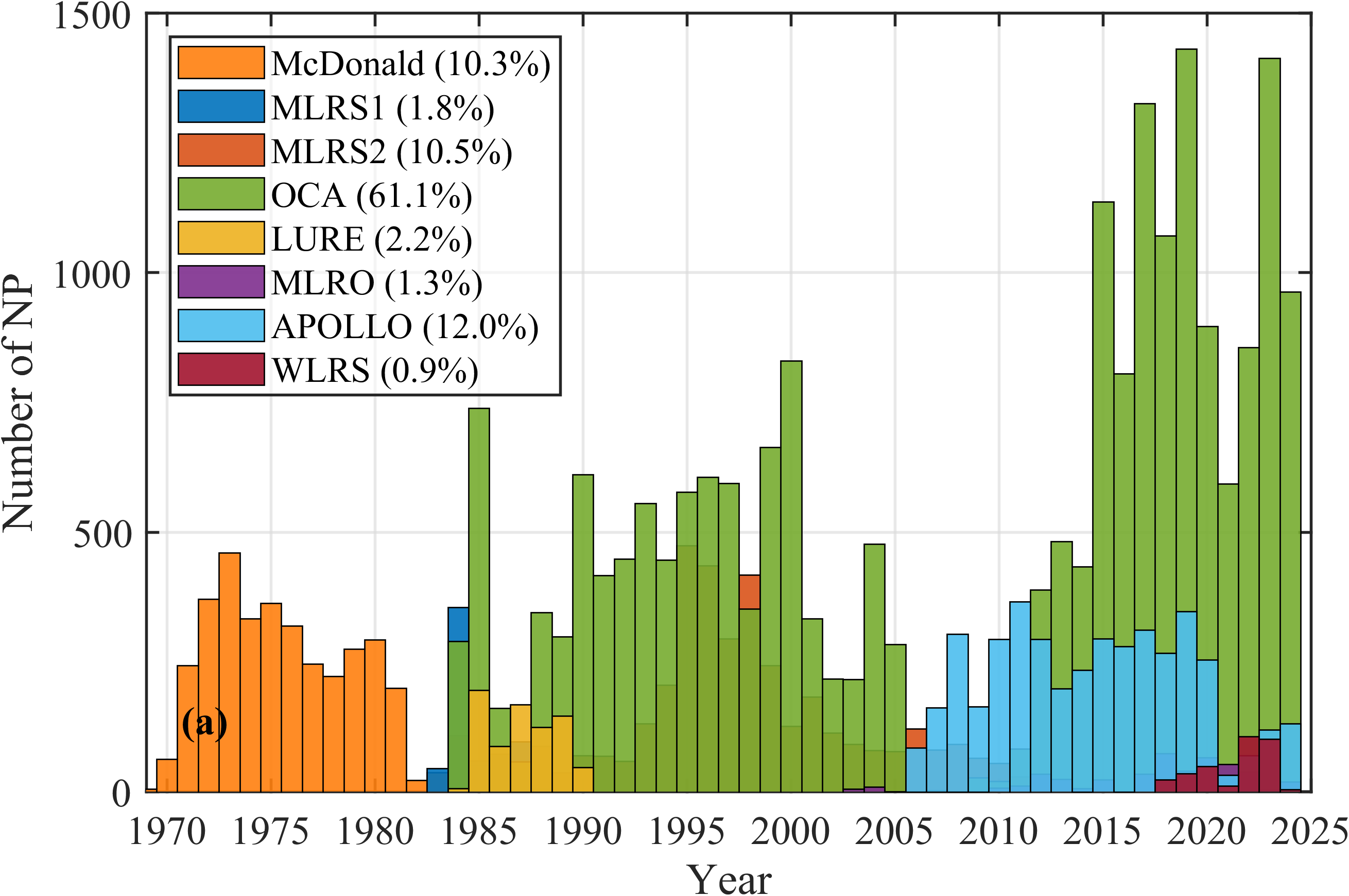}}
\quad
\subfigure{
    \includegraphics[width=3.27in]{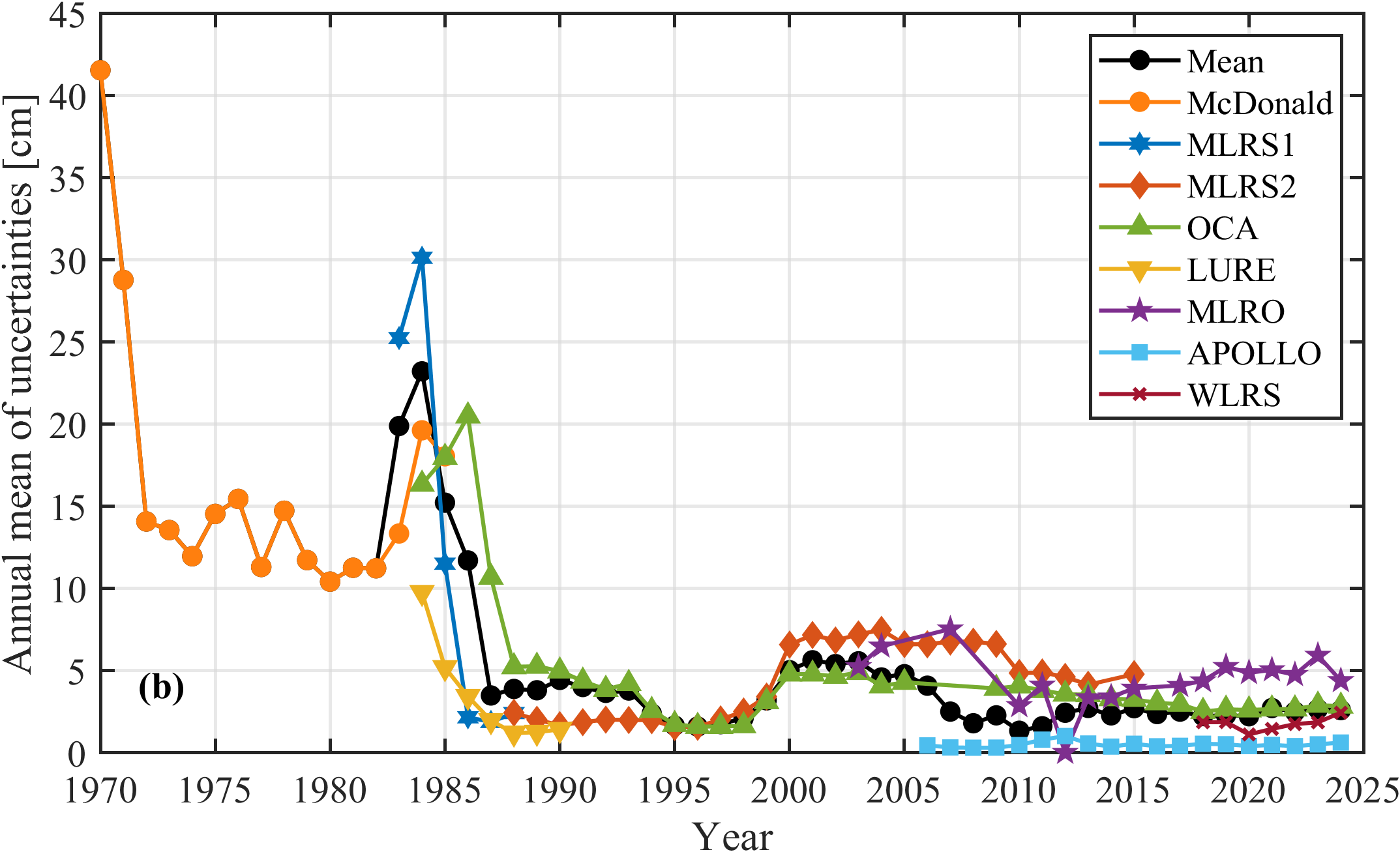}}
\caption{(a) Annual distribution of LLR normal points for different stations. (b) Annual mean one-way ranging uncertainties for different LLR stations.}
\label{Fig2}
\end{figure*}

The LLR normal point observations used in this study are publicly available through the Lunar Analysis Center of the Paris Observatory \cite{OCAwebsite}. All observations were provided with the standard “MINI” format (one line per normal point). Table \ref{T1} summarizes the number of processed normal points and the corresponding observation intervals for each station. The current data set contains 35069 normal points. Among them, the McDonald Laser Ranging Station (MLRS) contributes about 22.5${\%}$ of the observations, the Observatoire de la Côte d’Azur (OCA) contributes 61.1\%, the Apache Point Lunar Laser-ranging Operation (APOLLO) contributes 12.0\%, the Lunar Ranging Experiment (LURE) at the Haleakala Observatory contributes 2.2\%, the Matera Laser Ranging Observatory (MLRO) contributes 1.3\%, and the Wettzell Laser Ranging System (WLRS) contributes 0.9\%. The temporal distribution of this data set is shown in Fig. \ref{Fig2}(a). On the lunar side, the distribution of observations among the reflectors is also inhomogeneous. The Apollo 15 reflector dominates the data set due to its large size, accounting for about 61.7\% of all normal points. The Apollo 11 and Apollo 14 reflectors contribute 12.8\% and 12.0\%. The Lunokhod 1 and Lunokhod 2 reflectors account for 6.5\% and 7.1\%.

Fig. \ref{Fig2} (b) shows the annual distribution of the measurement uncertainties provided with the normal points, illustrating the continuous improvement in ranging precision over time. Since 2006, the annual measurement uncertainties have reached the level of 2 cm or better. The highest ranging precision is achieved by the APOLLO station, where the measurement uncertainties of most normal points are below 1 cm. It should be noted that the uncertainties shown in Fig. \ref{Fig2} (b) are the original values provided with the normal points and have not been rescaled. In fact, the uncertainties (or equivalently, the weights) need to be partially rescaled to eliminate discrepancies within the data set on individual stations and between the stations over time \cite{Viswanathan2018,Pavlov2016}.

\begin{table}[htbp]
\centering
\caption{Observation data sets from different LLR stations}
\label{T1}
\setlength{\tabcolsep}{8pt}   
\renewcommand{\arraystretch}{1.2}
\begin{tabular}{c c c}
\hline
{Station} & {Time span} & {Number of normal points} \\
\hline
McDonald  & 1969--1985 & 3604 \\
MLRS1     & 1983--1988 & 631  \\
MLRS2     & 1988--2013 & 3670 \\
OCA    & 1984--2024 & 21433 \\
LURE & 1984--1990 & 770  \\
MLRO    & 2003--2024 & 439  \\
APOLLO    & 2006--2024 & 4196 \\
WLRS  & 2018--2024 & 326  \\
\hline
\end{tabular}
\end{table}
\section{LLR Data Reduction}\label{section3}
\begin{figure*}[htbp]
\includegraphics[width=0.55\textwidth]{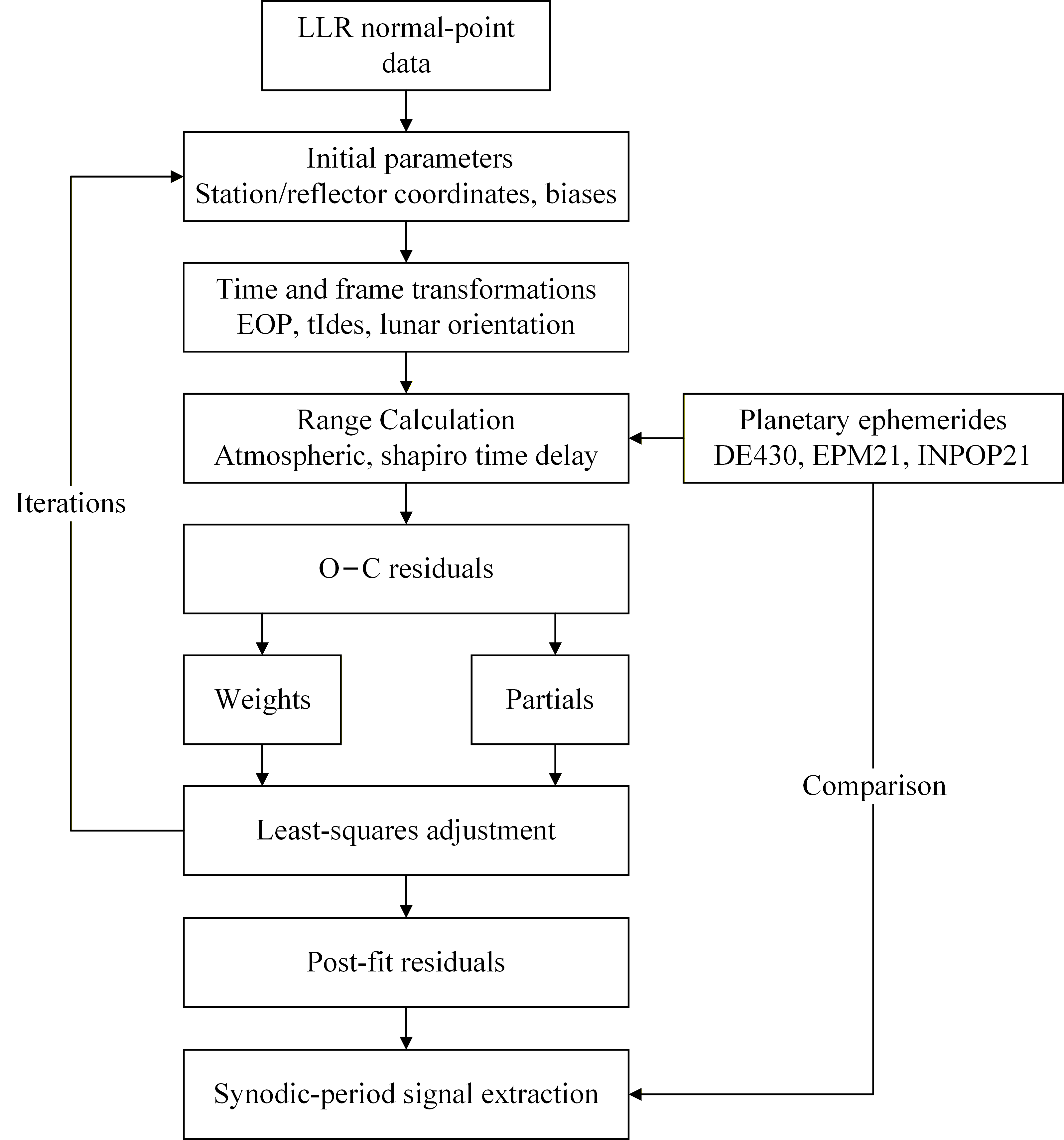}
\caption{Flowchart of the LLR data reduction procedure and subsequent equivalence-principle analysis.}
\label{Fig3}
\end{figure*}

Fig. \ref{Fig3} provides an overview of the LLR data-reduction and EP-analysis procedure adopted in this work. The same processing and parameter-estimation framework is applied to the three ephemerides to obtain post-fit O--C residuals, from which the characteristic synodic signature is subsequently extracted. Because the EP signal sought in this work corresponds to a millimeter-level synodic range perturbation, corrections capable of producing centimeter- or millimeter-level range effects must be included in the data reduction. Obtaining small and well-distributed post-fit residuals is therefore a prerequisite for a reliable extraction of the $\cos D$ coefficient. The main effects entering the light-time calculation and the station/reflector coordinate modeling are described in detail below.  

Each normal point in the standard “MINI” format contains the station identifier, reflector identifier, launch time, observed two-way light travel time (LTT), estimated uncertainty of the two-way LTT, and laser wavelength. Atmospheric pressure, ground temperature, and relative humidity at the station are also provided. Based on the LLR normal-point data, the International Earth Rotation and Reference Systems (IERS) 2010 recommended models \cite{IERS2010}, and the celestial positions provided by the planetary and lunar ephemerides, the computed two-way LTT C can be evaluated by solving the following system of iterative equations:
\begin{equation}\label{N8}
\left\{
  \begin{aligned}
    t_2 - t_1 &= \frac{|\mathbf{r}_{\rm ref}^{\rm BCRS}(t_2) - \mathbf{r}_{\rm sta}^{\rm BCRS}(t_1)|}{c} + \Delta\tau_{\rm grav}(t_1,t_2) + \Delta\tau_{\rm atm}(t_1,t_2) \\
    t_3 - t_2 &= \frac{|\mathbf{r}_{\rm sta}^{\rm BCRS}(t_3) - \mathbf{r}_{\rm ref}^{\rm BCRS}(t_2)|}{c} + \Delta\tau_{\rm grav}(t_3,t_2) + \Delta\tau_{\rm atm}(t_3,t_2)
  \end{aligned}
\right.
\end{equation}
where ${t_1}$, ${t_2}$, and ${t_3}$ are the times of emission, reflection, and reception of the laser pulse in the Barycentric Dynamical Time (TDB) timescale, respectively. ${\mathbf{r}_{{\rm{sta}}}^{{\rm{BCRS}}}({t_i})}$ and ${\mathbf{r}_{{\rm{ref}}}^{{\rm{BCRS}}}({t_i})}$ are the coordinates of the stations and reflectors at time $t_i$ in the Barycentric Celestial Reference System (BCRS). ${\Delta {\tau _{{\rm{grav}}}}({t_1},{t_2})}$ is the Shapiro time delay caused by the gravitational potential of the Sun and the Earth. ${\Delta {\tau _{{\rm{atm}}}}({t_1},{t_2})}$ is the atmospheric propagation delay. ${\varrho _{12}} = | \mathbf{r}_{{\text{ref}}}^{{\text{BCRS}}}({t_2}) -  \mathbf{r}_{{\text{sta}}}^{{\text{BCRS}}}({t_1})|$ denotes the distance between the station at time $t_1$ and reflector at time $t_2$, following the notation in Fig. \ref{Fig1}. 

The launch time recorded in the normal point data is given in Coordinated Universal Time (UTC). Since Eq. (\ref{N8}) is solved in the TDB timescale, the UTC timescale must be transformed into TDB timescale. The latter also serves as the time input of the planetary and lunar ephemerides. The station coordinates in the International Terrestrial Reference System (ITRS) are then corrected for solid Earth tides, ocean tides, polar motion, atmospheric pressure loading, and ocean pole tide loading. Similarly, the reflector coordinates in the lunar principal axis system (PAS) are corrected for lunar solid tides. The corrected station coordinates are transformed from the ITRS into the BCRS, while the corrected reflector coordinates are transformed from the PAS into the BCRS. After applying atmospheric and Shapiro time delay corrections, the epochs $t_2$ and $t_3$ in TDB can be determined iteratively using Eq. (\ref{N8}). In practice, the iterative solution converges within three to four iterations. Finally, $t_3$ is transformed back from TDB to UTC, and the computed two-way LTT C is obtained as ${\mathrm{C} = t_{3,UTC}- t_{1,UTC}}$.

\subsection{Transformation between time scales}
As mentioned before, the evaluation of Eq. (\ref{N8}) involves several time scales, including UTC, TDB, Terrestrial Time (TT), International Atomic Time (TAI), and Universal Time (UT1). The transformation procedures between these time scales are presented in this subsection. LLR measurements have been conducted since 1969. During the period from 1969 to 1972, the difference between UTC and TAI can be written as: 
\begin{equation}\label{N9}
\Delta AT = {\left( {TAI - UTC} \right)_s} = 4.21317 + 0.002592 \times ({\rm{MJD}} - 39126),
\end{equation}
indicating that UTC ran faster than TAI by a factor of ${(1+3 \times 10^{-8})}$. UTC has differed from TAI only by an integer number of leap seconds after 1972, which can be obtained from the IERS leap-second tables \cite{{IERS2010}}. 

\begin{figure*}[htbp]
\centering
\subfigure{%
    \includegraphics[width=3.05in]{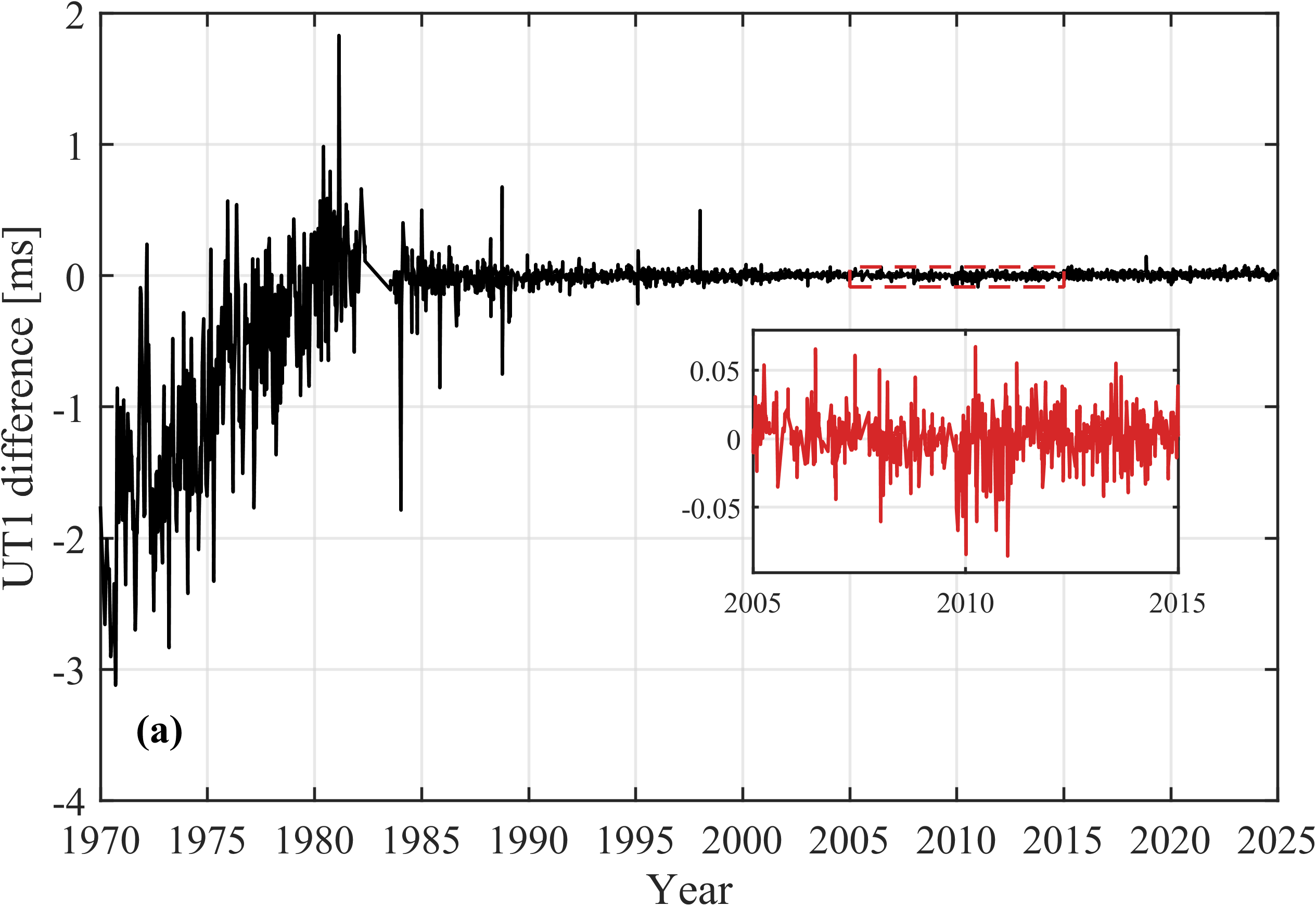}}
\quad
\subfigure{
    \includegraphics[width=3.15in]{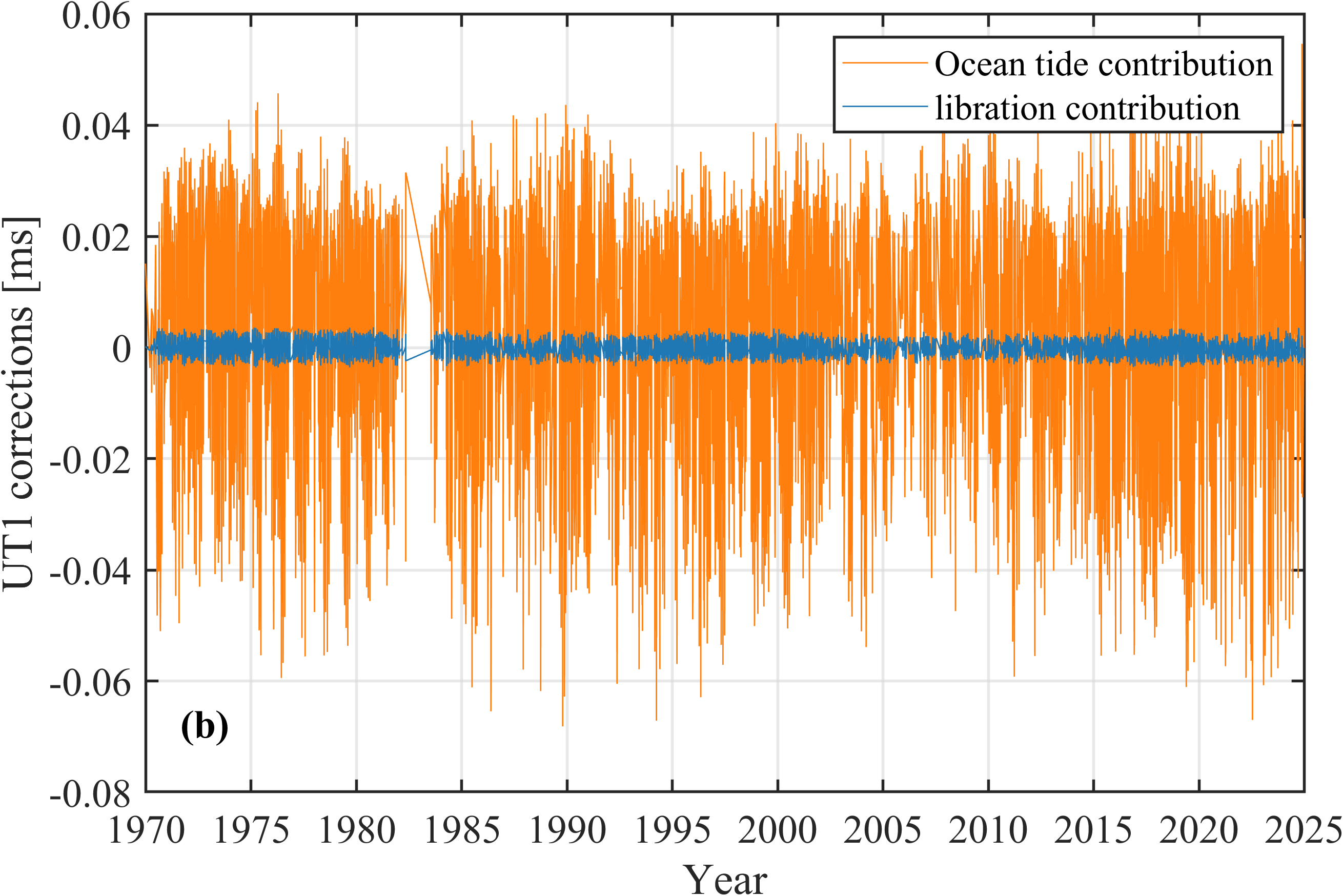}}
\caption{(a) Difference in UT1 between the JPL EOP2 and IERS EOP C04 series. (b) Corrections to UT1 due to ocean tides and libration effects.}
\label{Fig4}
\end{figure*}
For the difference between UT1 and TAI, the IERS Earth Orientation Parameters (EOP) C04 series \cite{Bizouard2019,C04} and the JPL EOP2 files \cite{Chin2009,eop2} both provide daily values of this difference since 1962. Therefore, we can obtain the corresponding values at the target times by using a 4th-order Lagrange interpolation. Similar to the results reported in Ref. \cite{Pavlov2016}, the JPL EOP2 series performs better than the IERS EOP C04 series in the processing of LLR data. This does not mean that EOP2 is more accurate than C04. Instead, the EOP2 solution appears to be more compatible with LLR observations. A major reason is that LLR observations were included in the construction of the JPL EOP2 series, particularly in the determination of the variation in latitude (VOL) and UT0 parameters. For data before 1984, the one-way LTT WRMS obtained with JPL EOP2 is about 10 cm smaller than that obtained with the IERS EOP C04 series. For observations after 1984, the improvement varies with station and observation interval, and the one-way LTT WRMS is typically reduced by about 0.2 mm to 1 cm. Based on these results, the JPL EOP2 series is adopted in this work. Fig. \ref{Fig4} (a) presents the differences between the UT1 values obtained from the JPL EOP2 and IERS EOP C04 series, with the horizontal axis representing the emission time of the normal points. The results further support the conclusion discussed above. In addition, the effects of ocean tides and libration on the UT1 should also be taken into account. Based on the IERS subroutines \cite{{IERS2010}}, these effects can be evaluated, as illustrated in Fig. \ref{Fig4} (b). It can be seen that the ocean-tide contribution is at the level of 40 \textmu s, comparable to the differences after 1984 shown in Fig. \ref{Fig4} (a), while the libration contribution is smaller than 3 \textmu s.

\begin{figure*}[htbp]
\centering
\subfigure{%
    \includegraphics[width=3.25in]{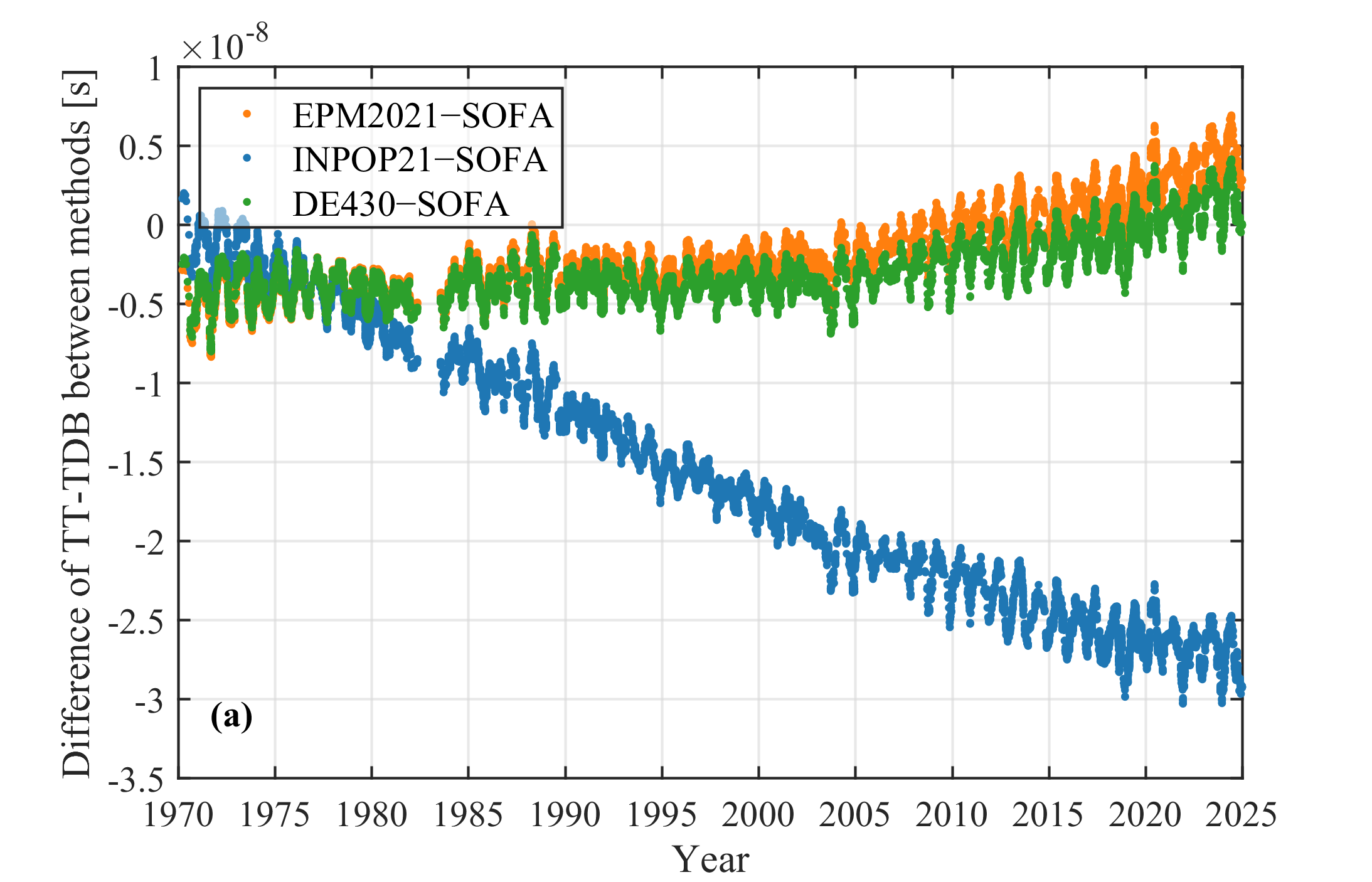}}
\quad
\subfigure{
    \includegraphics[width=3.25in]{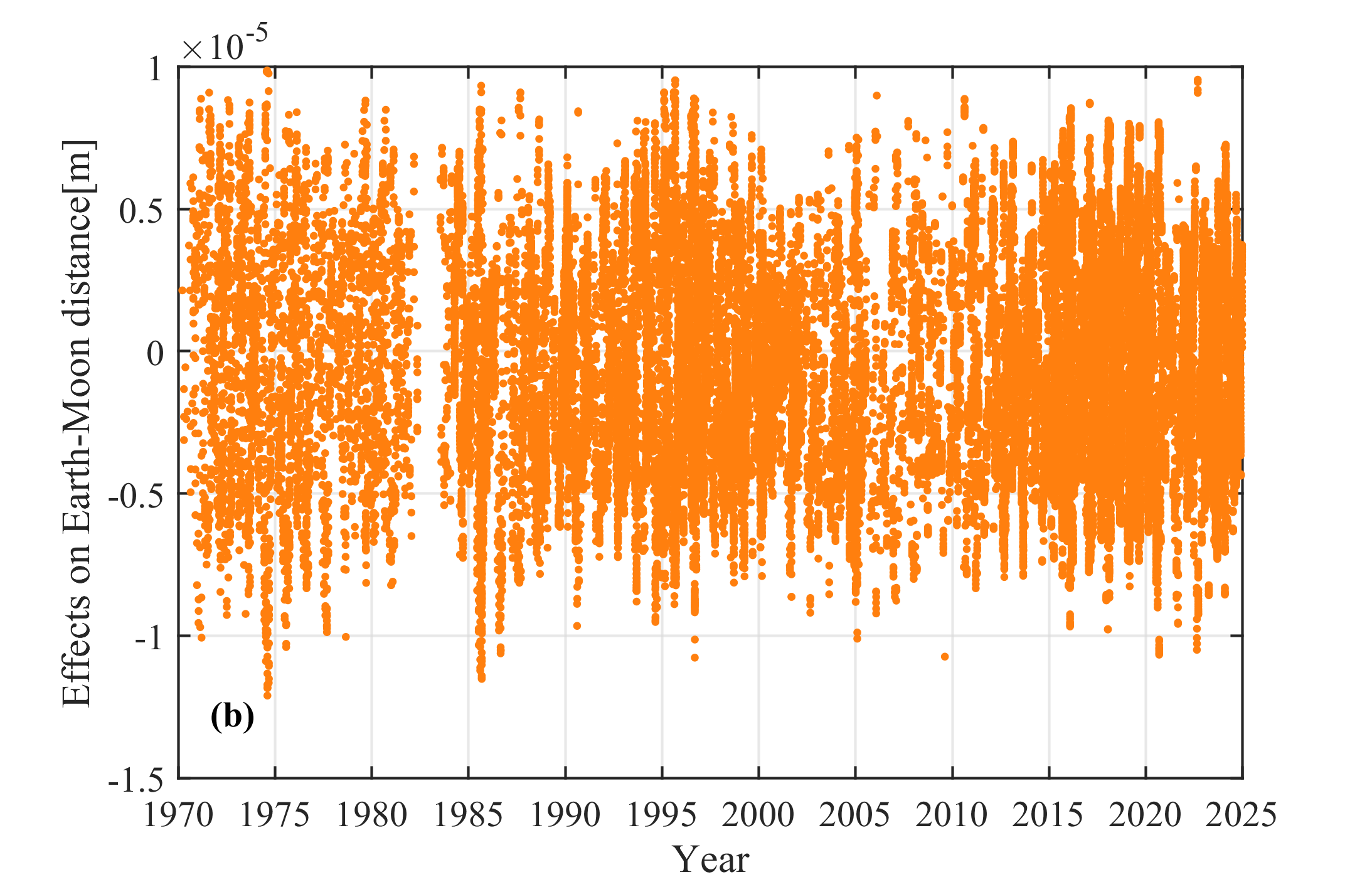}}
\caption{(a) Differences between the ephemeris-based and SOFA-based TT–TDB values. (b) Impact of different TT–TDB realizations on the Earth–Moon distance.}
\label{Fig5}
\end{figure*}

The difference between TT and TAI is fixed, i.e., ${\mathrm{TT}-\mathrm{TAI}=32.184}$ s. Two methods are used to compute the TT–TDB transformation. The first follows the numerical integration scheme adopted in planetary ephemerides (see Eq. (5) in Ref. \cite{Folkner2014}), while the second uses the empirical expression provided in the SOFA subroutine DTDB.F \cite{Hohenkerk2012}. For the first method, no additional ephemeris integration is performed here. The TT–TDB values are obtained directly from the Chebyshev coefficients stored in the ephemeris through interpolation. Note that the ephemeris only contains the geocentric part of the correction, so the topocentric contribution associated with the observing station still needs to be computed separately. If the transformation is involved implicitly in the equations, the corresponding time must be solved iteratively. The differences between the two methods for computing TT–TDB are shown in Fig. \ref{Fig5} (a). The EPM21 and DE430 results show similar long-term behavior and vary relatively slowly with time, whereas the INPOP21 result shows more rapid variations. Nevertheless, all three ephemerides differ from the SOFA routine by only about ${10^{-8}}$ s. Since the modeled LTT is determined from the difference between $t_1$ and $t_3$, these discrepancies will be canceled out. As shown in Fig. \ref{Fig5} (b), the impact of the different TT–TDB realization on the Earth–Moon distance is only at the level of about 10 \textmu m, and the results from the three ephemerides agree very well. For comparison, the effect of the difference TT–TDB on the Earth–Moon distance reaches an amplitude of up to 40 cm for the LLR observations from 1970 to 2024.

\subsection{Transformation between ITRS and BCRS}
\begin{figure*}[htbp]
\includegraphics[width=0.45\textwidth]{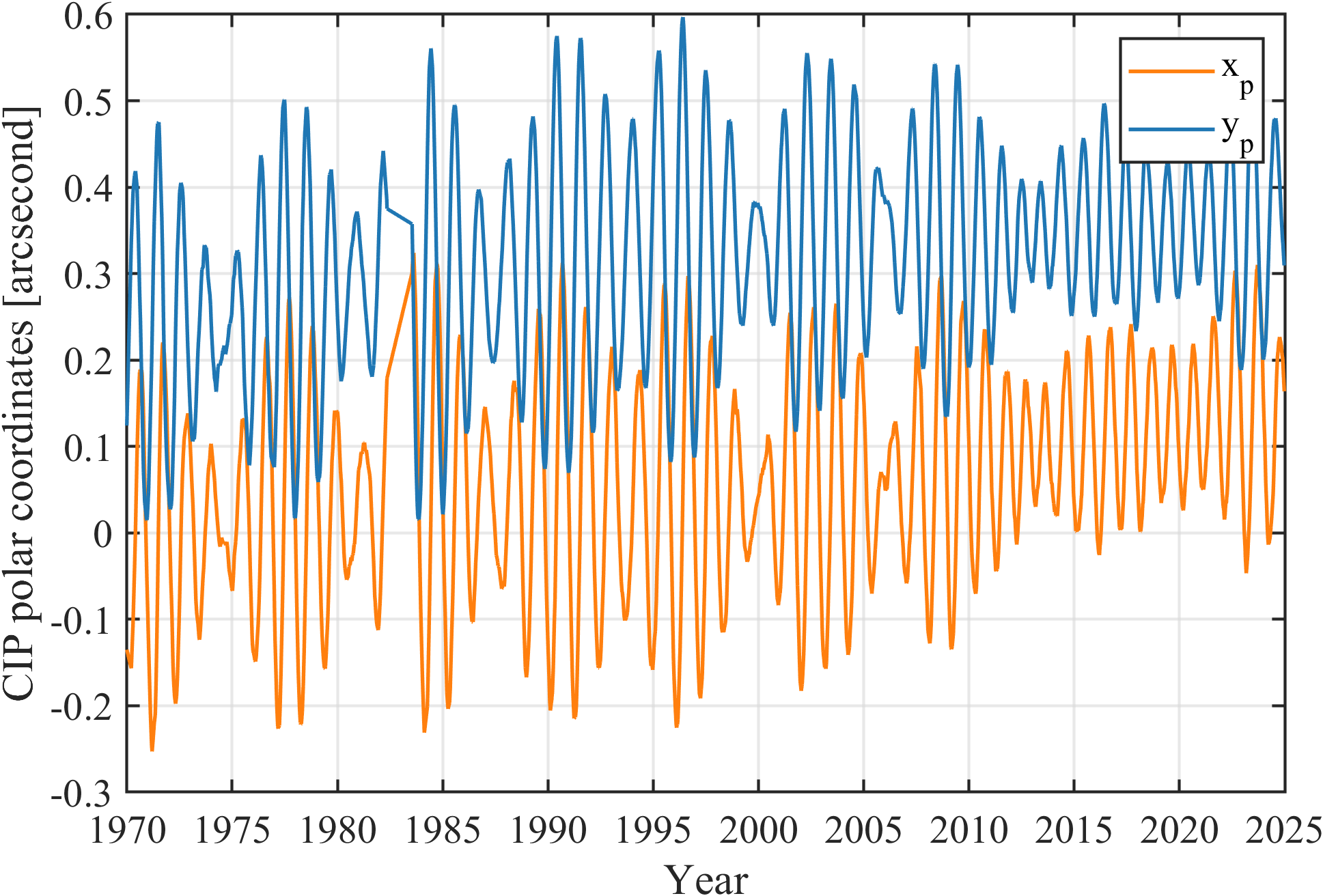}
\caption{Time series of the CIP polar coordinates from the JPL EOP2 series.}
\label{Fig6}
\end{figure*}

The station coordinates in the ITRS are first converted to a geocentric celestial reference system (GCRS). GCRS has its origin at the geocenter with its axes aligned with the BCRS. According to the IERS Conventions 2010 \cite{{IERS2010}}, the transformation between ITRS and GCRS can be performed using either the Celestial Intermediate Origin (CIO) based or equinox-based method. In fact, the difference between the two approaches is smaller than 0.3 ${\rm \mu as}$, which corresponds to about 10 ${\rm \mu m}$ in the station position and is negligible for current LLR analyses. In this study, the IAU 2006/2000A CIO-based transformation is adopted, and the corresponding transformation matrix is expressed as:
\begin{equation}\label{N10}
\mathbf{r}_{{\rm{sta}}}^{{\rm{GCRS}}} = \mathbf{Q}(t)\mathbf{R}(t)\mathbf{W}(t)\left[ {\mathbf{r}_{{\rm{sta}}}^{{\rm{ITRS}}} + \mathbf{\Delta} } \right],
\end{equation}
where ${\mathbf{\Delta}}$ denotes the station displacement correction vector discussed in Sec. \ref{subsec:3.3}. $\mathbf{W}(t)$ is the transformation matrix from the ITRS to the Terrestrial Intermediate Reference System (TIRS) associated with polar motion, and can be expressed as:
\begin{equation}\label{N11}
\mathbf{W}(t) = {\bm{\mathcal{R}}_z}\left( {\begin{array}{*{20}{c}}
{ - s'}
\end{array}} \right){\bm{\mathcal{R}}_y}\left( {\begin{array}{*{20}{c}}
{{x_p}}
\end{array}} \right){\bm{\mathcal{R}}_x}\left( {\begin{array}{*{20}{c}}
{{y_p}}
\end{array}} \right),
\end{equation}
with ${x_p}$ and ${y_p}$ denoting the polar coordinates of the Celestial Intermediate Pole, and $s'$ being the Terrestrial Intermediate Origin locator. $\bm{\mathcal{R}}_x$, $\bm{\mathcal{R}}_y$, and $\bm{\mathcal{R}}_z$ denote the right-handed rotation matrices about the $x$, $y$, and $z$ axes, respectively. $s'$ can be obtained through ${-47 \times T_u/36525}$ ${\rm \mu as}$, with ${T_u = \mathrm{JD_{UT1}}-2451545}$.

We can also obtain the corresponding values of $x_p$ and $y_p$ from the JPL EOP2 series using 4th-order Lagrange interpolation, as shown in Fig. \ref{Fig6}. Both $x_p$ and $y_p$ show obvious periodic variations, together with a slow long-term drift. Similarly, the corrections to the pole coordinates due to ocean tides and libration are given by: 
\begin{equation}\label{N12}
({x_p},{y_p}) = (x_p^{interp},y_p^{interp}) + (\Delta x_p^{ocean{\kern 1pt} tides},\Delta y_p^{ocean{\kern 1pt} tides}) + (\Delta x_p^{libration},\Delta y_p^{libration}),
\end{equation}
where the values of ${(\Delta x_p^{ocean{\kern 1pt} tides},\Delta y_p^{ocean{\kern 1pt} tides})}$ and ${(\Delta x_p^{libration},\Delta y_p^{libration})}$ can be obtained through the IERS subroutines; see Section 5.5.1 of Ref. \cite{{IERS2010}} for more details. The corrections to ${x_p}$ and ${y_p}$ caused by ocean tides are about three orders of magnitude smaller than the principal values, while the libration induced corrections are even smaller, at roughly four orders of magnitude below the principal terms. 

${\mathbf{R}(t)}$ is the transformation matrix from the rotation of the Earth around the axis associated with the pole:
\begin{equation}\label{N13}
\mathbf{R}(t) = {\bm{\mathcal{R}}_z}( - {\rm{ERA}}),
\end{equation}
with
\begin{equation}\label{N14}
ERA({T_u}) = 2\pi (0.7790572732640 + 1.00273781191135448{T_u}),
\end{equation}

The matrix ${\mathbf{Q}(t)}$ represents the transformation from the GCRS to the Celestial Intermediate Reference System (CIRS). In the CIO-based framework, ${\mathbf{Q}(t)}$ is evaluated using the celestial pole coordinates $X$ and $Y$ and the CIO locator $s$, yielding
\begin{equation}\label{N15}
\mathbf{Q}(t) = \left( {\begin{array}{*{20}{c}}
{1 - a{X^2}}&{ - aXY}&X\\
{ - aXY}&{1 - a{Y^2}}&Y\\
{ - X}&{ - Y}&{1 - a\left( {{X^2} + {Y^2}} \right)}
\end{array}} \right){R_z}(s),
\end{equation}
with ${a = 1/\left[ {1 + \sqrt {1 - \left( {{X^2} + {Y^2}} \right)} } \right]}$. In practice, the celestial pole coordinates $X$ and $Y$ are computed using either the Fukushima–Williams parameterization or the IAU 2006/2000A semi-analytical expansion, and the resulting differences are below 1 ${\rm \mu as}$. In the current implementation, both methods are supported and may be chosen by the user. Similarly, the CIO locator s can be obtained from the semi-analytical series compatible with the IAU 2006/2000A precession–nutation model. Unmodeled celestial pole offsets $dX$ and $dY$ were taken from JPL EOP2 series. 

The station coordinates are then transformed from the GCRS to the BCRS, during which the following relativistic effects must be taken into account \cite{{IERS2010}}:
\begin{equation}\label{N16}
\mathbf{r}_{{\rm{sta}}}^{{\rm{BCRS}}} = \mathbf{r}_{{E}}^{{\rm{BCRS}}} + \mathbf{r}_{{\rm{sta}}}^{{\rm{GCRS}}}\left( {1 - \frac{{{U_E}}}{{{c^2}}} - {L_C}} \right) - \frac{1}{2}\left( {\frac{{\mathbf{v}_{{E}}^{{\rm{BCRS}}} \cdot \mathbf{r}_{{\rm{sta}}}^{{\rm{GCRS}}}}}{{{c^2}}}} \right)\mathbf{v}_{{E}}^{{\rm{BCRS}}},
\end{equation}
where ${\mathbf{r}_{{E}}^{{\rm{BCRS}}}}$ and ${\mathbf{v}_{{E}}^{{\rm{BCRS}}}}$ are the barycentric position and velocity of the Earth in the BCRS, respectively. ${U_E}$ denotes the external gravitational potential evaluated at the geocenter. ${L_C}$ is the scale constant associated with the relativistic transformation between TCB and TCG, with a value of ${1.48082686741 \times 10^{-8}}$.
\subsection{Station displacement corrections}\label{subsec:3.3}
To obtain high-precision instantaneous station coordinates, station displacement corrections must be taken into account before transforming the station positions into the GCRS. Following the recommendations of the IERS Conventions 2010 \cite{{IERS2010}}, the corrections considered in this work include solid Earth tides, ocean tide loading, pole tides, atmospheric pressure loading, and ocean pole tide loading. 
\subsubsection{Displacement due to solid Earth tide}
The solid Earth tide is the elastic deformation of the solid Earth under the gravitational potential of external bodies, resulting in a displacement of the station coordinates \cite{Wahr1981,Williams2015}. The tidal displacement can be described by the Love and Shida numbers, which describe the radial and horizontal deformation of the solid Earth, respectively:
\begin{equation}\label{N17}
\Delta {\mathbf{r}_\mathrm{solid}} = \sum\limits_{i = 2,3} {\sum\limits_{n \ge 2} {\frac{{G{M_i}R_E^{n + 2}}}{{G{M_E}R_i^{n + 1}}}\left\{ {{h_n}{P_n}\left( {{{\widehat \mathbf{R}}_i} \cdot \widehat \mathbf{r}} \right)\widehat \mathbf{r} + {l_n}P_n^\prime \left( {{{\widehat \mathbf{R}}_i} \cdot \widehat \mathbf{r}} \right)\left[ {{{\widehat \mathbf{R}}_i} - \left( {{{\widehat \mathbf{R}}_i} \cdot \widehat \mathbf{r}} \right)\widehat \mathbf{r}} \right]} \right\}} } ,
\end{equation}
where ${GM_i}$ is the gravitational parameter of the Moon (${i = 2}$) or the Sun (${i = 3}$), $GM_E$ is the gravitational parameter of the Earth, ${R_E}$ is the Earth’s equatorial radius, ${{\widehat \mathbf{R}_i}}$ is the unit vector of the Moon/ Sun in ITRS with magnitude ${R}_i$, ${\widehat \mathbf{r}}$ is the unit vector of the station in ITRS, and ${P_n(x)}$ is the degree-$n$ Legendre polynomial, ${h_n}$ and ${l_n}$ are the degree-$n$ Love and Shida numbers, respectively. 

\begin{figure*}[htbp]
\includegraphics[width=0.85\textwidth]{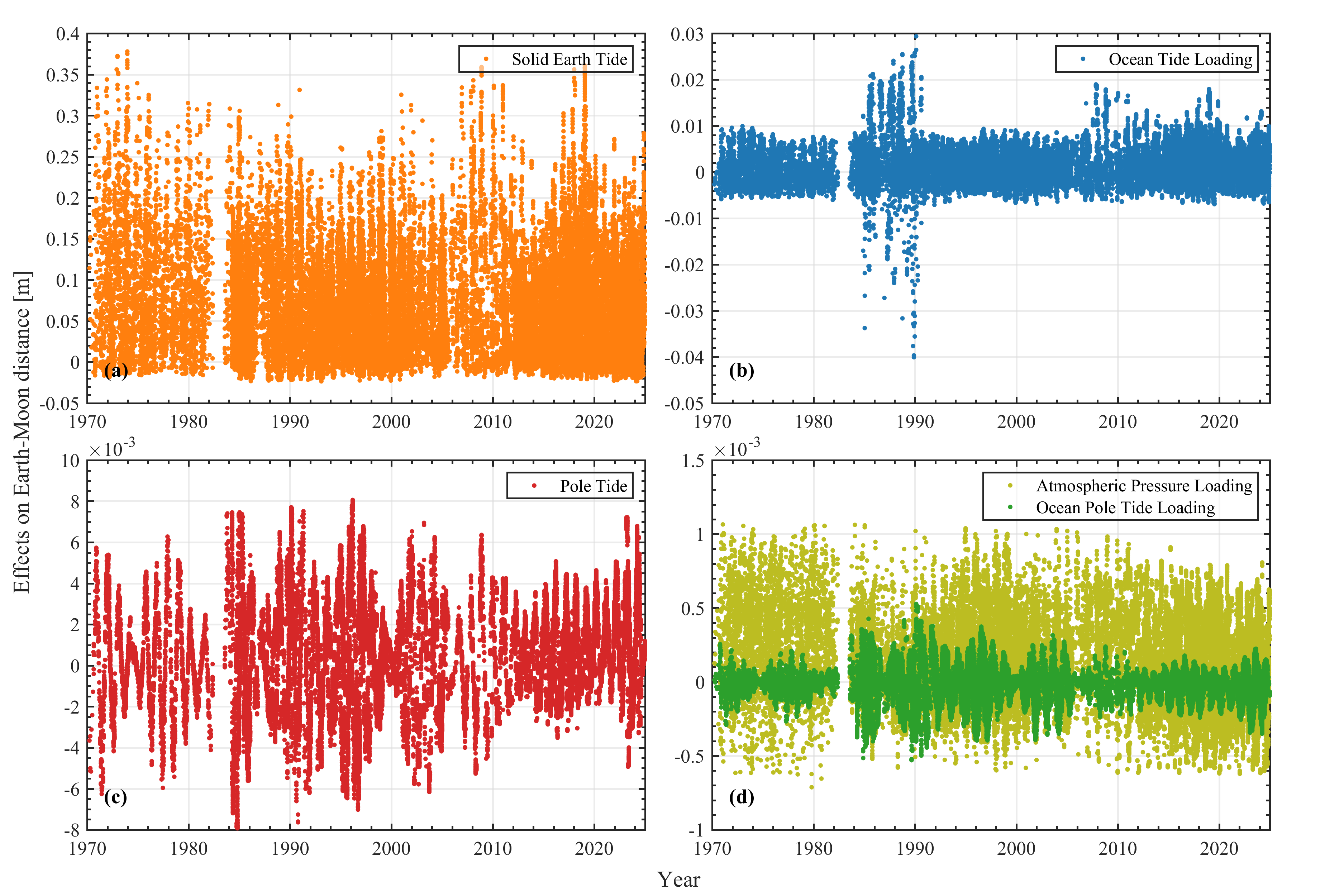}
\caption{Effects of geophysical station-displacement corrections on the modeled Earth–Moon distance: (a) solid Earth tides, (b) ocean tide loading, (c) pole tides, and (d) atmospheric pressure loading and ocean pole tide loading.}
\label{Fig7}
\end{figure*}

Here, the degree-2 and degree-3 solid Earth tide terms are used to account for the full tidal response, together with latitude-dependent Love numbers, out-of-phase corrections, and the frequency-dependent contributions in the diurnal and long-period bands. The resulting station displacement in the three ITRS components remains within 35 cm. The effect of the solid Earth tide on the Earth–Moon distance reaches an amplitude of up to 40 cm for the LLR observations from 1970 to 2024, as shown in Fig. \ref{Fig7} (a).

\subsubsection{Displacement due to ocean tide loading}
The ocean tides are produced by the gravitational attraction of the Moon and Sun. As ocean water moves in response to this tidal forcing, the distribution of ocean mass changes continuously with time \cite{Farrell1972}. These mass redistributions produce time-varying loads on the Earth's surface. Because the Earth is not completely rigid, it deforms elastically under these loads. This deformation is referred to as ocean tide loading. Due to the strong dependence of the ocean response on regional conditions, closed-form analytical expressions are not sufficient to describe it on a global scale. Therefore, gridded ocean tide models are commonly employed to represent the spatial variability of ocean tides. In practice, the station displacement due to ocean tide loading can be described as a sum of tidal harmonics:
\begin{equation}\label{N18}
\Delta {r_{\mathrm{ocean},c}} = \sum\limits_j {{A_{c,j}}\cos \left( {{\chi _j}(t) - {\phi _{c,j}}} \right)} ,
\end{equation}
where ${A_{c,j}}$ and ${\phi_{c,j}}$ are the amplitudes and phases of the loading response for the target station in direction $c$ (radial, west, south), respectively. ${{\chi _j}(t)}$ is the astronomical argument of the $j$-th tidal constituent. The 11 main tides are usually considered, including the semidiurnal waves $M_2, S_2, N_2, K_2$, the diurnal waves $K_1, O_1, P_1, Q_1$, and the long-period waves $M_f, M_m$, and $S_{sa}$. The corresponding amplitudes and phases for a specific site can be obtained from the Ocean Loading Provider \cite{Scherneck2026}. It should be noted that the correction must be applied to compensate for the periodic motion of the oceanic center of mass. The effect of the ocean tide loading on the Earth–Moon distance reaches an amplitude of up to 5 cm for the LLR observations from 1970 to 2024, as shown in Fig. \ref{Fig7} (b).
\vspace{-5pt}
\subsubsection{Displacement due to pole tide}
The Earth's rotation generates a centrifugal potential at any point on the Earth's surface. Since the orientation of the Earth's rotation axis varies with time, the centrifugal potential is perturbed from its mean state \cite{Wahr1985}. This perturbation induces elastic deformations of the Earth's crust and consequently leads to variations in station coordinates. This effect is commonly referred to as the pole tide. Using an approach analogous to that adopted for the solid Earth tide, the radial and horizontal displacement components due to the pole tide can be expressed as:
\begin{equation}\label{N19}
\begin{array}{l}
\Delta {r_\mathrm{pole,r}} =  - 0.033\sin 2\theta \left( {{m_1}\cos \lambda  + {m_2}\sin \lambda } \right),\\
\Delta {r_\mathrm{pole,\theta }} =  - 0.009\cos 2\theta \left( {{m_1}\cos \lambda  + {m_2}\sin \lambda } \right),\\
\Delta {r_\mathrm{pole,\lambda }} = 0.009\cos \theta \left( {{m_1}\sin \lambda  - {m_2}\cos \lambda } \right),
\end{array}
\end{equation}
where ${\lambda}$ and ${\theta}$ denote the longitude and the colatitude of the station, respectively. ${m_1}$ and ${m_2}$ are the time-dependent offsets of the instantaneous rotation pole:
\begin{equation}\label{N20}
{m_1} = {x_p} - {x_s},{m_2} =  - ({y_p} - {y_s}),
\end{equation}
with ${(x_s, y_s)}$ denoting the coordinates of the secular pole, which are given in IERS Conventions 2010 \cite{{IERS2010}}. The effect of the pole tide on the Earth–Moon distance reaches an amplitude of up to 1 cm for the LLR observations from 1970 to 2024, as shown in Fig. \ref{Fig7} (c).

\subsubsection{Displacements due to ocean pole tide loading and atmospheric pressure loading}

Since the displacements caused by ocean pole tide loading and atmospheric pressure loading are relatively small, the theoretical expressions for evaluating these two effects are not provided here \cite{Desai2002}. Analogous to the pole tide, variations in the centrifugal potential associated with polar motion also act on the ocean mass distribution. This perturbation leads to mass redistribution and associated loading effects, commonly referred to as ocean pole tide loading. The displacement at any given point on the Earth's surface can be evaluated using equation (7.29) in Ref.  \cite{{IERS2010}}. The real and imaginary parts of the ocean pole tide loading coefficients are obtained by two-dimensional interpolation of global gridded datasets. 

Variations in atmospheric pressure change the mass of the atmospheric column, inducing elastic deformation of the Earth's crust, commonly referred to as atmospheric pressure loading \cite{Petrov2004}. The corresponding displacement can be evaluated using an expression similar to Eq. (\ref{N18}). For a given station, the coefficients are obtained by interpolation of HARPOS-format data provided by the International Mass Loading Service \cite{Petrov2015,Petrov2015}. For the LLR observations from 1970 to 2024, the effects of these two displacements on the Earth–Moon distance reach amplitudes of about 0.5 mm and 1 mm, respectively, as shown in Fig. \ref{Fig7} (d).
\subsection{Transformation between LCRS and PA}
Similar to the solid Earth tide correction applied to terrestrial stations, the reflector coordinates in PAS should also be corrected for lunar solid tides. Here, only the dominant degree-2 contribution is included. The corrected reflector coordinates are then transformed from PAS to the Lunar Celestial Reference System (LCRS) using a rotation matrix parameterized by three Euler angles \cite{Folkner2014}:
\begin{equation}\label{N21}
\mathbf{r}_{{\rm{ref}}}^{{\rm{LCRS}}} = {\bm{\mathcal{R}}_z}( - {\phi _m}){\bm{\mathcal{R}}_x}( - {\theta _m}){\bm{\mathcal{R}}_z}( - {\psi _m})\left( {\mathbf{r}_\mathrm{ref}^{{\rm{PA}}} + \Delta \mathbf{r}_\mathrm{solid}^M} \right),
\end{equation}
where ${\Delta \mathbf{r}_\mathrm{solid}^M}$ is displacement due to the degree-2 lunar solid tide raised by the Earth and the Sun, computed using Eq. (\ref{N17}) with the corresponding lunar Love numbers. The effect of the solid Moon tide on the Earth–Moon distance reaches an amplitude of up to 65 cm for the LLR observations from 1970 to 2024.

\begin{figure*}[t!]
\centering
\subfigure{%
    \includegraphics[width=3.05in]{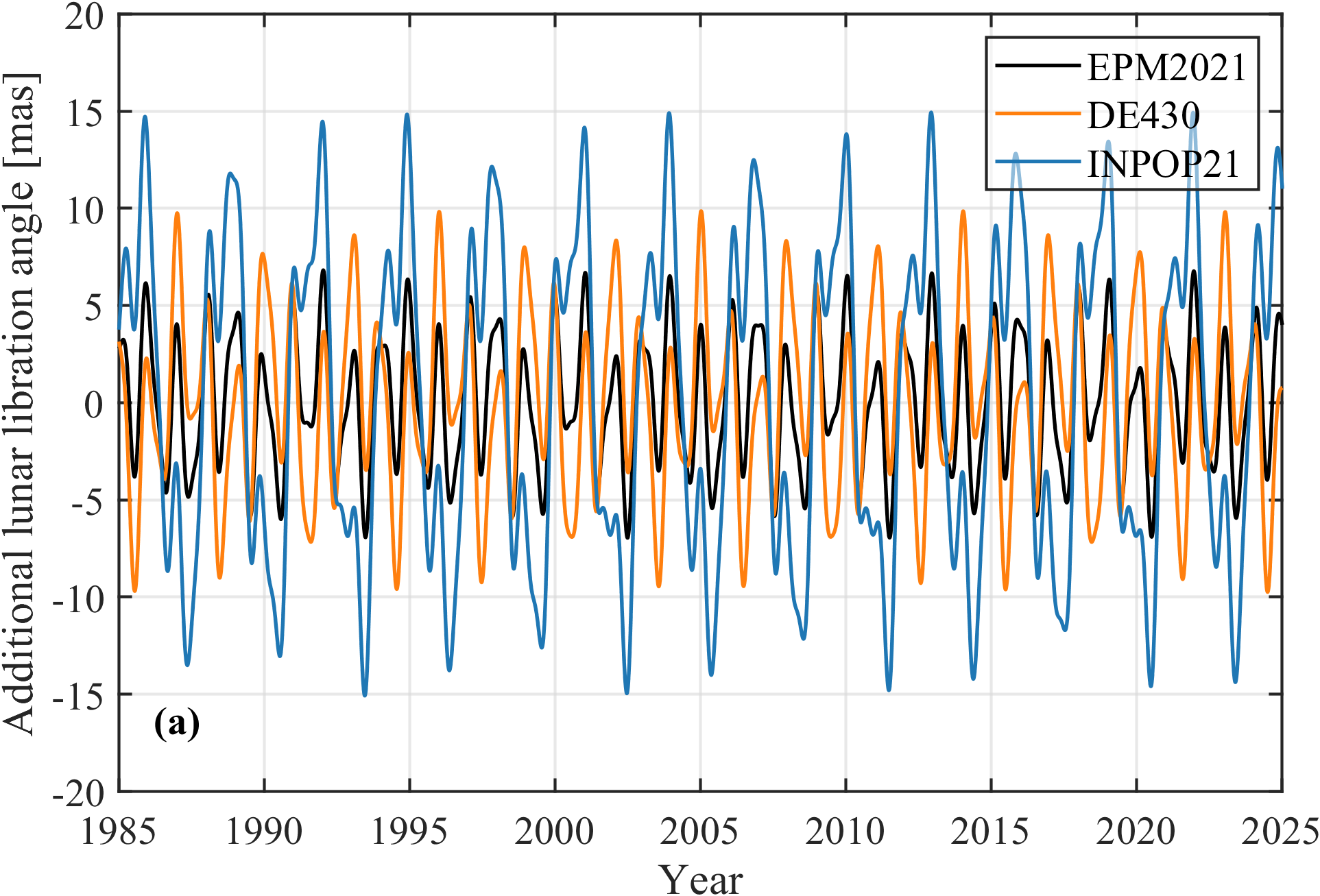}}
\quad
\subfigure{
    \includegraphics[width=3.15in]{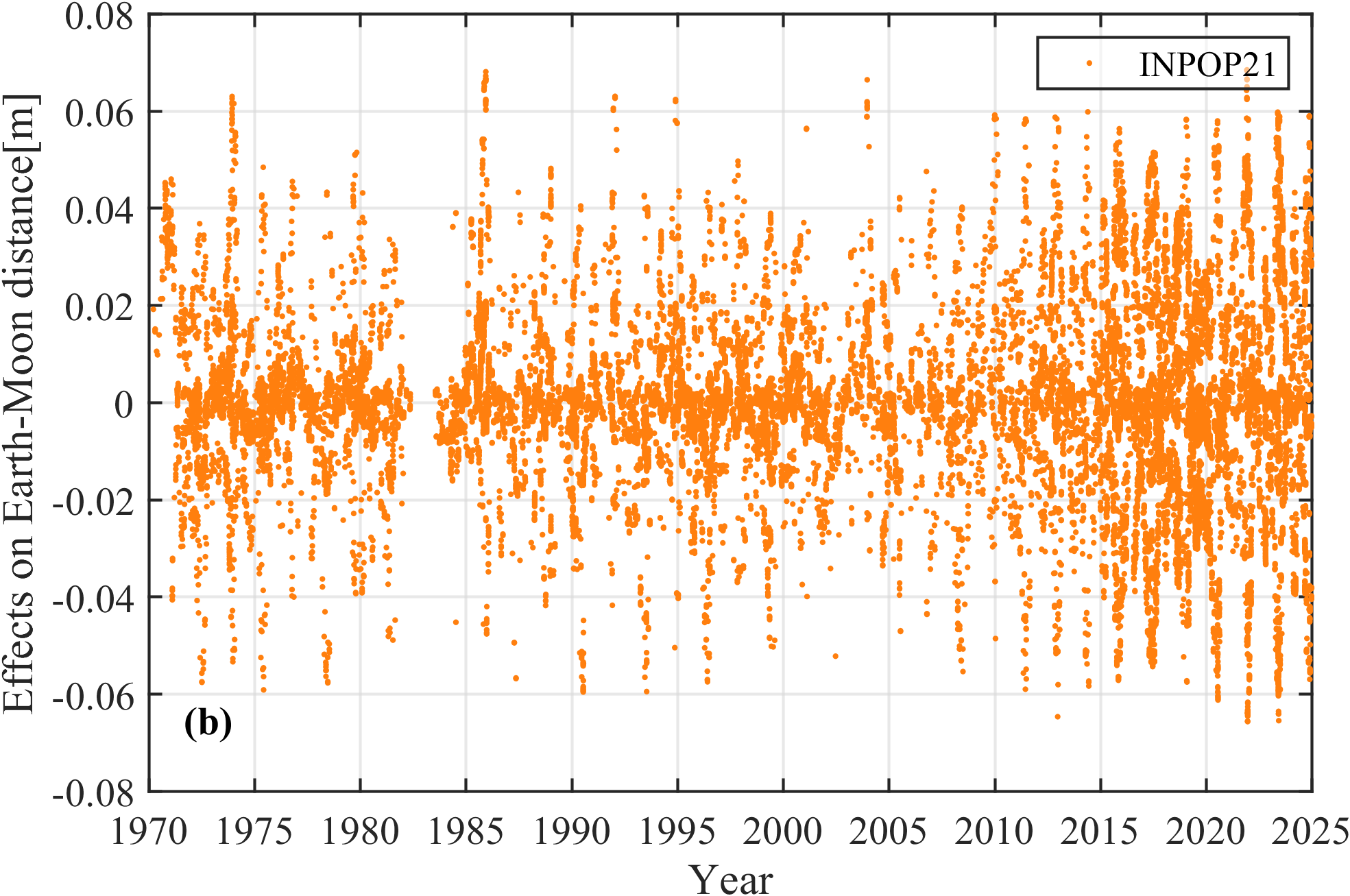}}
\caption{(a) Additional longitude-libration corrections and (b) the corresponding effects on the Earth–Moon distance.}
\label{Fig8}
\end{figure*}

$\phi_m, \psi_m$ and $\theta_m$ are the Euler angles describing the orientation of the Moon. The values of these angles are obtained by interpolation of the lunar orientation parameters provided in the ephemeris. In addition, small longitude-libration effects associated with frequency-dependent tidal dissipation are included through three periodic correction terms applied to the libration angle \cite{Pavlov2016, Williams2013}:
\begin{equation}\label{N22}
\Delta \Lambda  = {A_1}\cos l' + {A_2}\cos (2l - 2D) + {A_3}\cos (2F - 2l),
\end{equation}
where $l, l', F$, and $D$ are the relevant Delaunay arguments. The amplitudes $A_1, A_2$, and $A_3$ can be obtained from the parameters provided with the ephemeris or estimated from the observations through a weighted least-squares adjustment. Fig. \ref{Fig8} (a) presents the additional libration angle corrections obtained from the three ephemerides. The INPOP21 solution yields the largest correction, with amplitudes reaching about 15 mas. The corresponding effect on the Earth–Moon distance reaches up to 6.5 cm for the LLR observations from 1970 to 2024, as shown in Fig. \ref{Fig8} (b). This correction reduces the WRMS of one-way residuals by about 1 cm. 

Similar improvements are obtained for EPM21 and DE430. The reflector coordinates are then transformed from the LCRS to the BCRS \cite{{IERS2010}}:
\begin{equation}\label{N23}
\mathbf{r}_{{\rm{ref}}}^{{\rm{BCRS}}} = \mathbf{r}_{{M}}^{{\rm{BCRS}}} + \mathbf{r}_{{\rm{ref}}}^{{\rm{LCRS}}}\left( {1 - \frac{{{U_M}}}{{{c^2}}} - {L_C}} \right) - \frac{1}{2}\left( {\frac{{\mathbf{v}_{{M}}^{{\rm{BCRS}}} \cdot \mathbf{r}_{{\rm{ref}}}^{{\rm{LCRS}}}}}{{{c^2}}}} \right)\mathbf{v}_{{M}}^{{\rm{BCRS}}},
\end{equation}
where ${\mathbf{r}_{{M}}^{{\rm{BCRS}}}}$ and ${\mathbf{v}_{{M}}^{{\rm{BCRS}}}}$ are the barycentric position and velocity of the Moon in the BCRS, respectively. $U_M$ denotes the external gravitational potential at the Moon’s center, excluding the Moon’s mass.
\subsection{Atmospheric propagation delay correction}
When signals pass through the troposphere and stratosphere, a propagation delay is introduced by atmospheric refraction \cite{{IERS2010}}. For optical wavelengths used in LLR, the dominant contribution arises from the neutral atmosphere, while the ionospheric delay can be neglected at the millimeter level. The atmospheric delay is evaluated using the Mendes–Pavlis model \cite{Mendes2004,Jiang2026}, which expresses the slant delay as the product of the zenith delay and a dimensionless mapping function:
\begin{equation}\label{N24}
\Delta {\tau _{{\rm{atm}}}}({t_1},{t_2}) = \frac{{r_{{\rm{atm}}}^z(\lambda ,\varphi ,H,{P_s},{T_s}) \cdot m(\varepsilon )}}{c},
\end{equation}
where $\varepsilon$ is the elevation angle of the line of sight, and $\lambda$, $\varphi$, $H$, $P_s$ and $T_s$ denote the longitude, latitude, geodetic height, atmospheric pressure, and surface temperature of the station, respectively. All quantities required in Eq. (\ref{N24}) can be obtained from the LLR normal-point data or from Ref. \cite{{IERS2010}}. The effect of atmospheric propagation delay on the Earth--Moon distance is shown in Fig. \ref{Fig9} (a). For LLR observations from 1970 to 2024, the corresponding correction ranges mostly from 2 m to 10 m, reaching nearly 18 m for several low-elevation observations.
\subsection{Shapiro time delay correction}
\begin{figure*}[htbp]
\centering
\subfigure{%
    \includegraphics[width=3.05in]{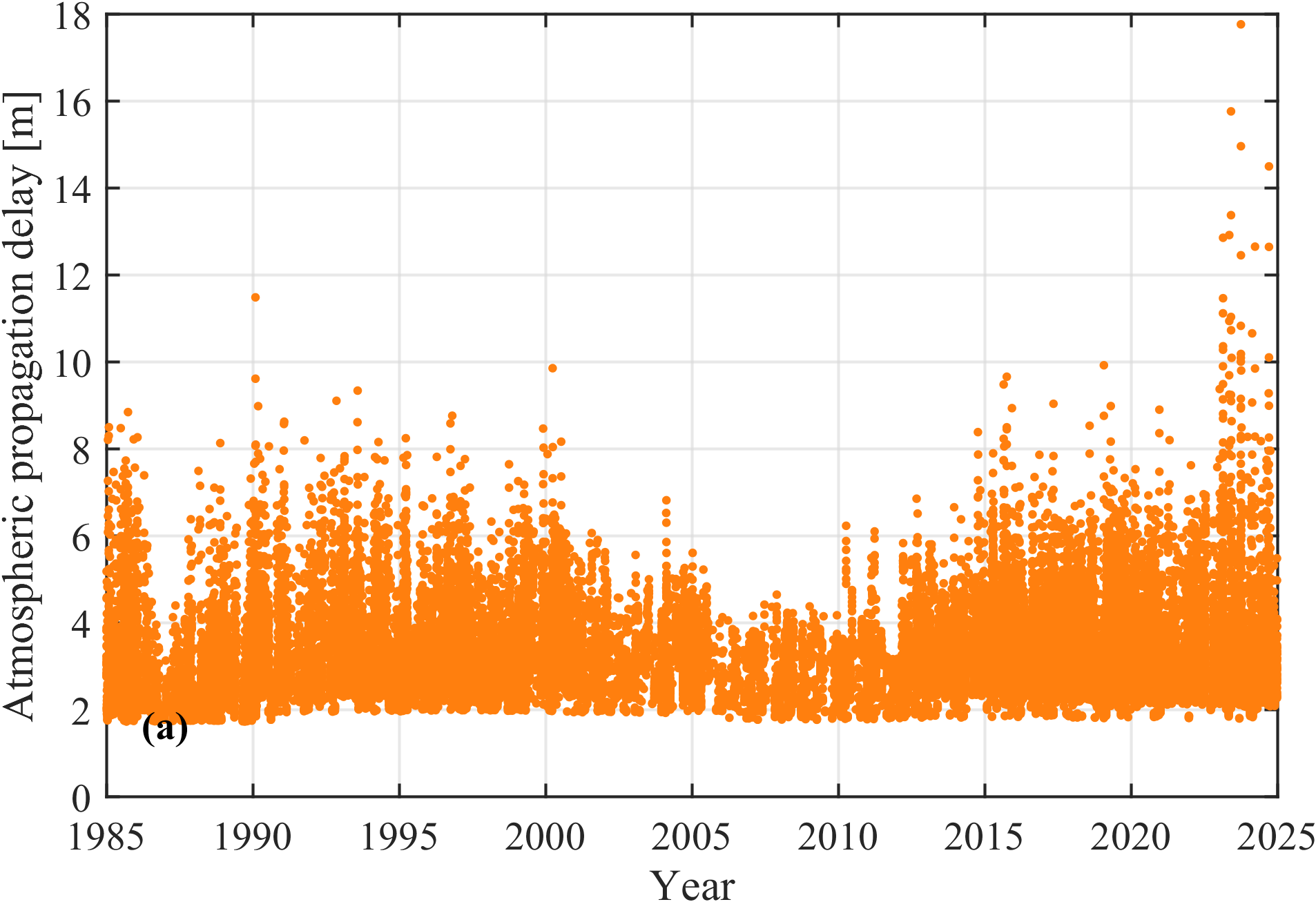}}
\quad
\subfigure{
    \includegraphics[width=3.07in]{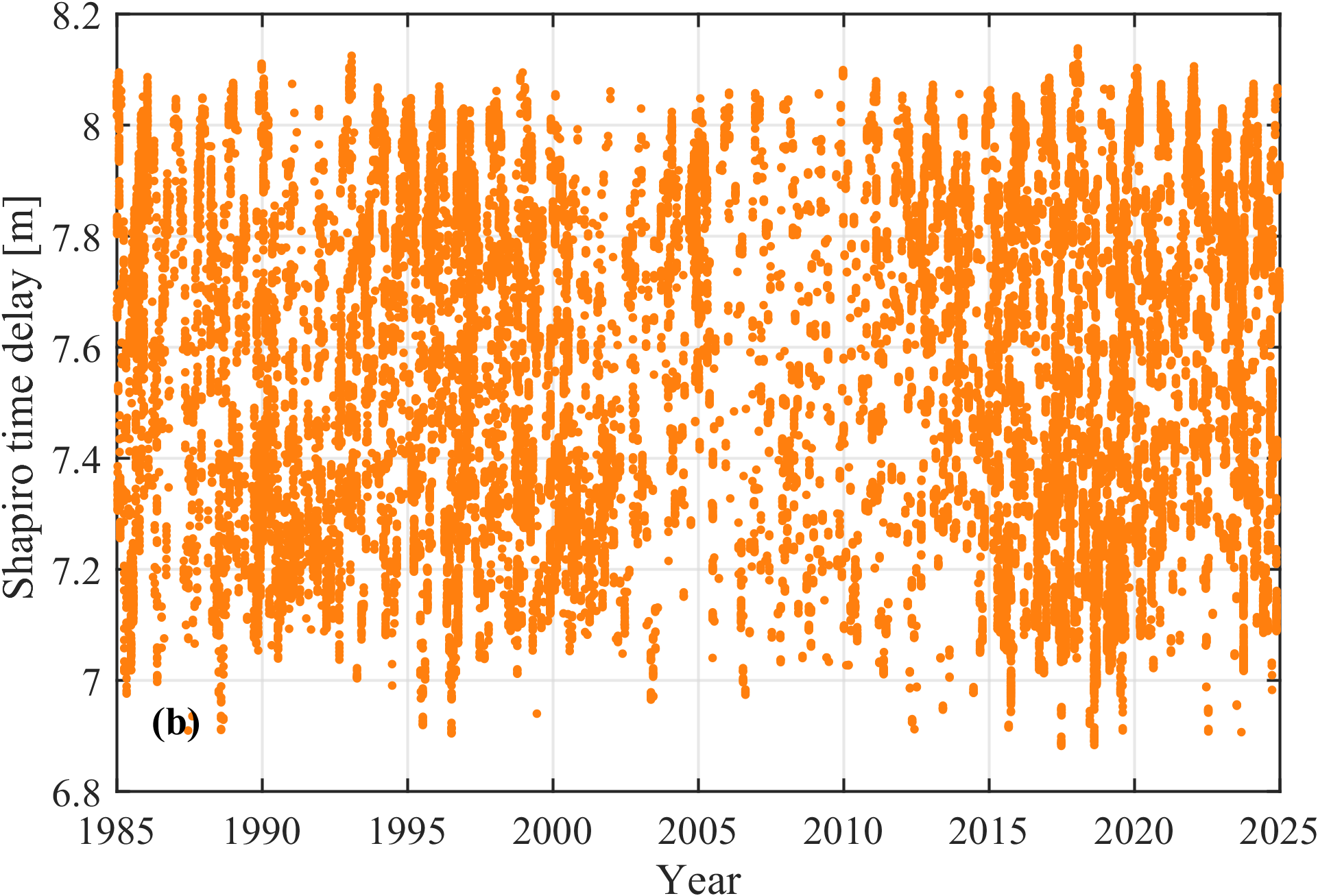}}

\caption{Effects of (a) atmospheric propagation delay and (b) gravitational propagation delay on the Earth–Moon distance.}
\label{Fig9}
\end{figure*}

According to general relativity, the propagation of light is affected by the gravitational potential along its path \cite{Shapiro1964}. As a result, an additional delay is introduced when the signal passes near a massive body. If a signal is emitted from point 1 at coordinate time $t_1$ and is received at point 2 at coordinate time $t_2$, the corresponding correction due to the Shapiro time delay is given by:
\begin{equation}\label{N25}
\Delta {\tau _{{\rm{grav}}}}({t_1},{t_2}) = \sum\limits_i {\frac{{2G{M_i}}}{{{c^3}}}} \ln \left( {\frac{{{r_{i1}} + {r_{i2}} + {r_{12}}}}{{{r_{i1}} + {r_{i2}} - {r_{12}}}}} \right),
\end{equation}
where $r_{i1}$ is the distance between gravitational body $i$ and point 1, $r_{i2}$ is the distance between gravitational body $i$ and point 2, $r_{12}$ is the distance between points 1 and 2. For the LLR observations from 1970 to 2024, the effect of the Shapiro time delay on the Earth–Moon distance reaches an amplitude of up to 8 m, as shown in Fig. \ref{Fig9} (b).

The principal corrections applied in the LLR data reduction are summarized in Table \ref{T2}. Their amplitudes show that several effects are much larger than the millimeter-level EP signature considered in this work, which justifies the detailed treatment described above.
\begin{table}[htbp]
\centering
\caption{Summary of the principal corrections applied in the LLR data reduction.}
\label{T2}
\renewcommand{\arraystretch}{1.2}
\setlength{\tabcolsep}{8pt}   
\begin{tabular}{c c c}
\hline
{Effects} & {Typical amplitude on the Earth--Moon distance} & {References} \\
\hline
TT--TDB transformation     & up to 40 cm      & \cite{Folkner2014} \\
Solid Earth tide           & up to 40 cm      & IERS 2010 \cite{{IERS2010}} \\
Ocean tide loading         & up to 5 cm       & IERS 2010 \cite{IERS2010}\\
Pole tide                  & up to 1 cm       & IERS 2010 \cite{IERS2010}\\
Ocean pole tide loading    & up to 0.5 mm     & IERS 2010 \cite{IERS2010}\\
Atmospheric pressure loading & up to 1 mm     & IERS 2010 \cite{IERS2010}\\
Lunar solid tide           & up to 65 cm      & IERS 2010 \cite{IERS2010}\\
Additional longitude libration terms & up to 6.5 cm & DE/INPOP/EPM \\
Atmospheric propagation delay & 2--18 m      & Mendes--Pavlis model \cite{Mendes2004} \\
Shapiro time delay         & 7--8 m           & \cite{Shapiro1964} \\
\hline
\end{tabular}
\end{table}
\section{LLR Data Processing Results and EP Parameter Estimation}\label{section4}
\subsection{O--C residual results}\label{section4.1}
Following the procedure described in Sec. \ref{section3}, the computed two-way LTT C can be obtained. A weighted least-squares (WLS) fit is then performed to minimize the O--C residuals by adjusting the model parameters $\mathbf{x}$. The corresponding corrections, $\Delta \mathbf{x}$, are given by:
\begin{equation}\label{N26}
\Delta \mathbf{x} = {\left( {{\mathbf{A}^T}\mathbf{W}\mathbf{A}} \right)^{ - 1}}{\mathbf{A}^T}\mathbf{W}\mathbf{y},
\end{equation}
where $\mathbf{y}$ is the vector of two-way O–C residuals, $\mathbf{W}$ is the weight matrix, and $\mathbf{A}$ is the design matrix constructed from the partial derivatives of C with respect to $\mathbf{x}$. The covariance matrix of the estimated parameters is
\begin{equation}\label{N27}
{\mathbf{Q}_{\hat {\boldsymbol{x}} \hat {\boldsymbol{x}}}} = \sigma _0^2{\left( {{\mathbf{A}^T}\mathbf{W}\mathbf{A}} \right)^{ - 1}},
\end{equation}
where ${\sigma _0^2}$ is the variance of unit weight. The formal uncertainties of the estimated parameters are obtained from the square roots of the diagonal elements of ${\mathbf{Q}_{\hat {\mathbf{x}}\hat {\mathbf{x}}}}$.
\begin{figure*}[htbp]
\includegraphics[width=0.60\textwidth]{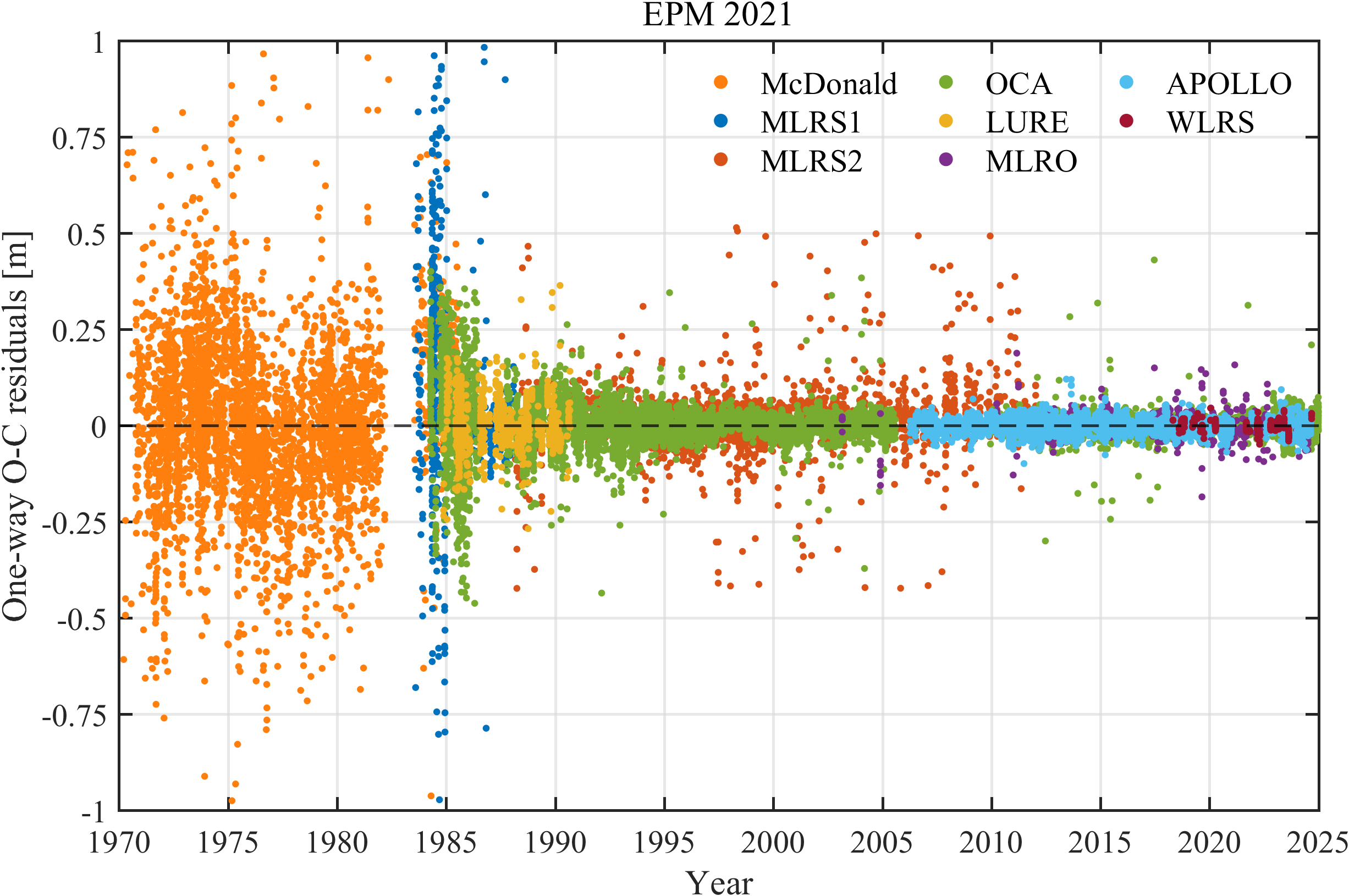}
\caption{One-way post-fit O--C residuals obtained with EPM21.}
\label{Fig10}
\end{figure*}

Here, the parameter vector $\mathbf{x}$ includes station coordinates, station velocities, reflector coordinates, and station biases. Station biases are modeled as piecewise-constant parameters, with separate values estimated for different stations and observation periods. The bias intervals are generally defined according to documented station configuration changes, such as equipment upgrades or optical fiber replacements. These periods are obtained from station log files whenever available. When such information is unavailable, the bias interval recommendations provided by Refs. \cite{Pavlov2016,ViswanathanPhd} are adopted. To ensure consistency among the ephemerides considered in this study, the same a priori station coordinates, station velocities, and reflector coordinates from Ref. \cite{Fienga2021} are used in all solutions. All station bias parameters are initialized to zero.

Assuming that the observation uncertainties are uncorrelated, the weight matrix $\mathbf{W}$ is simply the diagonal matrix with elements given by the inverse squared uncertainties of the observations. However, the uncertainties in NPs depend not only on the measurement precision but also on the station normal point generation and outlier rejection procedures. As a result, they do not always provide a homogeneous weighting scheme across different stations and observing periods. Therefore, following previous LLR analyses, recommended uncertainty rescaling is applied to several data sets with known weighting issues. In particular, the APOLLO observations are corrected using the publicly available scaling factors corresponding to different observing periods, while observations from Grasse (1998–1999), MLRS2 (1996), and Matera (2010–2012) are reweighted \cite{Pavlov2016,ViswanathanPhd}. For all remaining observations, the uncertainties provided in the normal point files are used directly.

\begin{figure*}[htbp]
\includegraphics[width=0.6\textwidth]{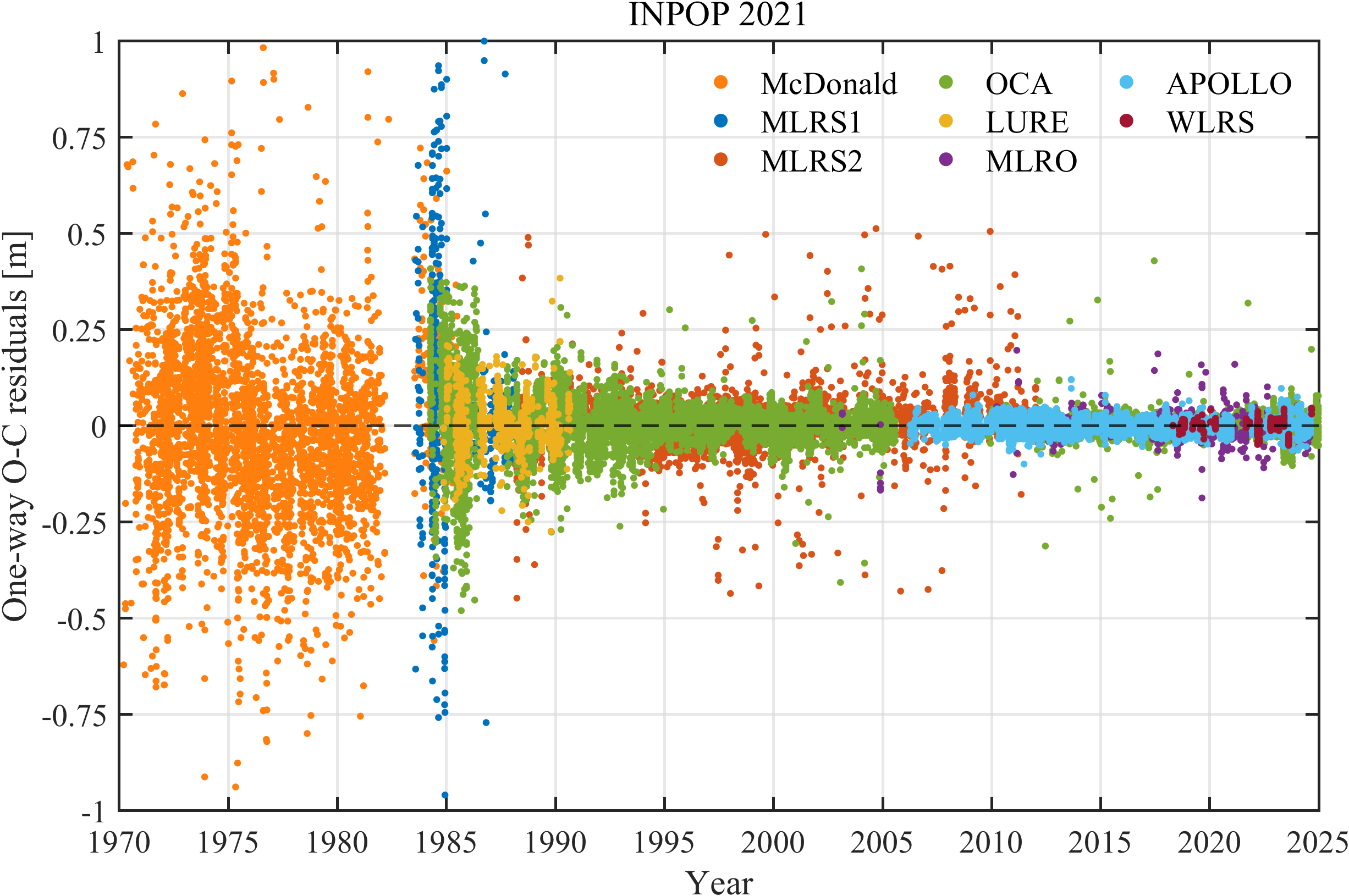}
\caption{One-way post-fit O–C residuals obtained with INPOP21.}
\label{Fig11}
\end{figure*}

\begin{figure*}[htbp]
\includegraphics[width=0.6\textwidth]{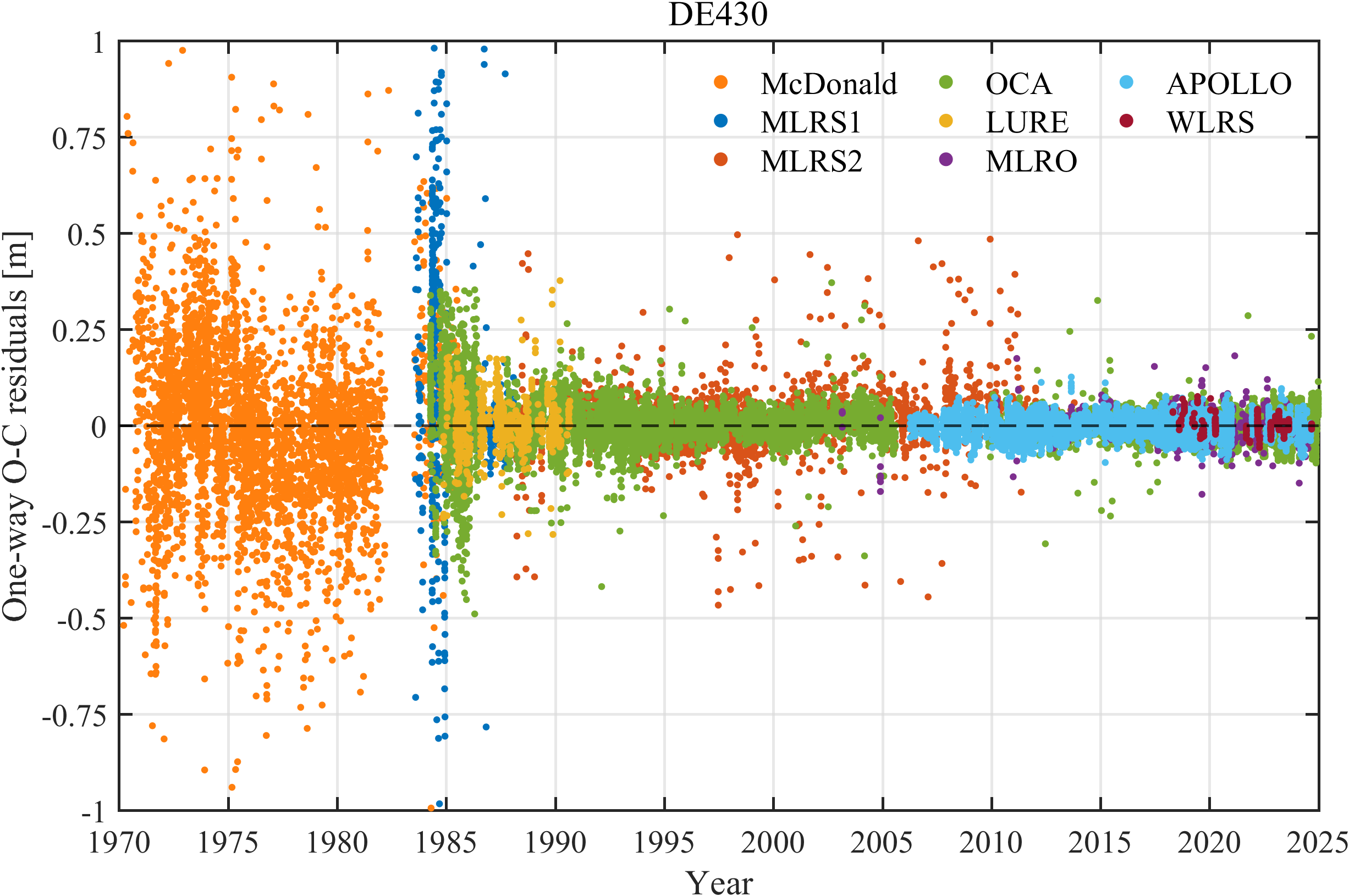}
\caption{One-way post-fit O–C residuals obtained with DE430.}
\label{Fig12}
\end{figure*}

\begin{figure*}[htbp]
\includegraphics[width=0.60\textwidth]{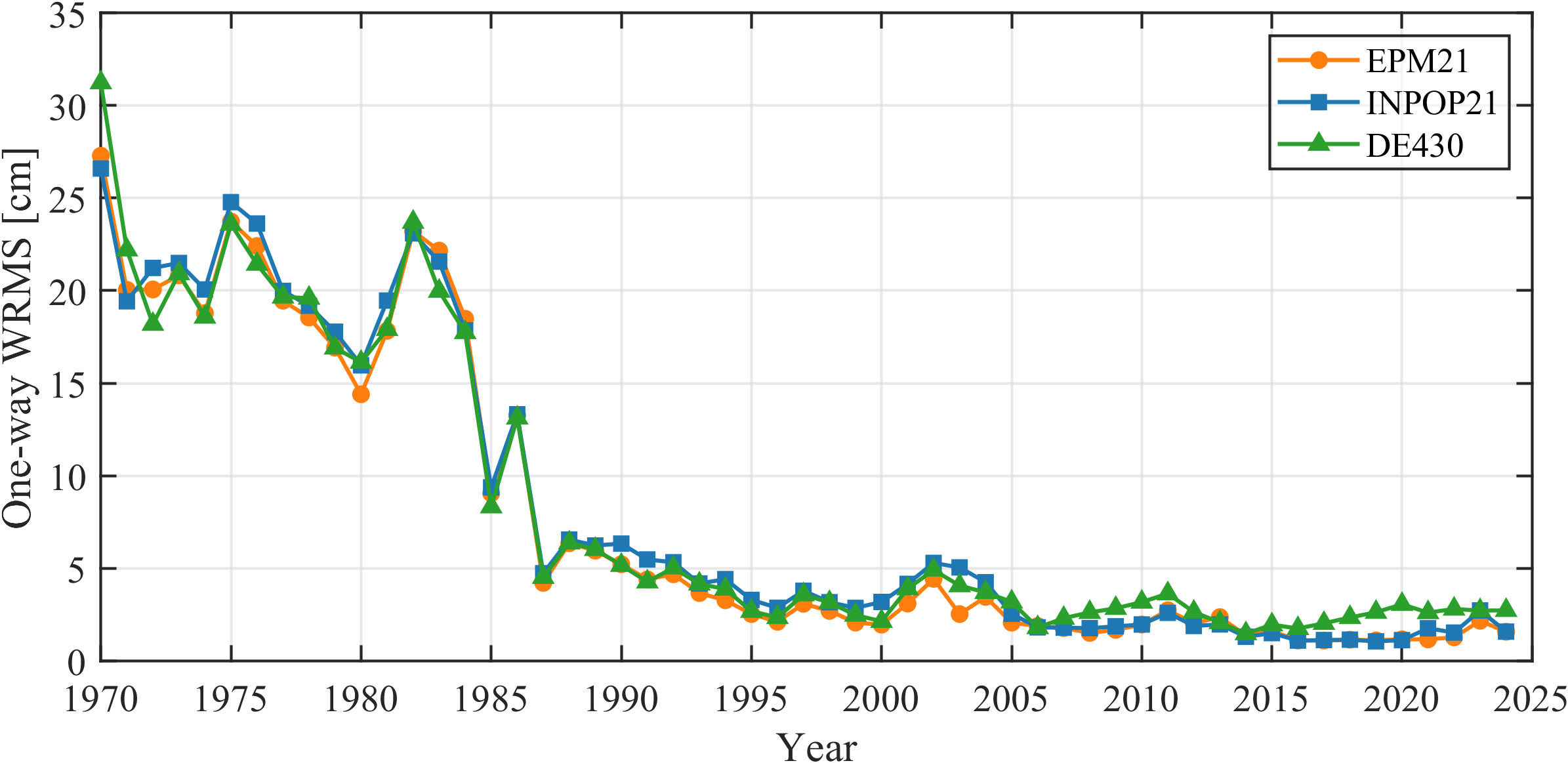}
\caption{Annual WRMS of the one-way post-fit O--C residuals obtained with EPM21, INPOP21, and DE430.}
\label{Fig13}
\end{figure*}

The WLS adjustment was performed iteratively until the change in the estimated parameter corrections between two consecutive iterations became negligible. In practice, convergence was usually reached within two to three iterations. Figs. \ref{Fig10}--\ref{Fig12} present the one-way post-fit O--C residuals obtained with EPM21, INPOP21, and DE430, respectively. The corresponding one-way WRMS values can be calculated by:
\vspace{-6pt}
\begin{equation}\label{N28}
\mathrm{WRMS_{one-way}} = \frac{1}{2}\sqrt {\frac{{\sum\limits_{i = 1}^N {{{\left( {{{\text{O}}_i} - {{\text{C}}_i}} \right)}^2}/\sigma _i^2} }}{{\sum\limits_{i = 1}^N {1/\sigma _i^2} }}},
\end{equation}
where $\sigma _i$ is the observation uncertainty of $\mathrm{O}_i$, and $N$ is the number of NPs. The results are summarized in Table \ref{T3}, with about 2\% of the points were rejected after residual-based outlier rejection. Fig. \ref{Fig13} compares the annual WRMS values of the one-way post-fit O--C residuals obtained with the three ephemerides, whose overall trend is broadly consistent with that of the normal-point uncertainties shown in Fig. \ref{Fig2} (b). Overall, INPOP21 and EPM21 show very similar performance over the full observation span. For both ephemerides, the annual WRMS of the one-way post-fit O--C residuals is below 2 cm after 2006 and remains close to 1 cm between 2016 and 2021. They increased again after 2023, indicating a gradual degradation in extrapolation accuracy as the LLR observations extended beyond the data spans used in constructing the ephemerides. 


\begin{table}[htbp]
\centering
\caption{WRMS of one-way post-fit LLR residuals obtained with EPM21, INPOP21 and DE430}
\label{T3}
\renewcommand{\arraystretch}{1.25}
\setlength{\tabcolsep}{6pt}   
\begin{tabular}{c c c c c c c}
\hline
\noalign{\vspace{3pt}}
Station & Time Span & NPs & \makecell[c]{EPM21\\(1970--2024) [cm]}& \makecell[c]{INPOP21\\(1970--2024) [cm]} & \makecell[c]{DE430\\(1970--2024) [cm]} & \makecell[c]{DE430\\(1970--2016) [cm]} \\
\hline
McDonald  & 1970--1985 & 3503  & 19.468 & 20.243 & 19.316 & 19.405 \\
MLRS1     & 1983--1988 & 592   & 16.538 & 17.280 & 16.497 & 16.399 \\
MLRS2     & 1988--2013 & 3244  & 4.117  & 4.598 & 4.297   & 4.135  \\
\multirow{2}{*}{OCA} 
          & 1984--2016 & 12812 & 2.283  & 2.755 & 2.747  & 2.495  \\
          & 2017--2024 & 8549  & 1.315  & 1.431  &2.902  & -        \\
LURE      & 1984--1990 & 746   & 5.538  & 5.613 & 5.635  & 5.405  \\
\multirow{2}{*}{MLRO}
          & 2003--2016 & 102   & 3.425  & 3.439  & 3.469 & 3.393  \\
          & 2017--2024 & 311   & 3.188  & 3.380  &  3.895 &  -    \\
\multirow{2}{*}{APOLLO}
          & 2006--2016 & 2615  & 1.448  & 1.406 & 1.957    & 1.349  \\
          & 2017--2024 & 1521  & 1.291  & 1.396  & 2.535 &   -    \\
WLRS      & 2018--2022 & 216   & 0.924  & 1.088  &  2.535 &  -    \\
\hline
\end{tabular}
\end{table}

In contrast, DE430 exhibits a noticeable degradation after 2016. This behavior is consistent with the fact that DE430 was constructed using LLR observations only up to 2012 \cite{Folkner2014}. Moreover, the weighted least-squares adjustment is performed over the full observation span. The recent high-precision measurements hence exert a stronger influence on the estimated parameters. The results represent a compromise between different periods and may partially reduce the WRMS of the later observations at the expense of an increase in the earlier-period WRMS. This effect also helps explain why the annual WRMS values obtained with DE430 are generally larger than those obtained with EPM21 and INPOP21 after 2006. To reduce the influence of this adjustment-induced redistribution, the DE430 solution is restricted to observations up to 2016 rather than using the full data set. With this restriction, the annual WRMS values over the corresponding period are comparable to those obtained with EPM21 and INPOP21, as shown in Table \ref{T3}. The residual statistics also reveal differences among the observing stations. APOLLO, OCA and WLRS achieve the best performance, with WRMS values at the 1-2 cm level. The other stations exhibit larger WRMS values and greater residual scatter. We also compared the WRMS values of the post-fit residuals obtained in this work with those published for the corresponding ephemerides \cite{Kan2021,Fienga2021}. The close agreement demonstrates the feasibility of the developed LLR data-reduction framework.
\subsection{EP Parameter Estimation}
According to the model in Sec. \ref{section2.1}, an EP violation in the Earth–Moon system would lead to a periodic variation in the Earth–Moon range, appearing as a $\cos D$ signature. It should be noted that standard planetary and lunar ephemerides are generally constructed under the assumption that the EP is valid \cite{Fienga2024}. Consequently, if such a signal exists and is not completely absorbed by the parameters adjusted during the construction of the ephemeris or the subsequent LLR adjustment described in Sec. \ref{section4.1}, it may be detectable in the post-fit O–C residuals. The essential task is therefore to estimate the amplitude of this signature. 

However, the LLR observations are not uniformly distributed over the lunar synodic angle $D$, which may affect the separation of the fundamental and higher-order harmonic components, leading to a bias in the estimated amplitude of the $\cos D$ term \cite{Muller1998, Williams2009}. In addition, we also noticed that intrinsic synodic differences exist among the three adopted ephemerides in the Earth--Moon distance, which are primarily attributable to differences in the treatment of solar radiation pressure. Therefore, we first examine the impact of these two effects before deriving the final result.
\vspace{-5pt}
\subsubsection{Harmonic leakage induced by nonuniform synodic-angle sampling}\label{section4.2.1}
\begin{figure*}[t!]
\includegraphics[width=0.95\textwidth]{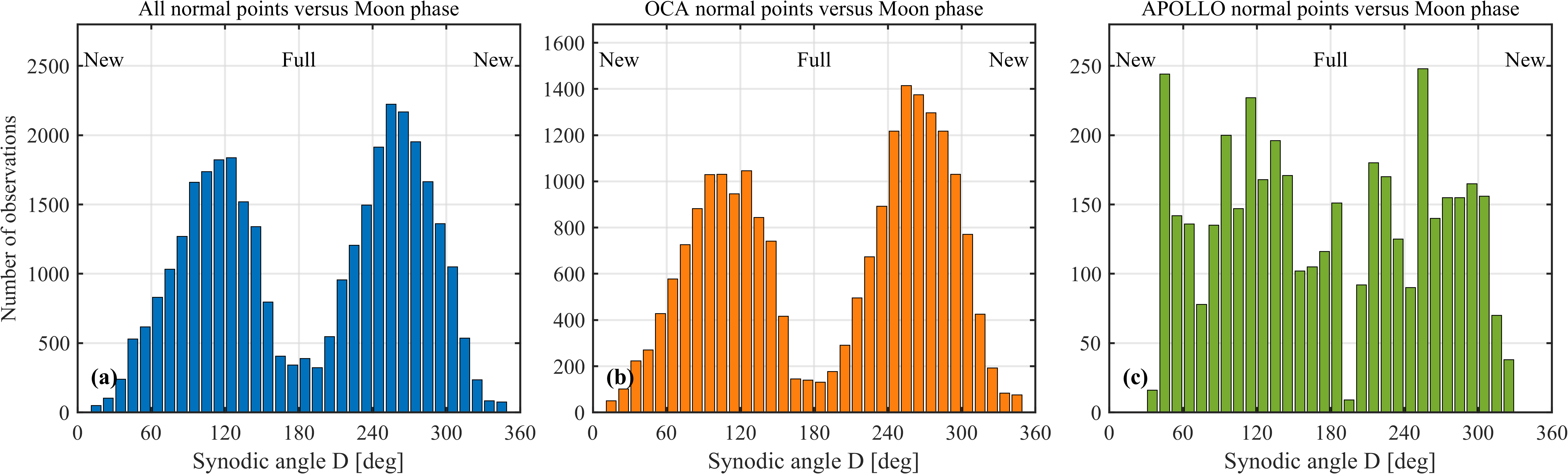}
\caption{Distribution of the LLR normal points used in this study over the lunar synodic angle \(D\) for the ALL data set, OCA, and APOLLO.}
\label{Fig14}
\end{figure*}

The lunar synodic angle $D$ is calculated following the algorithm proposed in Ref. \cite{Montenbruck2000}, not a Delaunay argument expression. Fig.~\ref{Fig14} shows the distributions of normal points with respect to the lunar synodic angle. It can be seen that fewer observations are available around new and full Moon in all three cases. For a uniform distribution over $D$, the different synodic harmonic terms are mutually orthogonal and can therefore be estimated independently. The nonuniform distribution in Fig. \ref{Fig14}  breaks this orthogonality and introduces correlations among them. If the one-way O--C residuals are fitted using only a $\cos D$ term, Eq. (\ref{N26}) can be used directly:
\begin{equation}\label{N29}
{\hat a_D} = {\left( {{\mathbf{h}^T}\mathbf{W}\mathbf{h}} \right)^{ - 1}}{\mathbf{h}^T}\mathbf{W} \bm{\upvarrho},
\end{equation}
where $\bm{\upvarrho}$ is the vector of one-way O--C residuals, and the design matrix $\mathbf{h} = \left[ \cos D_1,\; \cos D_2,\; \ldots,\; \cos D_N \right]^\mathrm{T}$. The corresponding formal uncertainty is
\begin{equation}\label{N30}
{\sigma _{{{\hat a}_D}}} = \sqrt {\frac{{\sigma _0^2}}{{{\mathbf{h}^T}\mathbf{W}\mathbf{h}}}},
\end{equation}
However, if the residual signal contains higher synodic harmonic terms:
\begin{equation}\label{N31}
{\varrho} (D) = {a_D}\cos D + \sum\limits_{n = 2}^m {{a_{nD}}} \cos (nD) + \zeta ,
\end{equation}
where $a_{nD}$ is the corresponding amplitude of $n$th-order harmonic, $\zeta$ is the unmodeled noise or signal. The leakage coefficient of an unmodeled $n$th-order harmonic into the estimated $\cos D$ coefficient can be written as:
\begin{equation}\label{N32}
{L_n}({\mathbf{q}_n} \to \cos D) = \frac{{{\mathbf{h}^T}\mathbf{W}{\mathbf{q}_n}}}{{{\mathbf{h}^T}\mathbf{W}\mathbf{h}}}.
\end{equation}
with
\begin{equation}\label{N33}
{\mathbf{q}_n} = {\left[ {\begin{array}{*{20}{c}}
  {\cos n{D_1}}&{\cos n{D_2}}& \cdots &{\cos n{D_N}} 
\end{array}} \right]^T},
\end{equation}
Thus, the estimated amplitude bias is:
\begin{equation}\label{N34}
\delta {\hat a_D} = \sum\limits_{n = 2}^m {{a_{nD}}{L_n}({\mathbf{q}_n} \to \cos D)} .
\end{equation}

In fact, this result can be generalized to a model that includes both cosine and sine harmonics up to a given order. If Eq. \ref{N31} is fitted using harmonics up to a higher order, the leakage coefficient associated with a remaining unmodeled term can be written as:
\begin{equation}\label{N35}
{L_n}({\mathbf{q}_{n,u}} \to \cos D) = \frac{{{{[(\mathbf{I} - {\mathbf{P}_{n,f}})\mathbf{h}]}^T}\mathbf{W}[(\mathbf{I} - {\mathbf{P}_{n,f}}){\mathbf{q}_{n,u}}]}}{{{{[(\mathbf{I} - {\mathbf{P}_{n,f}})\mathbf{h}]}^T}\mathbf{W}[(\mathbf{I} - {\mathbf{P}_{n,f}})\mathbf{h}]}},
\end{equation}
with weighted projection matrix
\begin{equation}\label{N36}
{\mathbf{P}_{n,f}} = {\mathbf{q}_{n,f}}{(\mathbf{q}_{n,f}^T\mathbf{W}{\mathbf{q}_{n,f}})^{ - 1}}\mathbf{q}_{n,f}^T\mathbf{W}.
\end{equation}
where the subscripts $f$ and $u$ denote the fitted and unmodeled terms, respectively. The formal uncertainty of the estimated amplitude is also increased to:
\begin{equation}\label{N37}
{{\sigma _{{{\hat a}_{D,f}}}}} \approx {\sigma _{{{\hat a}_D}}}\frac{1}{{\sqrt {1 - R_Z^2} }},
\end{equation}
with
\begin{equation}\label{N38}
R_Z^2 = \frac{{{\mathbf{h}^T}\mathbf{W}{\mathbf{P}_{n,f}}\mathbf{h}}}{{{\mathbf{h}^T}\mathbf{W}\mathbf{h}}}.
\end{equation}

\begin{figure*}[htbp]
\includegraphics[width=0.43\textwidth]{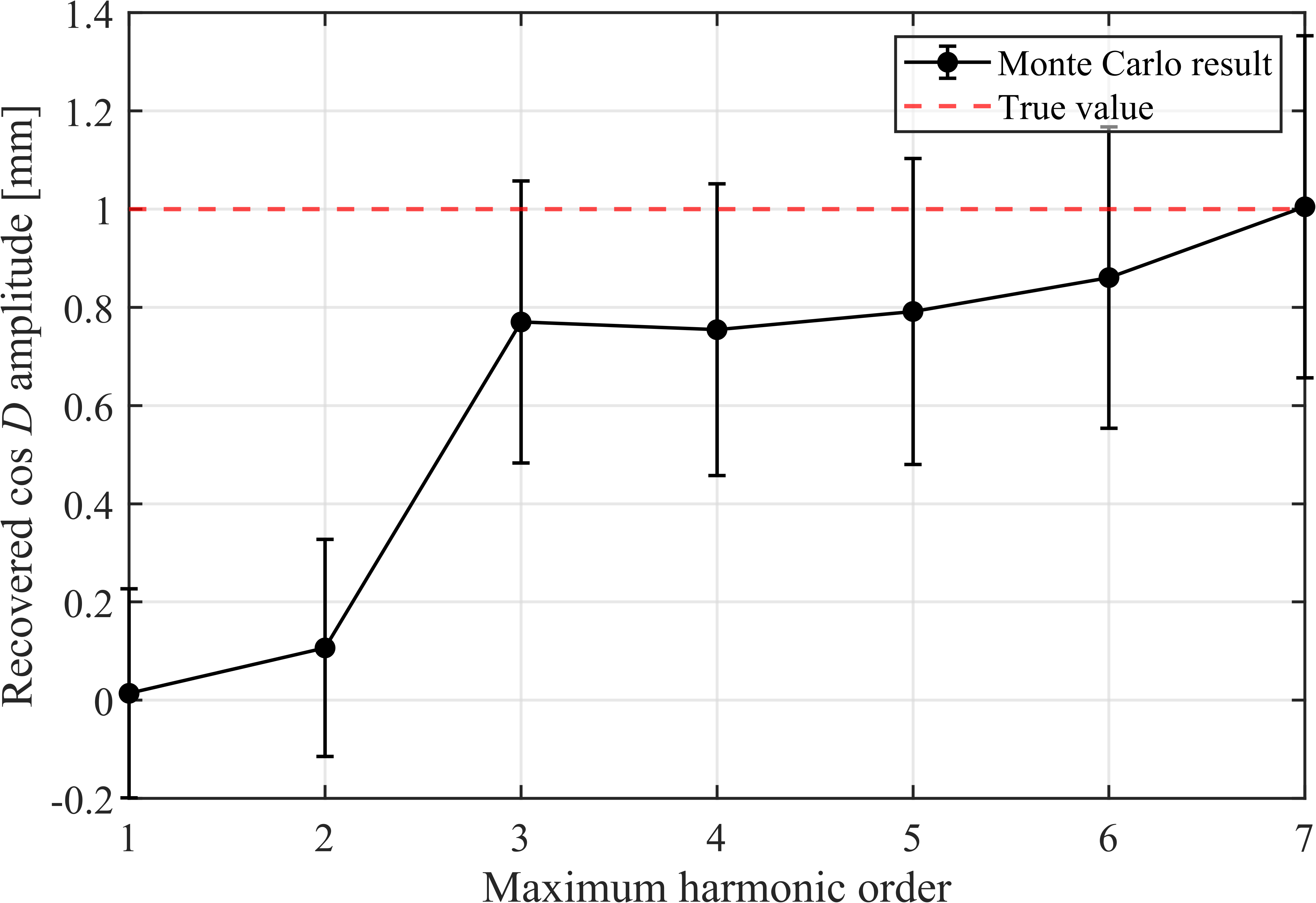}
\caption{ Recovery of the $\cos D$ amplitude in the OCA harmonic leakage simulation.}
\label{Fig15}
\end{figure*}

When only the $\cos D$ term and a constant are estimated, the leakage coefficients of the higher-order harmonics can be calculated using Eq. \ref{N35}. For all LLR observations from 1970 to 2024,  the largest leakage coefficients were found for the $2D$ and $3D$ harmonics, with values of $-0.388$ and $-0.761$, respectively, while the absolute values of other coefficients were smaller than 0.1. For the OCA observations, the corresponding coefficients are $-0.165$ and $-0.850$, respectively, whereas for APOLLO they are $-0.610$ and $-0.630$. To illustrate this effect more directly, we generated a simulated signal using the OCA observations. The simulated signal contained harmonic terms up to the $7D$. The amplitudes of the first three harmonics were set to 1 mm, while the remaining harmonics were set to 0.3 mm. Gaussian noise was added according to the measured uncertainty of each observation. The simulated data were fitted with harmonic models of increasing maximum order, and the results are shown in Fig. \ref{Fig15}. The recovered $\cos D$ amplitude changes when only very low-order models are used, showing the leakage from unmodeled harmonics. After harmonics up to about the $3D$ order are included, the estimate becomes stable, because both the coefficients and leakage coefficients of the higher-order terms are small in this example. The formal uncertainty also covers the true value, indicating that the remaining leakage does not produce a significant bias in this test. This example provides a useful guide for the analysis of the real residuals.

\vspace{-10pt}
\subsubsection{Differences in the Earth--Moon distance among ephemerides}
\begin{figure*}[htbp]
\includegraphics[width=0.85\textwidth]{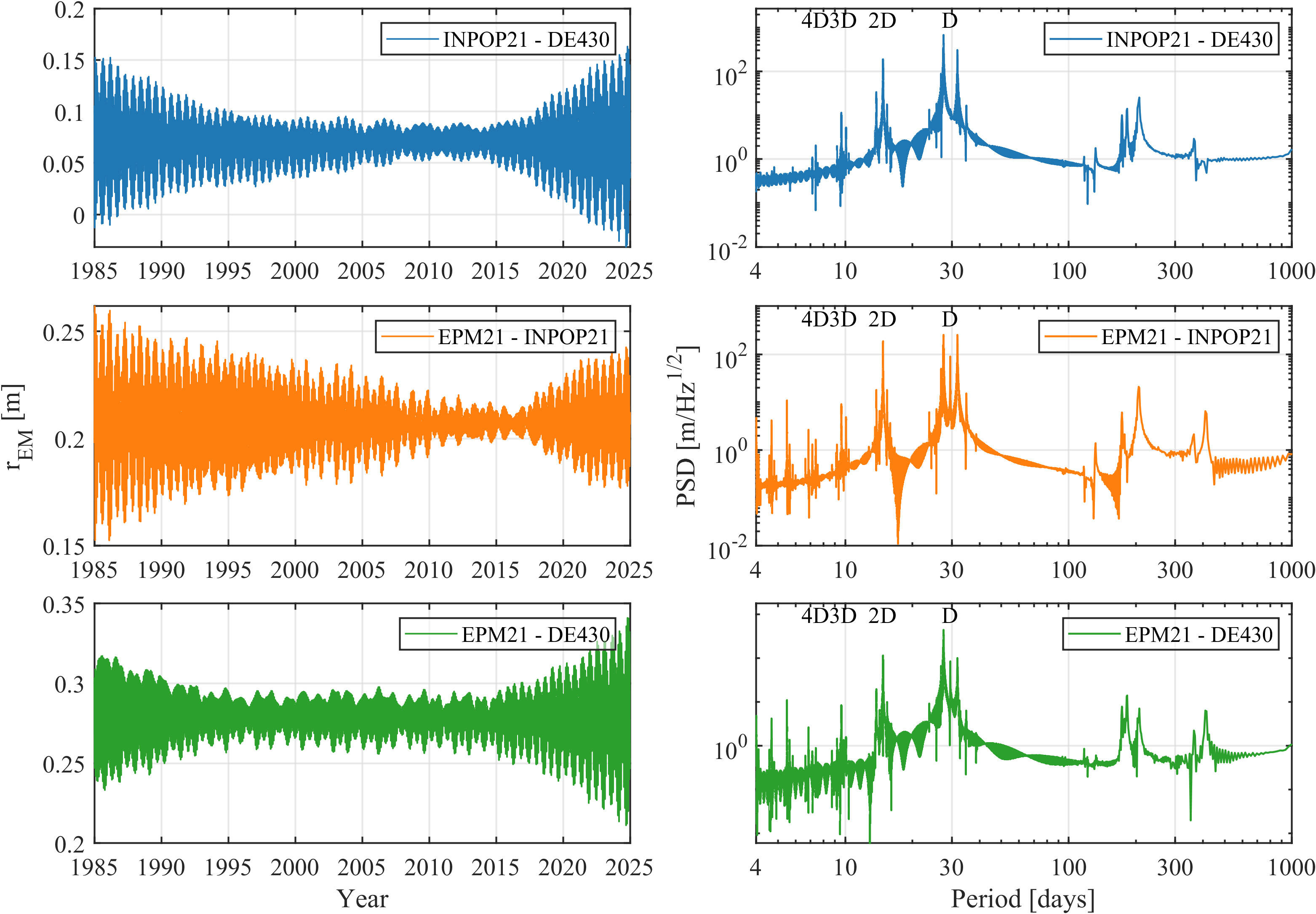}
\caption{ Time-domain and frequency-domain differences in the Earth--Moon distance among  EPM21, INPOP21, and DE430.}
\label{Fig16}
\end{figure*}

During the analysis, we found that the different dynamical models adopted by the three ephemerides can introduce synodic signatures in the modeled Earth--Moon distance. To investigate this effect, Earth--Moon distances were computed separately using EPM21, INPOP21, and DE430 over the period from 1985 to 2025, with a sampling interval of 2 days. The differences between the simulated Earth--Moon distances were calculated, and the results are shown in Fig. \ref{Fig16}. In the time domain, these differences vary by less than 0.2 m. In the frequency domain, a spectral peak at the synodic period appears in the EPM21--INPOP21 and EPM21--DE430 differences, whereas no comparable peak is found in the INPOP21--DE430 difference. We then fitted these differences with $\cos nD$ terms. The simulated data were uniformly sampled, so harmonic leakage is negligible in this test. The dominant fitted terms are listed in Table \ref{T4}.

\vspace{-5pt}
\begin{table}[htbp]
\centering
\caption{Fitted harmonic coefficients of the Earth--Moon distance differences among  EPM21, INPOP21, and DE430.}
\label{T4}
\setlength{\tabcolsep}{10pt}   
\renewcommand{\arraystretch}{1.5}
\begin{tabular}{c c c c}
\hline
$r_{EM}$ difference & $\cos D$ coefficient [mm] & $\cos 2D$ coefficient [mm] & Constant term [mm] \\
\hline
INPOP21$-$DE430  & $0.236 \pm 0.436$ & $-0.354 \pm 0.434$ & $72.003 \pm 0.307$ \\
EPM21$-$INPOP21  & $-3.511 \pm 0.232$ & $-1.346 \pm 0.231$ & $207.106 \pm 0.164$ \\
EPM21$-$DE430    & $-3.275 \pm 0.284$ & $-1.700 \pm 0.282$ & $279.109 \pm 0.200$ \\
\hline
\multicolumn{4}{l}{\footnotesize The given uncertainties are 1$\sigma$ standard deviations from the WLS.} \\
\end{tabular}
\end{table}
\vspace{-5pt}
According to Ref. \cite{Pavlov2026}, the EPM21 includes the contribution from the solar radiation pressure (SRP) on the Earth and the Moon. In contrast, this effect is not modeled in INPOP21 and DE430 based on the discussions in Refs. \cite{Williams2009, ViswanathanPhd}. The expected range signature caused by SRP is about $-3.65 \pm 0.08$ mm, which is comparable to the fitted synodic amplitudes found in the EPM21--INPOP21 and EPM21--DE430 differences. This suggests that the synodic-period difference involving EPM21 is likely related to whether lunar SRP is modeled in the ephemeris construction. Nevertheless, because the ephemerides also differ in other modeling choices and data weighting, SRP should be considered a plausible source of the difference rather than a definitive explanation.

\vspace{-10pt}
\subsubsection{EP Parameter Estimation from the post-fit residuals}\label{section4.2.3}
\begin{figure*}[htbp]
\includegraphics[width=0.43\textwidth]{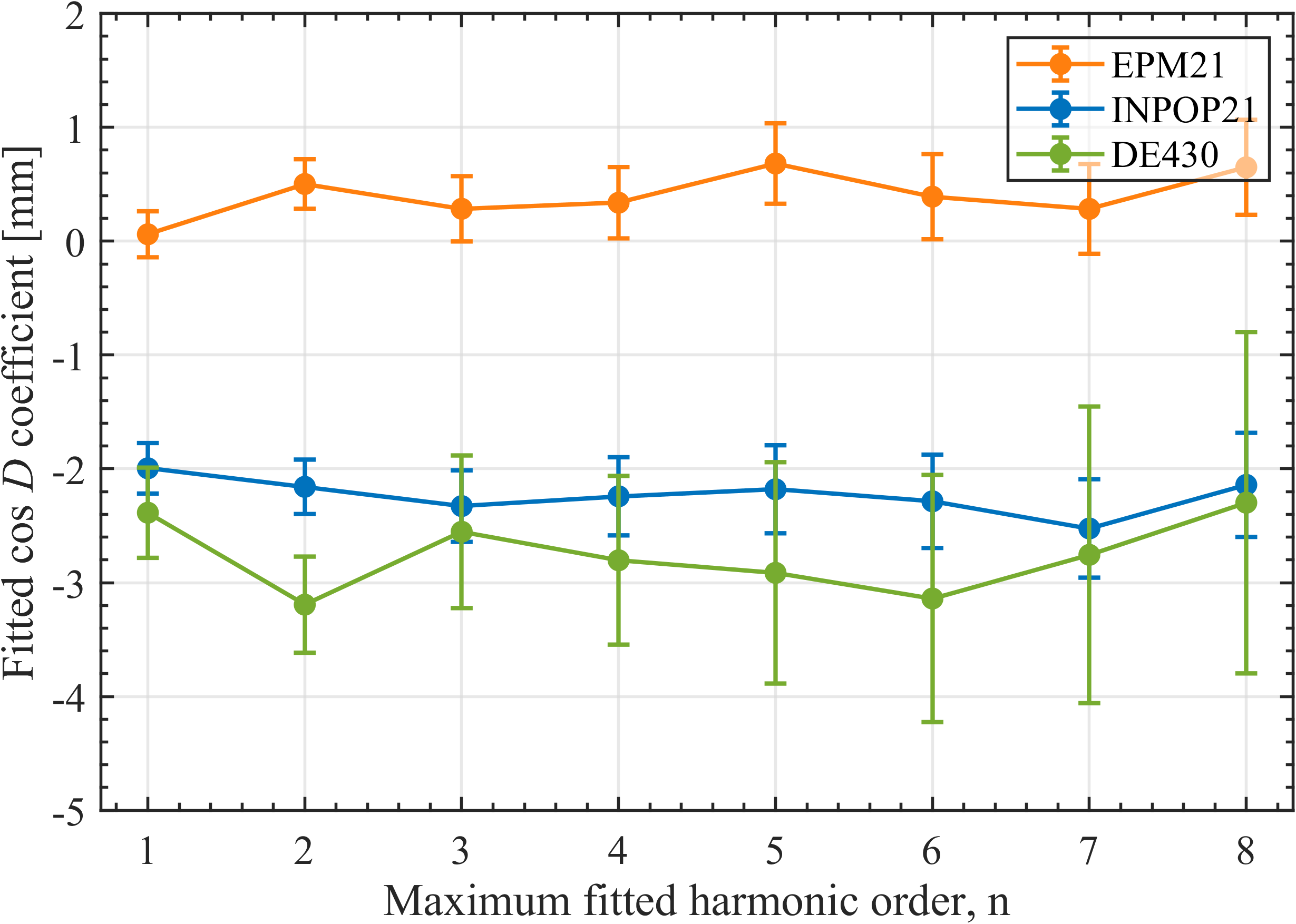}
\caption{Fitted $\cos D$ coefficient as a function of the maximum harmonic order $n$ for EPM21 and INPOP21 using observations through 2024, and for DE430 using observations through 2016. Error bars indicate the formal uncertainties from the WLS.}
\label{Fig17}
\end{figure*}

\begin{figure*}[htbp]
\centering
\subfigure{%
    \includegraphics[width=3.15in]{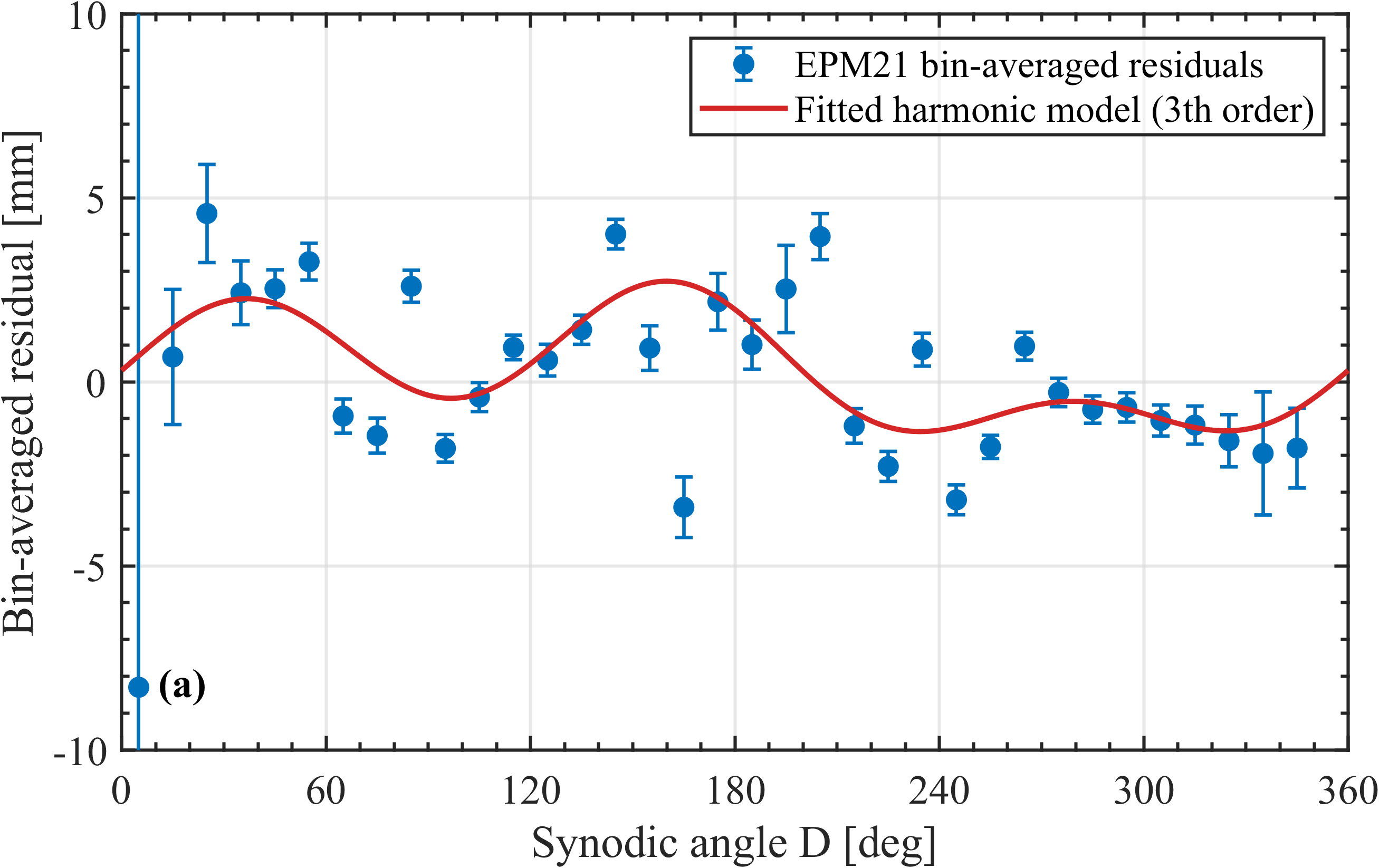}}
\hspace{0.25in}
\subfigure{%
    \includegraphics[width=3.15in]{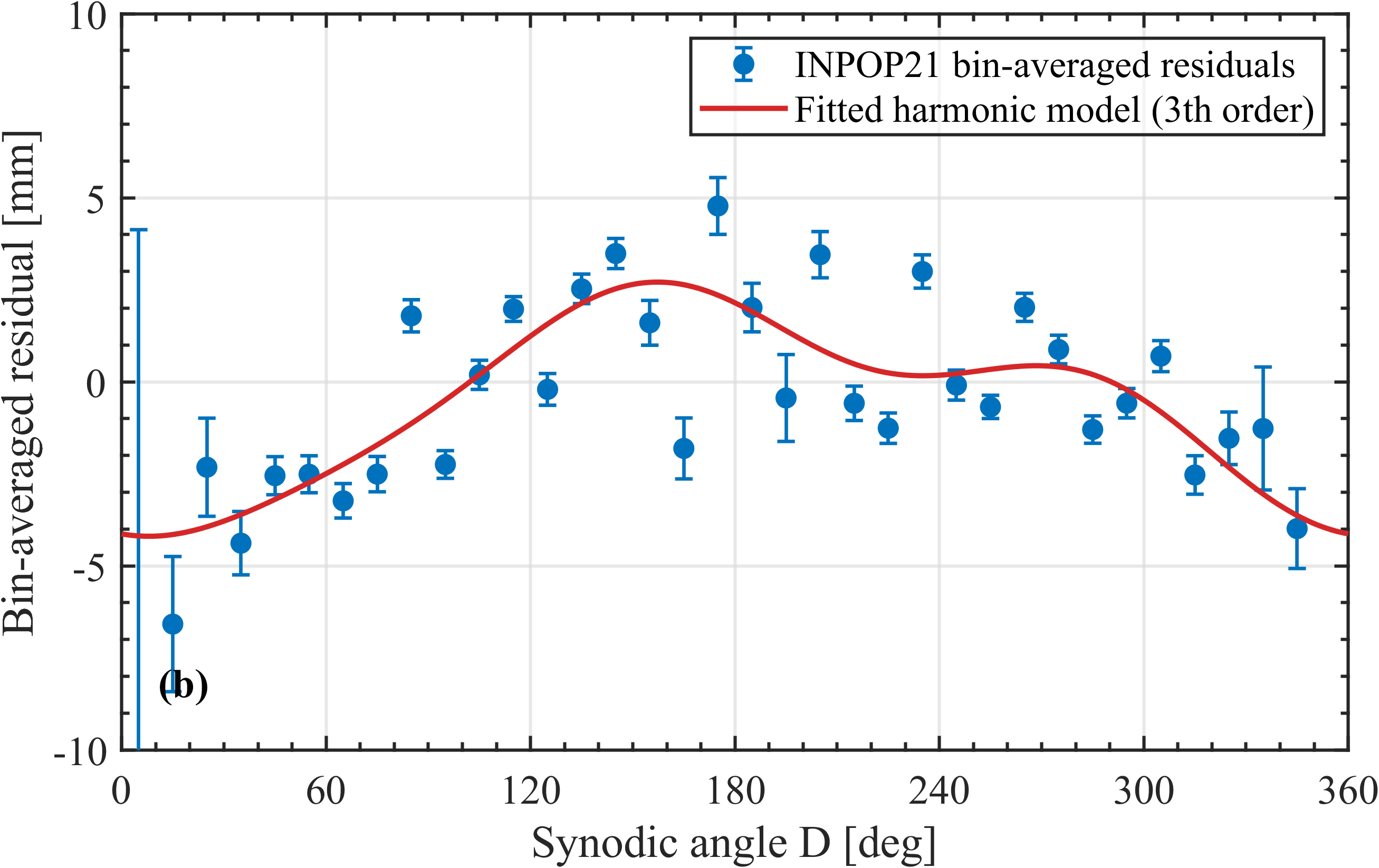}}
\vspace{0.25cm}
\subfigure{%
    \includegraphics[width=3.15in]{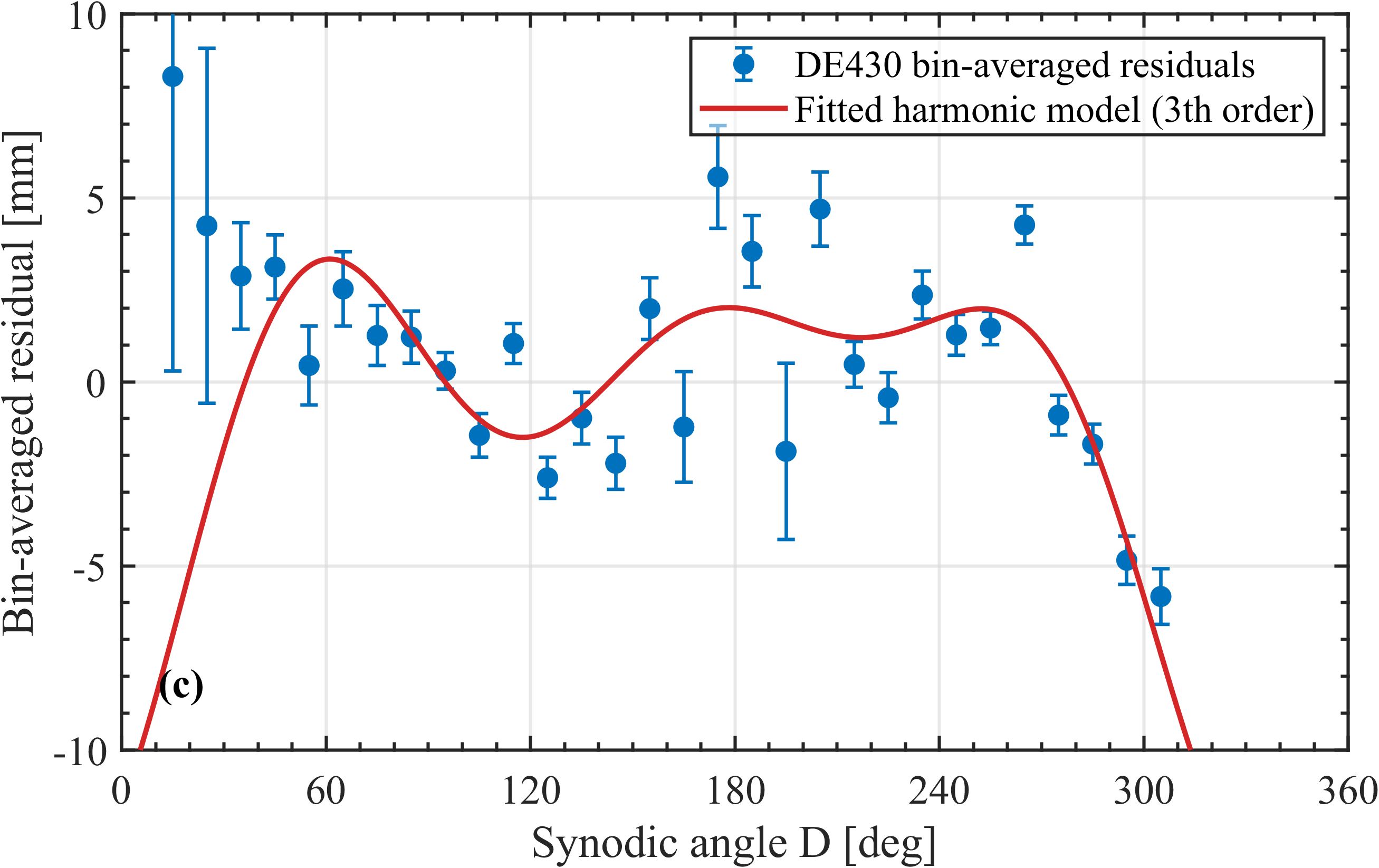}}
\caption{Bin-averaged residuals with their uncertainties and fitted harmonic curves corresponding to the $n=3$ solutions shown in Fig. 17 for (a) EPM21, (b) INPOP21, and (c) DE430.}
\label{Fig18}
\end{figure*}

Based on the post-fit O--C residuals presented in Sec. \ref{section4.1}, EPM21 and INPOP21 are adopted as the primary ephemerides in the following analysis because they provide comparable residual levels over all of the observation span. DE430 is retained as a supplementary solution to assess the ephemeris dependence of the results. Following the method described in Sec. \ref{section4.2.1}, the post-fit residuals were fitted with both cosine and sine harmonics up to the $n$th order, together with a constant term. Fig. \ref{Fig17} shows the variation of the fitted $\cos D$ coefficient with the maximum harmonic order $n$. The EPM21 and INPOP21 results were obtained using the full data set, whereas the DE430 was restricted to observations up to 2016. When the maximum fitted harmonic order is varied from 3 to 8, the coefficients of $\cos D$ obtained with EPM21 and INPOP21 remain stable. In contrast, because the restricted data span contains fewer observations and provides weaker constraints on the higher-order harmonics, the DE430 result shows larger variations. Therefore, we adopt the lowest order within this stable range, namely $n = 3$, which includes the dominant leakage terms while avoiding unnecessary high-order correlations. The variations obtained with higher orders will be considered in the final uncertainty estimate. For a clearer comparison between the residuals and the fitted harmonic models, we also performed a bin-averaged analysis. The synodic angle $D$ was divided into 36 bins of width $10^\circ$, and the inverse-variance-weighted mean residual was calculated within each bin. The representative $D$ values were taken as the bin center, i.e., $5^\circ, 15^\circ, 25^\circ,\ldots$, and the uncertainties of the bin-averaged residuals were determined from the measurement uncertainties of the included observations. The corresponding results for the three ephemerides are shown in Fig. \ref{Fig18}. The INPOP21 residuals show a larger $\cos D$ component than the EPM21 residuals. It should be emphasized that the fitted curves in Fig. \ref{Fig18} were reconstructed from the $n = 3$ coefficients shown in Fig. \ref{Fig17}, rather than obtained by fitting the bin-averaged residuals directly. 


\begin{table}[htbp]
\centering
\caption{Fitted $\cos D$ coefficients obtained from the post-fit residuals using harmonics up to the third order ($n=3$).}
\label{T5}
\renewcommand{\arraystretch}{1.5}
\setlength{\tabcolsep}{12pt}   %
\begin{tabular}{c c c c c}
\hline
Data set & Time span & EPM21 [mm] & INPOP21 [mm] & DE430 [mm] \\
\hline
\multirow{2}{*}{ALL} 
         & 1970--2024 & $0.285 \pm 0.264$ & $-2.328 \pm 0.314$ &  $-$\\
         & 1970--2016 &$0.192 \pm 0.662$  & $-2.378 \pm 0.764$ & $-2.612 \pm 0.671$ \\
\multirow{2}{*}{OCA} 
         & 1984--2024 & $-0.186 \pm 0.272$ & $-2.414 \pm 0.315$ & $-$ \\
         & 1984--2016 &$0.389 \pm 0.706$  &$-2.629 \pm 0.831$  & $-2.553 \pm 0.696$ \\
\multirow{2}{*}{APOLLO}
         & 2006--2024 & $-0.271 \pm 0.721$ & $-2.450 \pm 0.716$ & $-$ \\
         & 2006--2016 &$0.149\pm 1.118$  &$-3.052 \pm 1.106$  & $-2.669 \pm 1.081$ \\
\hline
\multicolumn{5}{l}{\footnotesize The given uncertainties are 1$\sigma$ standard deviations from the WLS.} \\
\end{tabular}
\end{table}

\begin{table}[htbp]
\centering
\caption{Final corrected $\cos D$ coefficients and their conservative combined uncertainties. The central coefficients are corrected for signal attenuation and known non-EP synodic effects, whereas the uncertainties are calculated using Eq.~(\ref{N39}).}
\label{T6}
\renewcommand{\arraystretch}{1.5}
\setlength{\tabcolsep}{12pt}   %
\begin{tabular}{c c c c c}
\hline
Data set & Time span & EPM21 [mm] & INPOP21 [mm] & DE430 [mm] \\
\hline
\multirow{2}{*}{ALL} 
         & 1970--2024 & $-0.265 \pm 0.937$ & $-0.146 \pm 1.066$ &  $-$\\
         & 1970--2016 &$-0.372 \pm 2.048$  & $-0.446 \pm 2.346$ & $-0.786 \pm 2.074$ \\
\multirow{2}{*}{OCA} 
         & 1984--2024 & $-0.863 \pm 0.957$ & $0.238 \pm 1.069$ & $-$ \\
         & 1984--2016 &$-0.158 \pm 2.176$  &$-0.328 \pm 2.543$  & $-0.232 \pm 2.147$ \\
\multirow{2}{*}{APOLLO}
         & 2006--2024 & $-1.134 \pm2.220$ & $-1.375 \pm 2.205$ & $-$ \\
         & 2006--2016 &$-0.406\pm 3.391$  &$-2.003 \pm 3.355$  & $-1.375 \pm 3.281$ \\
\hline
\end{tabular}
\end{table}

The resulting $\cos D$ coefficients obtained from the third-order harmonic fits are listed in Table \ref{T5}. In addition to the results obtained using all available LLR normal points, the results for OCA and APOLLO are also presented. OCA provides the largest number of normal points, while APOLLO offers the highest ranging precision. Among the station-specific subsets, OCA gives the smallest formal uncertainty because of its large number of observations and relatively good distribution over the angle $D$. 
In contrast, APOLLO has a larger formal uncertainty than OCA. This does not reflect its ranging precision. As shown in Fig.~\ref{Fig14}, APOLLO contains fewer normal points and provides sparser coverage over \(D\) than OCA. In particular, APOLLO has almost no observations near new Moon (\(D<30^\circ\) or \(D>330^\circ\)) and a limited number around full Moon (\(D\simeq180^\circ\)). Although the OCA observations also have a nonuniform distribution, their coverage is denser and more continuous. This difference leads to stronger correlations between the \(\cos D\) and higher-order harmonic terms for APOLLO, thereby increasing the formal uncertainty of the fitted coefficient.
 To facilitate a direct comparison among the ephemerides, Table \ref{T5} also presents the fitted results obtained over different data spans. For all data sets in Table \ref{T5}, the EPM21 solutions yield a $\cos D$ coefficient close to zero, whereas the INPOP21 and DE430 give coefficients of about $-2.5$ mm. This discrepancy is comparable with the $\cos D$ coefficients of the modeled Earth--Moon distance differences listed in Table \ref{T4}. Since the residuals are defined as O--C, the differences in the residual-based coefficients are expected to have the opposite sign to the corresponding differences in the modeled distances. 

To quantify the absorption of a possible EP signal by the preliminary adjustment of the other parameters described in Sec. \ref{section4.1}, we performed injection tests by adding $\cos D$ signals with amplitudes of \(-1,-2,-3,-4,\) and \(-5\) mm before repeating the adjustment. The recovered amplitudes show that the adjustment partially absorbs the injected signal. For the complete 1970--2024 data set, approximately 74.1\% of the injected amplitude is retained, corresponding to an absorption of about 25.9\%. A weighted-projection analysis of the attenuation and its comparison with the injection results are presented in Appendix~\ref{app:injection}. Accordingly, the fitted coefficients listed in Table \ref{T5} should be corrected for signal attenuation by dividing them by the corresponding recovery factor. 

In addition, the formal uncertainties from the WLS fit are likely to underestimate the realistic uncertainties because remaining modeling deficiencies and correlations are not fully represented by the formal WLS uncertainties. An exact estimation of the realistic uncertainty is difficult, since it would require identifying and propagating the uncertainty of each individual mismodeled effect~\cite{Muller2007}. A common treatment in LLR EP analysis is to construct a realistic uncertainty by multiplying the WLS formal uncertainty by an empirical factor and then adding other identified uncertainty contributions in quadrature~\cite{Williams1996}. Based on estimates of modeling discrepancies and empirical studies, Müller et al adopted a factor of 3 and applied it to the uncertainties reported in their analysis ~\cite{Muller2012a}. Subsequently, this factor has been adopted in many LLR EP studies~\cite{Viswanathan2018,Muller2014c,Hofmann2017,Hofmann2018,Biskupek2021, Pavlov2023}. Following this treatment and to maintain comparability with previous LLR results, we multiply the formal uncertainties by a factor of 3. In fact, this scaling also covers variations arising from the adopted maximum harmonic order \(n\) and other remaining modeling uncertainties. The attenuation correction discussed above is applied only to the estimated central coefficients, and the uncertainties are not additionally divided by the recovery factor.

Additional corrections are required to remove known non-EP synodic effects. Previous studies reported a SRP contribution of $-3.65 \pm 0.08$ mm to the modeled $\cos D$ range term \cite{Fienga2024}. When the lunar reflector thermal effect is included, the combined thermal and SRP contribution becomes $-3.0 \pm 0.5$ mm \cite{Williams2009}. EPM21 includes the lunar SRP contribution \cite{Pavlov2026}, whereas INPOP21 and DE430 do not \cite{Williams2009,ViswanathanPhd}. The lunar reflector thermal effect is not included in any of the three ephemerides used here. Taking into account the O--C sign convention, a correction of $-0.65 \pm 0.5$ mm is therefore applied to the EPM21 coefficient, while a correction of $3.0 \pm 0.5$ mm is applied to the INPOP21 and DE430 coefficients. The uncertainty can be written as:
\begin{equation}\label{N39}
{\sigma _{{\text{final}}}} = \sqrt {{{(3 \times {\sigma _{{\text{formal}}}})}^2} + {{(0.50 \; \mathrm{mm})}^2}} .
\end{equation}

The final corrected $\cos D$ amplitudes corresponding to the fitted coefficients in Table \ref{T5} are presented in Table \ref{T6}. The central coefficients have been corrected for signal attenuation and the known non-EP synodic effects. Their uncertainties are calculated using Eq. (\ref{N39}) and should be regarded as conservative estimates. For each ephemeris, the most suitable data span is adopted: 1970--2024 for EPM21 and INPOP21, and 1970--2016 for DE430. 
The corresponding values of \(\Delta (m_g/m_i)_{\rm EM}\) and the Nordtvedt parameter \(\eta\) are listed in Table \ref{T7}. The conversion from the corrected $\cos D$ amplitude to \(\Delta (m_g/m_i)_{\rm EM}\) is made using Eq. (\ref{N4}), and \(\eta\) is then derived from Eq. (\ref{N7}) under the assumption that the signal is produced entirely by an SEP violation.

Table~\ref{T7} also compares the present results with recent global LLR solutions reported in Refs.~\cite{Hofmann2018,Viswanathan2018,Biskupek2021,Pavlov2023}. For a consistent comparison, the results are expressed in terms of the $\cos D$ coefficient, $\Delta(m_g/m_i)_{\rm EM}$, and $\eta$. Quantities not explicitly given in the original references are converted using Eqs.~(\ref{N4}) and (\ref{N6}). We also list the adopted uncertainty scale and whether the SRP and reflector thermal corrections were applied. Our present results are based on post-fit residuals obtained with fixed ephemerides, whereas the previous results were obtained from global LLR adjustments.
After applying the non-EP synodic corrections, the three ephemerides used in this work give mutually consistent results. All values of $\Delta(m_g/m_i)_{\rm EM}$ and $\eta$ are consistent with zero within their uncertainties, and no significant EP-violating $\cos D$ signal is detected in the post-fit residuals. Since the adopted ephemerides were not refitted with an explicit EP-violation parameter, these results should be interpreted as ephemeris-conditioned, residual-based constraints. As summarized in Table~\ref{T7}, our best residual-based result, $\Delta(m_g/m_i)_{\rm EM}=(0.900\pm3.183)\times10^{-14}$, has an uncertainty of the same order as recent global LLR determinations. Although the two approaches differ in their parameterization and uncertainty assessment, the residual-based method provides a complementary, ephemeris-conditioned test of the EP.

\vspace{-5pt}
\begin{table*}[htbp]
\begingroup
\centering
\caption{Comparison of the present results with previous LLR EP tests. The present results are obtained from a residual-based analysis using fixed ephemerides, whereas the previous results are derived from global LLR adjustments. An asterisk denotes treatments inferred from the earlier IfE solution~\cite{Hofmann2018}. A dagger denotes quantities converted in this work using Eqs.~(\ref{N4}) and (\ref{N6}).}
\label{T7}
\footnotesize
\renewcommand{\arraystretch}{1.75}
\setlength{\tabcolsep}{3.0pt}
\begin{tabular}{lccccccc}
\hline
\shortstack[c]{Solution or reference} & \shortstack[c]{Data span} & \shortstack[c]{Uncertainty\\scale} & \shortstack[c]{Corrected $\cos D$\\$[\mathrm{mm}]$} & \shortstack[c]{$\Delta(m_g/m_i)_{\rm EM}$\\$[10^{-14}]$} & \shortstack[c]{$\eta$\\$[10^{-4}]$} & \shortstack[c]{SRP\\correction} & \shortstack[c]{Reflector thermal\\correction} \\
\hline
This work, EPM21 & 1970--2024 & 3$\times \sigma$ & $-0.265\pm0.937$ & $0.900\pm3.183$ & $-0.202\pm0.715$ & $\surd$ & $\surd$ \\
This work, INPOP21 & 1970--2024 & 3$\times \sigma$ & $-0.146\pm1.066$ & $0.496\pm3.624$ & $-0.111\pm0.814$ & $\surd$ & $\surd$ \\
This work, DE430 & 1970--2016 & 3$\times \sigma$ & $-0.786\pm2.074$ & $2.669\pm7.048$ & $-0.600\pm1.583$ & $\surd$ & $\surd$ \\
Hofmann  et al.~\cite{Hofmann2018} & 1969--2016 & 3$\times \sigma$ & $0.9\pm1.4$ & $-3.0\pm5.0$ & $0.7\pm1.1$ & $\surd$ & $\times$ \\
INPOP17a~\cite{Viswanathan2018} & 1969--2017 & 3$\times \sigma$ & $1.1\pm2.1$ & $-3.8\pm7.1$ & $0.85\pm1.59$ & $\surd$ & $\surd$ \\
Biskupek et al.~\cite{Biskupek2021} & 1970--2020 & 3$\times \sigma$\textsuperscript{*} & $0.62\pm0.71^{\dagger}$ & $-2.1\pm2.4$ & $0.47\pm0.54^{\dagger}$ & $\surd^{*}$ & $\times^{*}$ \\
Pavlov et al.~\cite{Pavlov2023} & 1970--2021 & 3$\times \sigma$ & $-0.29\pm0.88^{\dagger}$ & $1.0\pm3.0$ & $-0.22\pm0.67^{\dagger}$ & $\surd$ & $\times$ \\
\hline
\end{tabular}
\endgroup
\end{table*}
\section{Discussion and Conclusions}\label{section5}
In this work, we developed an independent LLR data-reduction and residual-analysis framework to investigate possible EP-violating signatures using three planetary and lunar ephemerides: EPM21, INPOP21, and DE430. The same observation model, weighting strategy, and parameter-adjustment procedure were applied to all three solutions. The effects of the nonuniform distribution of LLR observations over the synodic angle $D$ were examined through a multi-harmonic analysis, while signal-injection tests were used to quantify the attenuation of a possible $\cos D$ signal during the parameter adjustment. After correcting for signal attenuation, solar radiation pressure, and the thermal response of the lunar retroreflectors, the preferred EPM21, INPOP21, and DE430 solutions give $\Delta(m_g/m_i)_{\rm EM}=(0.900\pm3.183)\times10^{-14}, (0.496\pm3.624)\times10^{-14}$, and  $ (2.669\pm7.048)\times10^{-14} $, respectively. The EPM21 and INPOP21 results were obtained using observations from 1970 to 2024, whereas the DE430 result was obtained using observations through 2016. All three results are consistent with zero within their uncertainties and are mutually consistent despite the differences among the adopted ephemerides. The uncertainty of our best estimate is of the same order as those of the recent global LLR EP constraints, as summarized in Table~\ref{T7}.

The present results should nevertheless be interpreted as ephemeris-conditioned, residual-based constraints. An important assumption is that a possible EP-violation signal was not substantially absorbed by the parameters adjusted during the original construction of the adopted ephemerides. The injection tests performed in this work quantify the additional absorption introduced by our own parameter adjustment, but they cannot determine the amount of absorption that may already have occurred in the ephemeris construction. The residual-based approach is particularly sensitive to the distribution of observations over \(D\). Uneven synodic-angle coverage increases the correlations between the fundamental $\cos D$ term and higher-order harmonics  \cite{Muller1998}. Therefore, for the same observational data, a residual-based analysis may yield a somewhat larger uncertainty than a global solution in which the EP parameter is estimated simultaneously with the relevant dynamical and observational parameters. Nevertheless, the residual approach remains useful because it is independent, transparent, and relatively straightforward to apply to multiple ephemerides. It also provides a direct means of identifying ephemeris-dependent signatures and remaining unmodeled effects in the post-fit residuals.

DE440 was not included in the final analysis because it did not produce a satisfactory post-fit residual distribution within the common reduction framework adopted in this work. This may be related to improvements in the dynamical, observational, and reference-frame models introduced during the construction of DE440 that are not yet fully represented in our present implementation  \cite{Park2021, Williams2020}. This should not be regarded as a limitation of DE440 itself, but rather as an indication that the consistent application of a more recent ephemeris may require corresponding updates to the complete LLR reduction model. The next step will therefore be to extend the present framework to a full global adjustment in which an explicit EP-violation parameter is estimated together with the relevant lunar dynamical parameters, station and reflector coordinates, and other correlated parameters. Such an analysis will allow a direct comparison between the residual-based and global-fit approaches under the same observational and physical models. The present framework also provides a basis for analyzing future high-precision LLR observations and for assessing ephemeris-dependent effects in lunar mission applications.
\section*{ACKNOWLEDGMENTS}
Current LLR data were collected, archived, and distributed under the auspices of the International Laser Ranging Service (ILRS) \cite{Pearlman2002}. The Paris Observatory Lunar Analysis Centre website also carefully collected LLR observations from different sources \cite{OCAwebsite}. We acknowledge with thanks that more than 55 years of processed LLR data has been obtained under the efforts of the personnel at the Observatoire de la Côte d'Azur in France, the LURE Observatory in Maui, Hawaii, the McDonald Observatory in Texas, the Apache Point Observatory in New Mexico, the Matera Laser Ranging station in Italy and the Wettzell Laser Ranging Station in Germany. We are grateful to Dmitry A. Pavlov who provided many valuable suggestions. This work is supported by the National Key R\rm \&D Program of China (Grant No. 2022YFC2204602), the National Natural Science Foundation of China (Grant Nos. 12505072 and 12150012), the China Postdoctoral Science Foundation-Hubei Joint Support Program (Grant No. 2025T005HB),  the China Postdoctoral Science Foundation (Grant Nos. 2024M760994 and GZB20250771).

\appendix
\begingroup
\section{Calibration of the residual-based EP analysis}
\label{app:injection}

This appendix is intended to discuss whether the WLS parameter adjustment described in Sec.~\ref{section4.1} can absorb a possible EP-violation signal. One can assume that the two-way O--C residuals contain an unknown EP-violation component with a one-way amplitude \(a_D^{\rm true}\), then the residual vector can be written as:
\begin{equation}
\mathbf{y}(a_D^{\rm true})=\mathbf{y}_0+2a_D^{\rm true}\mathbf{h},
\label{eq:app_observation}
\end{equation}
where \(\mathbf{y}_0\) denotes the part of the residuals excluding the EP-violation signal, and \(\mathbf{h}=[\cos D_1,\;\cos D_2,\;\ldots,\;\cos D_N]^{\mathrm{T}}\). The EP-violation signal is not included as a solve-for parameter in the WLS adjustment. Instead, only corrections to the parameters described in Sec.~\ref{section4.1} are estimated from \(\mathbf{y}\). After correcting these parameters, the post-fit residual vector is given by \(\mathbf{y}_{\rm post}=\mathbf{y}-\mathbf{A}\Delta\mathbf{x}\). Substituting Eq.~(\ref{N26}) into this expression gives \(\mathbf{y}_{\rm post}=\mathbf{M}_{A}\mathbf{y}\), where
\begin{equation}
\mathbf{M}_{A}=\mathbf{I}-\mathbf{A}(\mathbf{A}^{\mathrm{T}}\mathbf{W}\mathbf{A})^{-1}\mathbf{A}^{\mathrm{T}}\mathbf{W}
\label{eq:app_projection}
\end{equation}
is the weighted residual-forming matrix, and \(\mathbf{I}\) is the \(N\times N\) identity matrix. Therefore, the difference between the one-way post-fit residual vectors obtained with and without the EP-violation signal is:
\begin{equation}
\begin{aligned}
\bm{\upvarrho}_{\rm post}^{(D)}=\frac{1}{2}\mathbf{M}_{A}\left[\mathbf{y}(a_D^{\rm true})-\mathbf{y}(0)\right] =a_D^{\rm true}\mathbf{M}_{A}\mathbf{h}.
\end{aligned}
\label{eq:app_postfit}
\end{equation}
Here, \(\mathbf{M}_{A}\mathbf{h}\) is the part of the EP-violation signal retained in the post-fit residuals, whereas \((\mathbf{I}-\mathbf{M}_{A})\mathbf{h}\) is the part absorbed by the adjusted parameters.

The basic principle of this absorption is the loss of orthogonality caused by the actual sampling. For continuous and uniform coverage of \(D\), different synodic harmonics satisfy \(\int_{0}^{2\pi}\cos D\cos(nD)\,dD=0\) for \(n\ne1\). For the discrete and weighted LLR observations, the inner product \(\sum_iw_i\cos D_i\cos(nD_i)\) is generally nonzero. Each column \(\mathbf{a}_j\) of \(\mathbf{A}\) represents the range signature produced by a change in the corresponding adjusted parameter. When \(\mathbf{a}_j^{\mathrm{T}}\mathbf{W}\mathbf{h}\ne0\), this signature has a nonzero weighted projection onto the \(\cos D\) signal, resulting in partial absorption of the signal. The station and reflector coordinate partial derivatives vary with the instantaneous line-of-sight direction expressed in the ITRF and lunar PA frame, respectively. Consequently, they depend on the Earth orientation, lunar position and orientation, the observed reflector, and so on. Since these geometrical configurations and their statistical weights are correlated with \(D\) through the actual observation sampling, some columns of \(\mathbf{A}\) have nonzero weighted projections onto \(\mathbf{h}\). A more direct theoretical support can be found in Sec.~6.4 of Williams et al.~\cite{Williams2009}. Their Eq.~(22) shows that the partial derivative of the range with respect to the reflector’s lunar-fixed \(X\) coordinate includes a \(\cos 2D\) term, together with other periodic terms. When the observations are unevenly distributed over \(D\), the sampled \(\cos D\) and \(\cos 2D\) terms are no longer orthogonal, resulting in partial absorption of the \(\cos D\) signal through adjustment of the reflector \(X\) coordinate. This loss of orthogonality is the same sampling mechanism that produces the harmonic leakage described by Eqs.~(\ref{N32})--(\ref{N38}).

First, consider a fitting model containing only the \(\cos D\) term. The coefficient estimated from the one-way post-fit residuals is:
\begin{equation}
\hat a_D^{\rm post}=\left(\mathbf{h}^{\mathrm{T}}\mathbf{W}\mathbf{h}\right)^{-1}\mathbf{h}^{\mathrm{T}}\mathbf{W}\bm{\upvarrho}_{\rm post}.\label{eq:app_extraction}
\end{equation}

Substituting Eq.~(\ref{eq:app_postfit}) into Eq.~(\ref{eq:app_extraction}) shows that the change in the fitted coefficient caused by an EP-violation signal satisfies \(\Delta\hat a_D^{\rm post}=K a_D^{\rm true}\), where the transfer factor is:
\begin{equation}
K=\left(\mathbf{h}^{\mathrm{T}}\mathbf{W}\mathbf{h}\right)^{-1}\mathbf{h}^{\mathrm{T}}\mathbf{W}\mathbf{M}_{A}\mathbf{h}.\label{eq:app_K}
\end{equation}

The factor \(K\) represents the fraction of the \(\cos D\) signal retained in the post-fit residuals, whereas \(1-K\) represents the fraction absorbed during the parameter adjustment. For ideal sampling in which the columns of \(\mathbf{A}\) are orthogonal to \(\mathbf{h}\), \(\mathbf{A}^{\mathrm{T}}\mathbf{W}\mathbf{h}=0\). It then follows from the definition of \(\mathbf{M}_{A}\) that \(\mathbf{M}_{A}\mathbf{h}=\mathbf{h}\) and hence \(K=1\).

To verify that the attenuation arises from the actual observation sampling, we generated simulated observations for one station and one reflector, with equal weights and no range-bias. The observations were sampled at two-day intervals over 1980--2025, with the UTC time randomly selected within each sampled day. The resulting transfer factor was \(K=0.9992\), corresponding to an absorbed fraction of only \(0.08\%\). This result shows that the station and reflector parameters do not produce appreciable attenuation under well-distributed observations. Therefore, the attenuation found in the actual data mainly arises from the nonuniform distribution of the observation dates and statistical weights over \(D\), which produces correlations between the \(\cos D\) signal and the partial derivatives of the adjusted station and reflector coordinates.

Eq.~(\ref{eq:app_K}) applies when only the \(\cos D\) term is fitted in the residual analysis. If the adopted fitting model contains additional harmonic terms, its design matrix may be written as \(\mathbf{H}\), with \(\mathbf{h}\) as its first column. The corresponding transfer factor for the fitted \(\cos D\) coefficient becomes:
\begin{equation}
K_{H}^{\rm lin}=\left[\left(\mathbf{H}^{\mathrm{T}}\mathbf{W}\mathbf{H}\right)^{-1}\mathbf{H}^{\mathrm{T}}\mathbf{W}\mathbf{M}_{A}\mathbf{h}\right]_{1},
\label{eq:app_general_K}
\end{equation}
where the subscript \(1\) denotes the first element of the resulting coefficient vector. Thus, the same derivation applies to any residual fitting model by replacing the single-template extraction in Eq.~(\ref{eq:app_extraction}) with the adopted matrix \(\mathbf{H}\). For the ALL data set, fitting only the \(\cos D\) term gives \(K=0.776\), whereas using the third-order harmonic model adopted in Table~V, including both cosine and sine terms, gives \(K_H^{\rm lin}=0.753\).

\begin{figure*}[t!]
\centering
\includegraphics[width=0.90\textwidth]{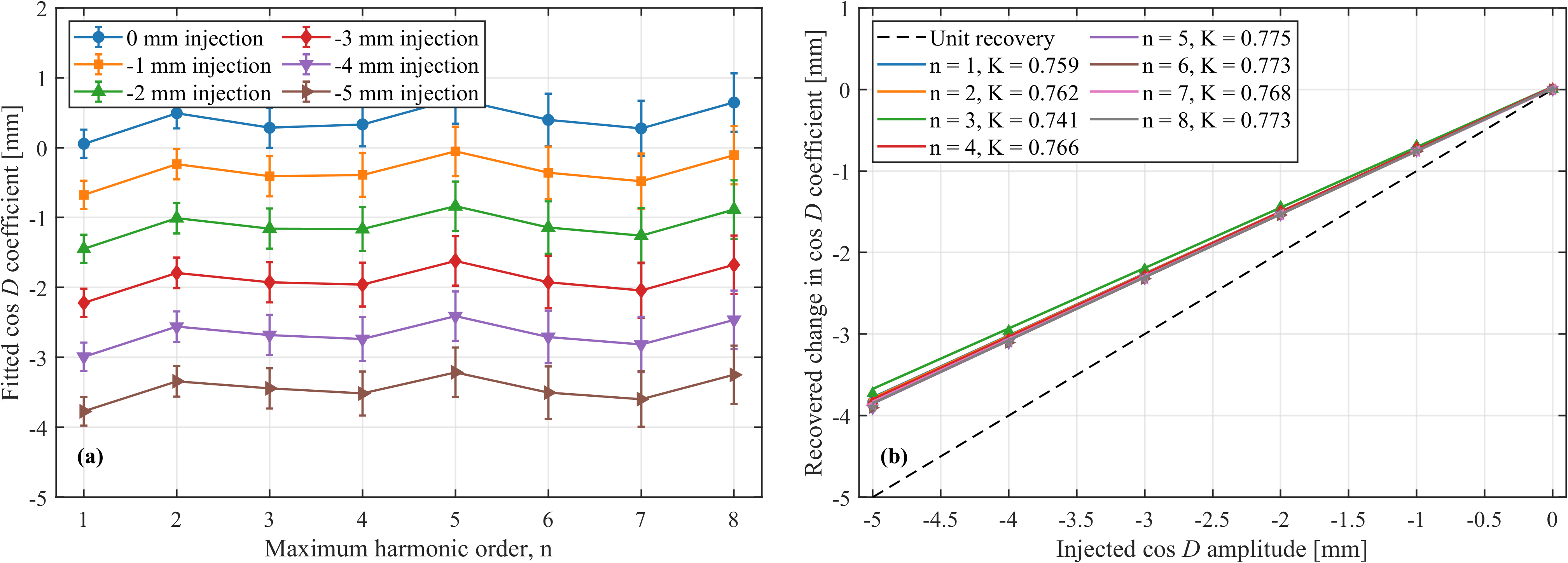}
\caption{Signal-injection tests for the residual-based EP analysis. (a) Fitted post-fit \(\cos D\) coefficients as functions of the maximum harmonic order for different injected one-way amplitudes. (b) Recovered changes relative to the solution without injection as functions of the injected amplitude. 
\label{FigAppInjection}
}
\end{figure*}

In fact, Eqs.~(\ref{eq:app_K}) and (\ref{eq:app_general_K}) describe only the first linearized adjustment. In the actual data reduction, the LLR adjustment was iterated until the parameter corrections became negligible. Therefore, the effective transfer factor for the complete iterative procedure should be determined from signal-injection tests. Signals with known one-way amplitudes \(a_D^{\rm inj}\cos D\) were added before the WLS adjustment. For each injected amplitude, the processing procedure was the same as in the nominal solution. The recovered amplitude change relative to the solution without injection can be calculated as:
\begin{equation}
\Delta\hat a_D^{\rm post}(a_D^{\rm inj})=\hat a_D^{\rm post}(a_D^{\rm inj})-\hat a_D^{\rm post}(0).
\label{eq:app_injection}
\end{equation}

The effective transfer factor was obtained from the slope of the linear relation between \(\Delta\hat a_D^{\rm post}\) and \(a_D^{\rm inj}\). Fig. \ref{FigAppInjection}(a) shows that the fitted \(\cos D\) coefficient varies moderately with the maximum harmonic order, whereas the shifts produced by injected signals remain nearly parallel. After subtracting the result without injection, Fig.~\ref{FigAppInjection}(b) shows a highly linear response. The dashed line indicates unit recovery, which was regarded as a reference. The slopes of the fitted lines give numerical recovery factors ranging from 0.741 to 0.775. For the third-order model, the fitted slope \(K_{eff}=0.741\) agrees well with the linearized analytical prediction of 0.753, which also provides an independent validation of the recovery analysis. Therefore, we adopt the injection-derived result as the transfer factor for the final calibration. Accordingly, the residual-based estimate is calibrated as:
\begin{equation}
\hat a_D^{\rm cal}=\frac{\hat a_D^{\rm post}}{K_{eff}}.
\label{eq:app_calibration}
\end{equation}
This correction is applied to the central value because \(K_{eff}\) describes the response of the WLS estimator to an EP-violation signal. The uncertainty is not determined by \(K_{eff}\) alone, but also depends on the observational noise, the adopted harmonic model, parameter correlations, and remaining modeling errors. The conservative uncertainty treatment adopted in this work is discussed in Sec.~\ref{section4.2.3}. The present calibration accounts only for attenuation introduced by WLS adjustment and does not quantify any absorption that may have occurred during the construction of the adopted ephemerides. 

\endgroup



\begin{thebibliography}{}\label{sec:TeXbooks}
\bibitem{Will2014} C. M. Will, The confrontation between general relativity and experiment, Living Reviews in Relativity. {\bf 17}, 4 (2014).
\bibitem{Fischbach1986} E. Fischbach, D. Sudarsky, A. Szafer, C. Talmadge, and S. H. Aronson, Reanalysis of the Eötvös Experiment, Physical Review Letters. {\bf 56}, 1427 (1986).
\bibitem{Damour2012} T. Damour, Theoretical aspects of the equivalence principle, Classical and Quantum Gravity. {\bf 29}, 184001 (2012).
\bibitem{Fayet2018} P. Fayet, MICROSCOPE limits for new long-range forces and implications for unified theories, Physical Review D. {\bf 97}, 055039 (2018).
\bibitem{Graham2016} P. W. Graham, D. E. Kaplan, J. Mardon, S. Rajendran, and W. A. Terrano, Dark matter direct detection with accelerometers, Physical Review D. {\bf 93}, 075029 (2016).
\bibitem{Carroll2009} S. M. Carroll, S. Mantry, M. J. Ramsey-Musolf, and C. W. Stubbs, Dark-Matter-Induced Violation of the Weak Equivalence Principle, Physical Review Letters. {\bf 103}, 011301 (2009).
\bibitem{Schlamminger2008} S. Schlamminger, K. Y. Choi, T. A. Wagner, J. H. Gundlach, and E. G. Adelberger, Test of the Equivalence Principle Using a Rotating Torsion Balance, Physical Review Letters. {\bf 100}, 041101 (2008).
\bibitem{Wagner2012} T. A. Wagner, S. Schlamminger, J. H. Gundlach, and E. G. Adelberger, Torsion-balance tests of the weak equivalence principle, Classical and Quantum Gravity. {\bf 29}, 184002 (2012).
\bibitem{Zhu2018} L. Zhu, Q. Liu, H.-H. Zhao, Q.-L. Gong, S.-Q. Yang, P. Luo, C.-G. Shao, Q.-L. Wang, L.-C. Tu, and J. Luo, Test of the Equivalence Principle with Chiral Masses Using a Rotating Torsion Pendulum, Physical Review Letters. {\bf 121}, 261101 (2018).
\bibitem{Ross2026} M. P. Ross, E. A. Shaw, C. Gettings, et al., An Improved Torsion Balance Test of the Equivalence Principle Towards the Sun, arXiv:2602.02815 (2026).
\bibitem{Touboul2017} P. Touboul et al., MICROSCOPE Mission: First Results of a Space Test of the Equivalence Principle, Physical Review Letters. {\bf 119}, 231101 (2017).
\bibitem{Touboul2022} P. Touboul, G. Métris, M. Rodrigues, et al., MICROSCOPE Mission: Final Results of the Test of the Equivalence Principle, Physical Review Letters. {\bf 129}, 121102 (2022).
\bibitem{Shapiro1976} I. I. Shapiro, C. C. Counselman, III, and R. W. King, Verification of the Principle of Equivalence for Massive Bodies, Physical Review Letters. {\bf 36}, 1068 (1976).
\bibitem{Williams1976} J. G. Williams, R. H. Dicke, P. L. Bender, et al., New Test of the Equivalence Principle from Lunar Laser Ranging, Physical Review Letters. {\bf 36}, 551 (1976).
\bibitem{Dickey1994} J. O. Dickey, P. L. Bender, J. E. Faller, X. X. Newhall, et al., Lunar Laser Ranging: A Continuing Legacy of the Apollo Program, Science. {\bf 265}, 482 (1994).
\bibitem{Williams1996} J. G. Williams, X. X. Newhall, and J. O. Dickey, Relativity parameters determined from lunar laser ranging, Physical Review D. {\bf 53}, 6730 (1996).
\bibitem{Anderson2001} J. D. Anderson and J. G. Williams, Long-range tests of the equivalence principle, Classical and Quantum Gravity. {\bf 18}, 2447 (2001).
\bibitem{Williams2004a} J. G. Williams, S. G. Turyshev, and D. H. Boggs, Progress in Lunar Laser Ranging Tests of Relativistic Gravity, Physical Review Letters. {\bf 93}, 261101 (2004).
\bibitem{Williams2004b} J. G. Williams, S. G. Turyshev, and T. W. Murphy, Improving LLR Tests of Gravitational Theory, International Journal of Modern Physics D. {\bf 13}, 567 (2004).
\bibitem{Williams2009} J. G. Williams, S. G. Turyshev, and D. H. Boggs, Lunar Laser Ranging Tests of the Equivalence Principle with the Earth and Moon, International Journal of Modern Physics D. {\bf 18}, 1129 (2009).
\bibitem{Williams2012} J. G. Williams, S. G. Turyshev, and D. H. Boggs, Lunar laser ranging tests of the equivalence principle, Classical and Quantum Gravity. {\bf 29}, 284004 (2012).
\bibitem{Muller1998} J. Müller and K. Nordtvedt, Lunar laser ranging and the equivalence principle signal, Physical Review D. {\bf 58}, 062001 (1998).
\bibitem{Hofmann2010} F. Hofmann, J. Müller, and L. Biskupek, Lunar laser ranging test of the Nordtvedt parameter and a possible variation in the gravitational constant, Astronomy and Astrophysics. {\bf 522}, L5 (2010).
\bibitem{Muller2012a} J. Müller, F. Hofmann, and L. Biskupek, Testing various facets of the equivalence principle using lunar laser ranging, Classical and Quantum Gravity. {\bf 29}, 184006 (2012).
\bibitem{Muller2014} J. Müller, L. Biskupek, F. Hofmann, et al., Lunar laser ranging and relativity, Frontiers in Relativistic Celestial Mechanics. {\bf 2}, 99 (2014).
\bibitem{Biskupek2015} L. Biskupek, Bestimmung der Erdorientierung mit Lunar Laser Ranging, Doctoral dissertation, Leibniz Universität Hannover, 2015.
\bibitem{Hofmann2017} F. Hofmann, Lunar Laser Ranging–Verbesserte Modellierung der Monddynamik und Schätzung Relativistischer Parameter, Doctoral dissertation, Leibniz Universität Hannover, 2017.
\bibitem{Hofmann2018} F. Hofmann and J. Müller, Relativistic tests with lunar laser ranging, Classical and Quantum Gravity. {\bf 35}, 035015 (2018).
\bibitem{Zhang2020} M. Zhang, J. Müller, and L. Biskupek, Test of the equivalence principle for galaxy’s dark matter by lunar laser ranging, Celestial Mechanics and Dynamical Astronomy. {\bf 132}, 25 (2020).
\bibitem{Muller2019} J. Müller, T. W. Murphy, U. Schreiber, et al., Lunar Laser Ranging: a tool for general relativity, lunar geophysics and Earth science, Journal of Geodesy. {\bf 93}, 2195 (2019).
\bibitem{Biskupek2021} L. Biskupek, J. Müller, and J. M. Torre, Benefit of New High-Precision LLR Data for the Determination of Relativistic Parameters, Universe. {\bf 7}, 34 (2021).
\bibitem{Zhang2022} M. Zhang, J. Müller, L. Biskupek, et al., Characteristics of differential lunar laser ranging, Astronomy and Astrophysics. {\bf 659}, A148 (2022).
\bibitem{Singh2023} V. V. Singh, J. Müller, L. Biskupek, et al., Equivalence of Active and Passive Gravitational Mass Tested with Lunar Laser Ranging, Physical Review Letters. {\bf 131}, 021401 (2023).
\bibitem{Zhang2024} M. Zhang, J. Müller, and L. Biskupek, Advantages of combining Lunar Laser Ranging and Differential Lunar Laser Ranging, Astronomy and Astrophysics. {\bf 681}, A5 (2024).
\bibitem{Fienga2015} A. Fienga, J. Laskar, P. Exertier, et al., Numerical estimation of the sensitivity of INPOP planetary ephemerides to general relativity parameters, Celestial Mechanics and Dynamical Astronomy. {\bf 123}, 325 (2015).
\bibitem{ViswanathanPhd} V. Viswanathan, Improving the dynamical model of the Moon using lunar laser ranging (LLR) and spacecraft data, Doctoral dissertation, Université Paris sciences et lettres, 2017.
\bibitem{Viswanathan2018} V. Viswanathan, A. Fienga, O. Minazzoli, et al., The new lunar ephemeris INPOP17a and its application to fundamental physics, Monthly Notices of the Royal Astronomical Society. {\bf 476}, 1877 (2018).
\bibitem{Fienga2024} A. Fienga and O. Minazzoli, Testing theories of gravity with planetary ephemerides, Living Reviews in Relativity. {\bf 27}, 1 (2024).
\bibitem{Mariani2024} V. Mariani, O. Minazzoli, A. Fienga, et al., Bayesian test of Brans–Dicke theories with planetary ephemerides: Investigating the strong equivalence principle, Astronomy and Astrophysics. {\bf 682}, A175 (2024).
\bibitem{Pitjeva2014} E. V. Pitjeva and N. P. Pitjev, Development of planetary ephemerides EPM and their applications, Celestial Mechanics and Dynamical Astronomy. {\bf 119}, 237 (2014).
\bibitem{Pitjeva2019} E. Pitjeva, D. Pavlov, D. Aksim, et al., Planetary and lunar ephemeris EPM2021 and its significance for Solar system research, Proceedings of the International Astronomical Union. {\bf 15}, 220 (2019).
\bibitem{Pavlov2020} D. Pavlov, Role of lunar laser ranging in realization of terrestrial, lunar, and ephemeris reference frames, Journal of Geodesy. {\bf 94}, 5 (2020).
\bibitem{Pavlov2023} D. Pavlov and I. Dolgakov, General relativity tests by the dynamics of the Solar system, presented at the Journées des systèmes de référence spatio-temporels, France, 2023.
\bibitem{Pavlov2024} D. Pavlov and I. Dolgakov, Studying the Properties of Spacetime with an Improved Dynamical Model of the Inner Solar System, Universe. {\bf 10}, 413 (2024).
\bibitem{Merkowitz2010} S. M. Merkowitz, Tests of Gravity Using Lunar Laser Ranging, Living Rev Relativ. {\bf 13}, 7 (2010).
\bibitem{Nordtvedt1968a} K. Nordtvedt, Equivalence Principle for Massive Bodies. II. Theory, Physical Review. {\bf 169}, 1017 (1968).
\bibitem{Nordtvedt1995} K. Nordtvedt, The Relativistic Orbit Observables in Lunar Laser Ranging, Icarus. {\bf 114}, 51 (1995).
\bibitem{Nordtvedt1968b} K. Nordtvedt, Testing Relativity with Laser Ranging to the Moon, Physical Review. {\bf 170}, 1186 (1968).
\bibitem{Folkner2014} W. M. Folkner, J. G. Williams, D. H. Boggs, et al., The Planetary and Lunar Ephemerides DE430 and DE431, Report No. 42-196, Jet Propulsion Laboratory, California Institute of Technology, Pasadena, 2014.
\bibitem{Park2021} R. S. Park, W. M. Folkner, J. G. Williams, et al., The JPL Planetary and Lunar Ephemerides DE440 and DE441, The Astronomical Journal. {\bf 161}, 105 (2021).
\bibitem{Pavlov2016} D. A. Pavlov, J. G. Williams, and V. V. Suvorkin, Determining parameters of Moon’s orbital and rotational motion from LLR observations using GRAIL and IERS-recommended models, Celestial Mechanics and Dynamical Astronomy. {\bf 126}, 61 (2016).
\bibitem{Kan2021} M. O. Kan and E. I. Yagudina, Parameters of EPM 2021a Lunar Ephemeris, Transactions of IAA RAS. {\bf 56}, 32 (2021).
\bibitem{Fienga2021} A. Fienga, P. Deram, A. Di Ruscio, et al., INPOP21a planetary ephemerides, Notes Scientifiques et Techniques de l'Institut de Mécanique Céleste. {\bf 110}, (2021).
\bibitem{Vokrouhlicky1997} D. Vokrouhlický, A Note on the Solar Radiation Perturbations of Lunar Motion, Icarus. {\bf 126}, 293 (1997).
\bibitem{Nordtvedt1998} K. Nordtvedt, Optimizing the observation schedule for tests of gravity in lunar laser ranging and similar experiments, Classical and Quantum Gravity. {\bf 15}, 3363 (1998).
\bibitem{Murphy2013} T. Murphy, Lunar laser ranging: the millimeter challenge, Reports on Progress in Physics. {\bf 76}, 076901 (2013).
\bibitem{Murphy2012} T. Murphy, E. Adelberger, J. Battat, et al., Apollo: millimeter lunar laser ranging, Classical and Quantum Gravity. {\bf 29}, 184005 (2012).
\bibitem{Huang2024} K. Huang, Y. Yang, R. Tang, et al., Review of the development of Lunar Laser Ranging, Astronomical Techniques and Instruments. {\bf 1}, 295 (2024).
\bibitem{Zhou2022} C. Zhou, Y. Jia, J. Liu, et al., Scientific objectives and payloads of the lunar sample return mission—Chang’E-5, Advances in Space Research. {\bf 69}, 823 (2022).
\bibitem{Williams2023} J. G. Williams, L. Porcelli, S. Dell’Agnello, et al., Lunar Laser Ranging Retroreflectors: Velocity Aberration and Diffraction Pattern, The Planetary Science Journal. {\bf 4}, 89 (2023).
\bibitem{Yu2026} W. Yu, X. Han, X. Lin, et al., Lunar laser ranging localization and independent validation of the NGLR-1 retroreflector coordinates, Measurement. {\bf 283}, 122094 (2026).
\bibitem{He2018} Y. He, Q. Liu, H. Z. Duan, et al., Manufacture of a hollow corner cube retroreflector for next generation of lunar laser ranging, Research in Astronomy and Astrophysics. {\bf 18}, 136 (2018).
\bibitem{Cao2024} J. Cao, R. Tang, K. Huang, et al., Analysis of the Effect of Tilted Corner Cube Reflector Arrays on Lunar Laser Ranging, Remote Sensing. {\bf 16}, 3030 (2024).
\bibitem{OCAwebsite} C. Barache, S. Bouquillon, A. Bourgoin, T. Carlucci, and G. Francou, Paris Observatory Lunar Analysis Centre, Observatoire de Paris (SYRTE) - CNRS UMR 8630, \url{http://polac.obspm.fr}.
\bibitem{IERS2010} G. Petit and B. Luzum, IERS conventions (2010), Technical report, Bureau International des Poids et Mesures, Sèvres, France, 2010.
\bibitem{Bizouard2019} C. Bizouard, S. Lambert, C. Gattano, et al., The IERS EOP 14C04 solution for Earth orientation parameters consistent with ITRF 2014, Journal of Geodesy. {\bf 93}, 621 (2019).
\bibitem{C04}International Earth Rotation Service, Earth Orientation Parameters Data, 2024, \url{https://hpiers.obspm.fr/iers/series/opa/eopc04}
\bibitem{eop2} Jet Propulsion Laboratory, Earth Orientation Files, 2026, \url{https://eop2-external.jpl.nasa.gov/}
\bibitem{Chin2009} T. M. Chin, R. S. Gross, D. H. Boggs, et al., Dynamical and Observation Models in the Kalman Earth Orientation Filter, The Interplanetary Network Progress Report. {\bf 42-176}, 1 (2009).
\bibitem{Hohenkerk2012} C. Y. Hohenkerk, SOFA and the algorithms for transformations between scales \& between systems, Proceedings of the Journées 2011 Systèmes de référence spatio-temporels, Vienna, 21 (2012).
\bibitem{Wahr1981} J. M. Wahr, Body tides on an elliptical, rotating, elastic and oceanless earth, Geophysical Journal. {\bf 64}, 677 (1981).
\bibitem{Williams2015} J. G. Williams and D. H. Boggs, Tides on the Moon: Theory and determination of dissipation, Journal of Geophysical Research: Planets. {\bf 120}, 689 (2015).
\bibitem{Farrell1972} W. E. Farrell, Deformation of the Earth by surface loads, Reviews of Geophysics. {\bf 10}, 761 (1972).
\bibitem{Scherneck2026} H. Scherneck and M. S. Bos, the Ocean Loading Provider, Chalmers University of Technology, 2026, \protect\url{https://barre.oso.chalmers.se/loading}.
\bibitem{Wahr1985} J. M. Wahr, Deformation induced by polar motion, Journal of Geophysical Research. {\bf 90}, 9363 (1985).
\bibitem{Desai2002} S. D. Desai, Observing the pole tide with satellite altimetry, Journal of Geophysical Research. {\bf 107}, 3186 (2002).
\bibitem{Petrov2004} L. Petrov and J.-P. Boy, Study of the atmospheric pressure loading signal in VLBI observations, Journal of Geophysical Research. {\bf 109}, B03405 (2004).
\bibitem{Petrov2015} L. Petrov, The International Mass Loading Service, arXiv:1503.00191 (2015).
\bibitem{Pearlman2002} M. R. Pearlman, J. J. Degnan, and J. Bosworth, The international laser ranging service, Advances in Space Research. {\bf 30}, 135 (2002).
\bibitem{Williams2013} J. G. Williams, D. H. Boggs, and W. M. Folkner, DE430 Lunar Orbit, Physical Librations and Surface Coordinates, JPL Interoffice Memorandum 335-JW,DB,WF-20130722-016, Jet Propulsion Laboratory,  2013.
\bibitem{Mendes2004} V. Mendes and E. C. Pavlis, High-accuracy zenith delay prediction at optical wavelengths, Geophysical Research Letters. {\bf 31}, L14602 (2004).
\bibitem{Jiang2026} Z. T. Jiang, C. G. Qin, W. S. Huang, et al., Impact of Atmospheric Delay on Equivalence Principle Tests Using Lunar Laser Ranging, Symmetry. {\bf 18}, 50 (2026).
\bibitem{Pavlov2026} D. A. Pavlov, private communication (2026).
\bibitem{Shapiro1964} I. I. Shapiro, Fourth Test of General Relativity, Physical Review Letters. {\bf 13}, 789 (1964).
\bibitem{Montenbruck2000} O. Montenbruck and T. Pfleger, Astronomy on the Personal Computer, Springer Verlag, Heidelberg, 4th edition, (2000).
\bibitem{Williams2020} J. G. Williams and D. H. Boggs, The JPL Lunar Laser Range Model 2020, JPL Interoffice Memorandum IOM 335N-20-01, 2020.
\bibitem{Muller2007} J. Müller and L. Biskupek, Variations of the gravitational constant from lunar laser ranging data, Classical and Quantum Gravity. {\bf 24}, 4533 (2007).
\bibitem{Muller2014c} J. Müller, L. Biskupek, F. Hofmann, and E. Mai, Lunar laser ranging and relativity, in Frontiers in Relativistic Celestial Mechanics: Volume 2 Applications and Experiments, edited by S. M. Kopeikin (De Gruyter, Berlin, 2014), pp. 103-156.
\end{thebibliography}
\end{document}